\documentclass[aps,prx,twocolumn,superscriptaddress,longbibliography]{revtex4-2}
\usepackage{amsmath,amssymb}
\usepackage[pdftex]{hyperref,graphicx}
\hypersetup{colorlinks = true, urlcolor = blue, linkcolor = blue, citecolor = blue}
\usepackage{physics}
\usepackage{xcolor}
\usepackage{bm}
\usepackage{tabularx}
\usepackage{tikz}
\usepackage{adjustbox}
\usepackage{makecell}
\usepackage{subfig}

\newcommand{\hl}[1]{#1}
\graphicspath{{/mnt/d/CMTC/bothlead/Rp56/}}

\begin{document}
\title{Disordered Majorana nanowires: Studying disorder without any disorder}
\author{Haining Pan}
\affiliation{Department of Physics and Astronomy, Center for Materials Theory, Rutgers University, Piscataway, New Jersey 08854 USA}
\author{Sankar Das Sarma}
\affiliation{Condensed Matter Theory Center and Joint Quantum Institute, Department of Physics, University of Maryland, College Park, Maryland 20742, USA}

\begin{abstract}
  The interplay of disorder and short finite wire length is the crucial physics hindering progress in the semiconductor-superconductor nanowire platform for realizing non-Abelian Majorana zero modes. 
  Disorder effectively segments the nanowire into isolated patches of quantum dots which act as subgap Andreev bound states often mimicking Majorana zero modes.
  In this work, we propose and develop a new theoretical approach to model the disorder, effectively a spatially varying effective mass model, which does not rely on incorporating the unknown microscopic details of disorder into the Hamiltonian. 
  This model effectively segments the wire into multiple quantum dots, characterized by highly enhanced effective mass at impurity sites leading to the segmentation of the wire into effective random quantum dots.
  We find that this model can qualitatively and quantitatively reproduce the disorder physics, providing a crystal clear way to understand the effects of disorder by comparing the mean free path to the superconducting coherence length.
  In addition, this model allows precise control over the disorder regime, enabling us to evaluate the reliability of topological invariants in predicting Majorana zero modes.
  We find that topological invariants alone may yield a significant false-positive rate as indicators for topology in the actual wire with increasing disorder strength.
  Therefore, we propose a set of new indicators to characterize the spatial distribution of the zero-energy state, emphasizing the key necessity for isolated Majorana zero modes localized at the wire ends. 
  Employing this set of new indicators for stringent characterizations, we explore their experimental relevance to the measured differential conductance spectra.
  Our findings highlight the critical role of isolated localized states, beyond the topological invariant, in identifying topological Majorana zero modes.
  We believe that this approach is a powerful tool for studying realistic Majorana nanowires where disorder and short wire length obfuscate the underlying topological physics.
\end{abstract}

\maketitle

\section{Introduction}

The semiconductor-superconductor (SM-SC) nanowire has been proposed as a platform for pursuing the Majorana zero modes (MZMs) a decade ago~\cite{lutchyn2010majorana,oreg2010helical,sau2010generic,sau2010robustness,dassarma2023search}. 
However, the experimental advances have been suffering from the ambiguity in the interpretation of the observed zero-bias conductance peak (ZBCP) since 2012~\cite{das2012zerobias,deng2012anomalous,mourik2012signatures,churchill2013superconductornanowire,finck2013anomalous,nichele2017scaling,zhang2017quantized,zhang2021large,yu2021nonmajorana,song2021large,microsoftquantum2023inasal}, although such conductance transport spectroscopy has remained the main experimental tool to search for MZMs in nanowires.
The main challenge is whether the observed ZBCP which manifests some levels of nearly quantized conductance can be attributed to the theoretically predicted Majorana zero modes, or other alternative trivial explanations such as inhomogeneous potential~\cite{kells2012nearzeroenergy,liu2017andreev,pan2020physical} and disorder~\cite{brouwer2000localization,brouwer2011topological,brouwer2011probability,liu2012zerobias,bagrets2012class,akhmerov2011quantized,sau2013density,takei2013soft,adagideli2014effects,pekerten2017disorderinduced,pan2020physical,ahn2021estimating,woods2021chargeimpurity,dassarma2021disorderinduced,dassarma2023density,dassarma2023spectral,dassarma2023search,taylor2023machine, brzezicki2017driving, barmanray2021symmetrybreaking}.
The fundamental reason for this challenging problem is the interplay of disorder and short wire length, both are present in the current experimental nanowires.
Therefore, in this paper, we aim to answer the heart of this challenging question: How to cleanly define the topology in disordered finite-size systems? 
This is what has hindered progress in the subject and has led to controversies~\cite{dassarma2023search}.

To answer this question, we first need to understand the disorder physics in the SM-SC nanowire in a controllable and transparent way.
Therefore, we propose a new model that reproduces the same disorder physics~\cite{sau2013density,pan2020physical,pan2021threeterminal,dassarma2021disorderinduced} without explicitly putting any disorder in the chemical potential.
The motivation is that we know that disorder plays a crucial role in the current experiments, and the disorder is likely to split the 1D nanowire system into isolated quantum dots (i.e., short wire segments without any coherence among them). 
Therefore, the microscopic (which is unknown anyway) details of the disorder (e.g., charged impurities or not, long range versus short range, correlated disorder or not) are no longer important--- the only physical effect of disorder is that it leads to a spatially-dependent random effective mass as the 1D wire is effectively segmented into a bunch of random dots.
We believe that this model is also very physical since the presence of quenched impurities is likely to lead to segmentation, dividing the wire into a system of short wires of random lengths and positions.
Therefore, we directly segment the wire using a spatially varying effective mass, where the effective mass is large at the sites of impurities, and uniform in the rest of the wire.
Note that the ``impurities" here have nothing to do with real impurities in the wire causing disorder--- here the ``impurities" are just a formal term implying the sites where the effective mass changes randomly, staying uniform in between the impurities. In other words, the number of impurities ($n$) defines how many segments the wire breaks into, namely $n+1$ segments for a wire of length $L$, implying also that the mean free path by definition is then $L/(n+1)$.
The biggest advantage of this method is the transparency in the underlying disorder physics without worrying about the microscopic details of the disorder itself, allowing us to directly parametrize the strength of the disorder, through the variations in the effective mass with no explicit disorder term in the Hamiltonian, which was a hard problem before because the disorder details are unknown in nanowires--- in fact, even the actual source of the disorder is unknown.

The hallmark of the current approach is that the disorder-induced mean free path (MFP), which is what matters, naturally emerges as it is the average distance that a particle can travel before hitting an impurity (and changing its mass suddenly). 
This distance is precisely the average length scale of the quantum dots that the wire is segmented into, and thus the essential ingredient of disorder, the MFP, is built into the model without getting bogged down in the unnecessary details of the actual disorder potential itself.
Therefore, we can directly compare this mean free path with the superconducting coherence length, which is also the length scale of the localization of Majorana zero modes.
If the mean free path is larger than the superconducting coherence length, the wire is then in the weak disorder regime, where the physics is expected to be qualitatively similar to the pristine wire.
If the mean free path is smaller than the superconducting coherence length, the wire is then in the strong disorder regime, where the topology is expected to be suppressed.
If the two length scales are comparable, the wire is then in the intermediate disorder regime, where the physics is complicated and depends on the details. 

We apply this new model to the recent experiment from Microsoft Quantum on InAs-Al nanowires~\cite{microsoftquantum2023inasal} and find that we can reproduce both the ``positive" (i.e., topological) and ``negative'' (i.e., trivial) results in the experiment by simply changing the number (or the density) of impurities (or segments), which directly controls the mean free path.
Our results show that the positive results belong to the weak disorder regime, while the negative results belong to the intermediate disorder regime, which is consistent with our previous analysis~\cite{dassarma2023spectral,dassarma2023density}.
Of course, for even stronger disorder with extremely short MFP, the system is completely Anderson localized, and there is no topology whatsoever~\cite{motrunich2001griffiths}.

Besides the success in reproducing the experiment (and introducing a physically transparent model eliminating any disorder details), we also use this model to explore other interesting physics, for example, we can make a nontopological wire topological by simply increasing the superconducting gap with other parameters fixed, which effectively decreases the superconducting coherence length, and thus satisfies the hierarchy of the length scale to enter the topological regime, happening when the SC coherence length decreases below the MFP.

After establishing the model, we then return to the fundamental question by studying the efficacy of using the topological invariant in order to predict the existence of the topological Majorana zero modes.
This is particularly important because most previous works are based on the topological invariant to determine the topology of a zero energy state~\cite{kitaev2001unpaired}.
We find that the false-positive rate of using the topological invariant to predict MZMs increases as the disorder strength increases, which is consistent with the previous work~\cite{dassarma2023spectral,dassarma2023density}.
To overcome this problem of false positives, we propose looking at the isolated Majorana zero modes at the ends of the wire from the local density of states (LDOS).
This is because, on the one hand, for any topological MZMs to be useful in fusion and braiding, the nonabelian state has to reside at the end of the wire; on the other hand, we will also establish, using statistical analysis, that the existence of the isolated localized state (i.e., absence of the bulk state) at zero energy has a direct correlation with the topology, and can serve as a definitive criterion to predict the topological states with isolated MZMs at ends of the wire in finite short wires.  
We emphasize that the topological invariant is a unique Majorana criterion only for very long wires with little disorder, and must be complemented by other operational topological indicators in finite disordered wires, a point first made in Ref.~\cite{dassarma2023spectral}.

With this additional criterion of isolated localized end states, allowing us to obtain a more accurate prediction of the topological localized MZMs, we study the connection between these `theoretical indicators' (e.g., topological invariant and other indicators based on the LDOS to ensure the isolated localized MZMs) to `experimental observables' (i.e., differential conductance).
We raise several questions of concern in the field, for example, `Does the quantized ZBCP at both ends always lead to the presence of topological localized MZMs?'--- If not, `is this absence of topological localized MZMs because of the absence of topology or the absence of localized state (i.e., multiple topological patches in the bulk of the wire)'; `Does the state with nontrivial topological invariant but without isolated localized state always lead to ZBCP at one end of the wire smaller than the quantized value?'
We provide answers to these questions using a rigorous statistical analysis.

The rest of the paper is organized as follows.
In Sec.~\ref{sec:model}, we introduce the model of the spatially varying effective mass in the nanowire.
Our main results start in Sec.~\ref{sec:results_varyingmass}, where we first present the pristine limit in Sec.~\ref{sec:pristine} (as a baseline result for comparison), and then simulate the recent experiment~\cite{microsoftquantum2023inasal} from Microsoft Quantum on the InAs-Al nanowire to demonstrate different disorder regimes in Sec.~\ref{sec:MSFT}.
We then study the dependence of the strength of the disorder in Sec.~\ref{sec:dependence_of_k}, and the wire length in Sec.~\ref{sec:dependence_of_L}.
To end this section, we show that we can bring a nontopological wire back to a topological wire by only increasing the proximitized SC gap in Sec.~\ref{sec:increase_SC_gap}.
In Sec.~\ref{sec:results_indicators}, we propose several new indicators to predict the topological localized MZMs. We begin with the definition for these indicators in Sec.~\ref{sec:definition_indicators} and then show their benchmark results ranging from the pristine limit to the strong disorder regime in Sec.~\ref{sec:benchmark}.
With these new indicators, we study the false-positive rate of only using the topological invariant to predict the topological localized MZMs in Sec.~\ref{sec:predictive_power}.
Finally, we study the connection of these `theoretical indicators' to the experimental observables in Sec.~\ref{sec:exp_relevance}.
Our discussion and conclusion are in Sec.~\ref{sec:conclusion}.
We also provide a detailed appendix.
In Appendix~\ref{app:MSFT}, we provide additional simulations for the experimental data.
In Appendix~\ref{app:dependence_of_k} and~\ref{app:dependence_of_L}, we provide more results for the dependence on the strength of the impurities, and the wire length.
In Appendixes~\ref{app:indicator}, we show more benchmark results using the new indicators.
In Appendixes~\ref{app:localized_MZM} and~\ref{app:quantized_ZBCP}, we provide more results that demonstrate the statistics of the special cases of interest, such as `localized MZMs without quantized ZBCP' and `quantized ZBCP without localized MZMs', respectively.

\section{Spatially varying effective mass model}\label{sec:model}

We start with the standard 1D single-band model of a superconductor-semiconductor heterostructure nanowire~\cite{dassarma2023search,lutchyn2010majorana,oreg2010helical,sau2010generic,sau2010robustness} with a large Rashba-type spin-orbit coupling (SOC) in the presence of a magnetic field. 
The standard Hamiltonian is described by
\begin{equation}\label{eq:H_NW}
    H = H_K + H_{\text{SOC}} + H_Z + H_{\text{SC}},
\end{equation}
where the kinetic term is
\begin{equation}\label{eq:H_K}
    H_K= \sum_{\sigma=\uparrow/\downarrow} \int_0^L dx \psi_{\sigma}^\dagger(x)\left( -\frac{\hbar^2}{2m^*}\frac{d^2}{dx^2} -\mu \right) \psi_{\sigma}(x),
\end{equation}
the Rashba SOC term is
\begin{equation}
    H_{\text{SOC}} =  \sum_{\sigma,\sigma'=\uparrow/\downarrow}\int_0^L dx \psi_{\sigma}^\dagger(x) \left( i\hbar \alpha \hat{\sigma}_y \frac{d}{dx} \right)_{\sigma,\sigma'} \psi_{\sigma'}(x),
\end{equation}
the Zeeman term is
\begin{equation}
    H_Z = \sum_{\sigma,\sigma'=\uparrow/\downarrow} \int_0^L dx  \psi_{\sigma}^\dagger(x) V_Z~ (\hat{\sigma}_z)_{\sigma,\sigma'} \psi_{\sigma'}(x),
\end{equation}
and the SC term is 
\begin{equation}
    H_{\text{SC}} = \int_0^L dx \left( \Delta \psi_{\uparrow}^\dagger(x) \psi_{\downarrow}^\dagger(x) + \text{h.c.} \right).
\end{equation}
Here, $m^*$, $\mu$, $\alpha$, $V_Z$, $\Delta$, and $L$ are effective mass, chemical potential, Rashba-type SOC strength, Zeeman energy, proximitized superconducting gap, and wire length, respectively. $\sigma$ is the spin index, $\hat{\sigma}_{x,y,z}$ are three Pauli matrices acting on the spin space, and $\psi_{\sigma}^\dagger(x)$ creates an electron with spin $\sigma$ at position $x$.

Based on the standard Hamiltonian, we modify the kinetic part Eq.~\eqref{eq:H_K} by introducing a spatially varying effective mass, which gives a natural way to segment the wire into a bunch of quantum dots. 
The effective `cut' in the nanowire is then modeled by the spike in the spatial profile of effective mass because it effectively reduces the hopping amplitude across both sides of the cuts.
Therefore, the original model in Eq.~\eqref{eq:H_NW} has a different kinetic term $H_K$ as per, 
\begin{equation}
    H_K=\sum_{\sigma=\uparrow/\downarrow}  \int dx  \psi_{\sigma}^\dagger(x)\left[ -\frac{\hbar^2}{2}\frac{d}{dx}\left( \frac{1}{m^*(x)}\frac{d}{dx}\right) -\mu\right] \psi_{\sigma}(x),
\end{equation}
where $m^*(x)$ is the spatial profile of the effective mass. Note that the effective mass $m^*(x)$ is inside the derivative to ensure the hermiticity of the kinetic term.

Depending on the details of the effective mass inhomogeneity (to be described below),  the wire of length $L$ will be subdivided into a number of incoherent segments, i.e., dots.
Intuitively, the fundamental physics arises from the fact that disorder can randomly create local impurities with large effective mass, which effectively reduce the electron mobility in their surroundings, and thus inhibit coherent electron transport across the regions (``dots'') separated by these impurities.
This mechanism of segmenting the wire due to the disorder provides a more physical way of incorporating disorder compared to manually putting disorder into the chemical potential term in the Hamiltonian, particularly since the detailed disorder is unknown (and the full details of the disorder are irrelevant for the Majorana physics).

To numerically solve the spatially varying nanowire, we construct the corresponding effective tight-binding model in the usual manner by discretizing the wire with a fictitious lattice constant $a$ (where $L=N a$, \hl{and $a=10$ nm such that $N=300$ for $L=3~\mu$m, and $N=1000$ for $L=10~\mu$m , which is large enough for numerical convergence}) and replacing the differential operator with the finite difference. The kinetic term Eq.~\eqref{eq:H_K}  then becomes 
\begin{equation}\label{eq:H_K_discrete}
    \begin{split}
        \tilde{H}_K=&\sum_{i,\sigma} -\frac{\hbar^2}{2m^*_{i+\frac{1}{2}} a^2} \left( c_{i,\sigma}^\dagger c_{i+1,\sigma} + \text{h.c.} \right) \\
        +& \left( \frac{\hbar^2}{2m^*_{i+\frac{1}{2}} a^2} + \frac{\hbar^2}{2m^*_{i-\frac{1}{2}} a^2} -\mu \right) c_{i,\sigma}^\dagger c_{i,\sigma},
    \end{split}
\end{equation}
Here, the $m^*_{i+\frac{1}{2}}$ is the discretized version of the effective mass defined at the bond connecting site $i$ and $i+1$, and $c_{i,\sigma}^\dagger$ creates an electron with spin $\sigma$ at site $i$.

Therefore, it becomes evident that the large local effective mass at $m^*_{i+\frac{1}{2}}$ results in a `cut' in the nanowire through the following two processes: 
(1) the suppression of hopping amplitude between site $i$ and $i+1$; 
(2) the increase of the onsite chemical potential at site $i$.
In the following calculations, we will model the spatially varying effective mass profile using two key parameters: (1) the number of the impurities $n$ scattered randomly in the wire, creating possibly $n +1$ segments or quantum dots; (2) the strength of these impurities $k$, which controls the magnitude of the effective mass at the impurities, i.e., $\delta m^*/m^*\sim U(0, k)$, where $U(a,b)$ is a continuous uniform distribution between $a$ and $b$.  
For simplicity, we assume that the impurities only occupy one site, i.e., the same length scale as the lattice constant $a$. Relaxing this criterion should not change any of our conclusions.

The SOC and Zeeman term in the Hamiltonian Eq.~\eqref{eq:H_NW} can be discretized similarly and remains the same form as the standard model in~\cite{dassarma2023search,lutchyn2010majorana,oreg2010helical,sau2010generic,sau2010robustness}. 
To incorporate the SC term, we integrate out the SC degrees of freedom and obtain the Green's function of the SM-SC system $G(\omega) = \left[ \omega +  i\eta - (H_K+H_{\text{SOC}}+H_Z +\Sigma(\omega)) \right]^{-1}$, where $\omega$ is the energy, $\eta$ is an infinitesimal value to ensure causality, and $\Sigma(\omega)$ is the self-energy of the interface of SC and SM. The self-energy expanded in the Bogoliubov-de Gennes (BdG) basis can be treated as $-\gamma \frac{\omega+\Delta_0 \tau_x}{\sqrt{\Delta_0^2-\omega}}$~\cite{stanescu2010proximity}, where $\gamma$ is the effective SC-SM coupling strength, $\Delta_0$ is the parent SC gap size, and $\tau_x$ is the Pauli matrix $x$ acting on the particle-hole space. Moreover, in order to simulate the suppression of the parent SC gap due to the magnetic field, we manually make the parent SC gap collapse following $\Delta_0(V_Z)= \Delta_0(V_Z=0)\sqrt{1-(V_Z/V_{Z0})^2}$.
These are all standard methodologies for studying SC-SM-based Majorana nanowires.   

In the following, unless otherwise stated, we choose the typical parameters close to InAs-Al SM-SC nanowire from Ref.~\onlinecite{microsoftquantum2023inasal}: the uniform effective mass $m^*$ is 0.032 $m_e$ (the impurities thus induce the effective mass of $k m^*$), the SOC strength $\alpha$ is 0.84 eV$\cdot$\AA, effective $g$-factor is 11.5 \hl{(which converts the magnetic field, $B$, to Zeeman field, $\frac{1}{2}\mu_BgB$, with $\mu_B$ being Bohr magneton)}, the parent SC gap at zero magnetic fields $\Delta_0(V_Z)$ is 0.3 meV with a critical field $V_{Z0}=2$ meV, the effective SC-SM coupling strength is 0.2 meV, the wire length $L$ is 3 $\mu$m (except for Sec.~\ref{sec:dependence_of_L}), and zero dissipation and zero temperature are assumed. Since the chemical potential $\mu$ is not directly reported in the experiment~\cite{microsoftquantum2023inasal}, we conventionally choose $\mu=0.5$ meV, which leads to a topological phase transition at $V_{Zc}=\sqrt{\mu^2+\gamma^2}\sim 0.54$ meV ($V_{Z}>V_{Zc}$ is the topological regime), and pristine superconducting coherence length $\xi_{\text{SC}}\sim0.36~\mu$m (estimated from the Majorana localization length~\cite{cheng2009splitting,dassarma2023density}) in the clean wire in the thermodynamic limit.

\section{Results for the spatially varying effective mass model}\label{sec:results_varyingmass}
In this section, we present the numerical results of the spatially varying effective mass model. 
We compute the following metrics for the numerical results:
\begin{enumerate}
    \item The local differential conductance spectrum $G_{\text{LL}}$ ($G_{\text{RR}}$) for the reflection at the left (right) lead, and the nonlocal conductance spectrum $G_{\text{LR}}$ ($G_{\text{RL}}$) for the transmission from the right to the left lead (left to the right lead).
    We first attach two SM leads to both ends of the wire, where the interface effectively creates a Schottky barrier whose potential height is 5 meV~\cite{liu2017andreev}. We then compute the S matrix of both leads with the help of a Python package KWANT~\cite{groth2014kwant} following Blonder-Tinkham-Klapwijk formalism~\cite{blonder1982transition}. 
    We refer readers to a series of earlier works~\cite{setiawan2015conductance,liu2017andreev,rosdahl2018andreev,pan2020physical,pan2021threeterminal} for the detailed calculation as it is a standard procedure. 
    This differential conductance spectrum is the only metric that can be directly measured and reported in current experiments.
    \item The local density of states (LDOS) defined as $\rho(\omega,x_i)=-\frac{1}{\pi}\text{Im}\left[\tr_{\sigma,\tau} G(\omega)\right]_{i,i}$. This metric can serve as a theoretical tool to study the spatial distribution of the state, though not necessarily directly accessible in the transport experiment, but, in principle, LDOS can be measured by scanning tunneling spectroscopy (STS).
    \item The topological visibility (TV)~\cite{fulga2011scattering,dassarma2016how} $Q=\det(S_{\text{LL}})=\det(S_{\text{RR}})$, where $S_{\text{LL}}$ ($S_{\text{RR}}$) is the block in S matrix at zero energy that describes the local reflection process at the left (right) lead, and $\det$ is the determinant. This metric theoretically probes the change from the normal reflection to the Andreev reflection, and thus serves as an indicator for topological quantum phase transition (TQPT), where $Q$ change from +1 (trivial regime) to -1 (topological regime) at the TQPT in the pristine wire in the thermodynamic limit. 
    \item The thermal conductance $\kappa=\kappa_0 \tr(S_{\text{LR}}S_{\text{LR}}^\dagger)=  \kappa_0 \tr(S_{\text{RL}}S_{\text{RL}}^\dagger)$, where $\kappa_0$ is the quantized thermal conductance $\kappa_0=\frac{\pi^2 k_B^2 T}{6h}$~\cite{akhmerov2011quantized,fulga2011scattering} ($T$ is the temperature and $k_B$ is the Boltzmann constant), and $S_{\text{LR}}$ ($S_{\text{RL}}$) is the block in S matrix at zero energy that describes the transmission from the right to the left lead (the left to the right lead).  This metric indicates the TQPT by a peak of $\kappa/\kappa_0=1$ in the pristine wire in the thermodynamic limit. In contrast to the TV, which is purely theoretical, this, in principle, can be measured.
\end{enumerate}

\subsection{Pristine wire}\label{sec:pristine}

\begin{figure*}[ht]
    \centering
    \includegraphics[width=6.8in]{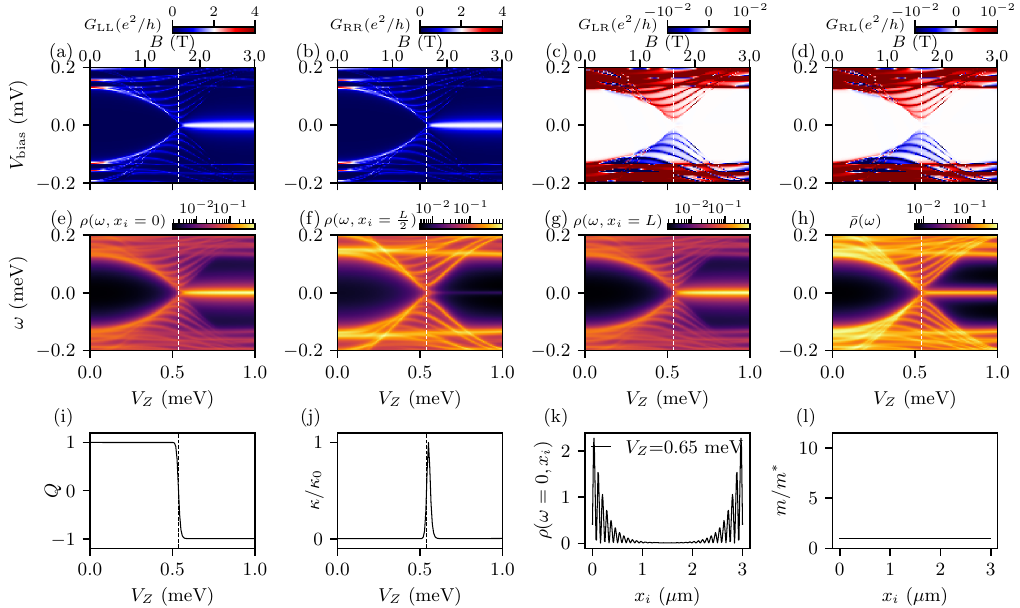}
    \caption{
    Pristine wire with $n=0$ and $L=3~\mu$m. 
    (a-d) Local ($G_{\text{LL}}$ and $G_{\text{RR}}$) and nonlocal ($G_{\text{LR}}$ and $G_{\text{RL}}$) conductance spectra. The bottom (top) axis shows the Zeeman (magnetic) field. The vertical dashed line indicates nominal TQPT.
    (e-g) LDOS at left, midpoint, and right end. 
    (h) Total DOS.
    (i-j) Topological visibility $Q$ and thermal conductance $\kappa$.
    (k) LDOS at zero energy for $V_Z=0.65$ meV.
    (l) Spatial profile of effective mass. 
    }
    \label{fig:pristine}
\end{figure*}
We start with a pristine wire to provide a standard reference (i.e., the baseline for comparison with realistic disordered systems) for the metrics mentioned above. 
In Fig.~\ref{fig:pristine}(a-b), we show the local conductance spectra $G_{\text{LL}}$ and $G_{\text{RR}}$ as a function of the Zeeman field and bias voltage, respectively, where the quantized ZBCPs at $2e^2/h$ appear beyond the TQPT critical field $V_{Zc}\sim 0.54$ meV at both ends.
In Fig.~\ref{fig:pristine}(c-d), we present the nonlocal conductance spectra $G_{\text{LR}}$ and $G_{\text{RL}}$, where the topological gap closes and reopens at the same critical field $V_{Zc}$.
In Fig.~\ref{fig:pristine}(e-g), we present the LDOS at the left (0 $\mu$m), midpoint (1.5 $\mu$m), and right end (3 $\mu$m), which show two zero energy states at the wire ends and no zero energy state in the bulk of the wire. 
In Fig.~\ref{fig:pristine}(h), we present the total DOS, the spatial integral of the LDOS defined above.
In Fig.~\ref{fig:pristine}(i-j), we present the topological visibility $Q$ and thermal conductance $\kappa$, respectively, which both show the topological phase transition at the same critical field $V_{Zc}\sim 0.54$, indicated by the abrupt change of TV from $+1$ to $-1$, and the peak in the thermal conductance quantized at $\kappa/\kappa_0=1$.
In Fig.~\ref{fig:pristine}(k), we present the line cut of the LDOS of a zero energy state in the topological regime at $V_Z=0.65$ meV. The two Majorana modes are localized at the wire ends, where the envelope of oscillation can be used to estimate the lower bound of Majorana localization length (which is also the SC coherence length) $\xi_{\text{SC}}$.
We note that $V_Z=0.65$ meV has the largest topological gap. The estimated $\xi_{\text{SC}}\sim 0.36 ~\mu$m. 
In Fig.~\ref{fig:pristine}(l), we present the spatial profile of the effective mass, which is uniform in the pristine wire.

In order to make the wire enter the topological regime, one should ensure the following hierarchy of length scales~\cite{dassarma2023density,dassarma2023spectral},
\begin{equation}\label{eq:hierarchy}
    \xi_{\text{SC}} < \xi_{\text{MFP}} \lesssim L,
\end{equation} 
where $\xi_{\text{SC}}$ is the SC coherence length and $\xi_{\text{MFP}}$ is the mean free path. 
Note that the inequality with respect to the wire length simply cuts off the MFP at $L$ since the electrons cannot travel a distance larger than the wire length by definition, so $L$ is an effective stringent upper limit on the MFP.
In the pristine wire, since there is no inhomogeneity along the wire, the mean free path can be at most the wire length $L=3~\mu$m, which manifestly satisfies the hierarchy as $\xi_{\text{SC}}\sim 0.36 ~\mu$m.
It is obvious that, for short disordered wires, the inequality defined by Eq.~\eqref{eq:hierarchy} becomes increasingly unachievable, making it a huge challenge to enter the topological regime.  We mention that the inequality defined by Eq.~\eqref{eq:hierarchy} is strongly violated by all the early Majorana nanowire experiments except possibly the recent Microsoft experiment, where the two length scales are likely to be comparable allowing some minimal topological gap to emerge~\cite{microsoftquantum2023inasal,dassarma2023spectral,dassarma2023density}.

This relation provides a very simple and intuitive rule of thumb to characterize the strength of effective disorder, which was not easy to do in the earlier models of using an explicit disorder potential in the chemical potential before. 
Namely, in the previous model~\cite{pan2020physical,pan2021threeterminal,dassarma2021disorderinduced,dassarma2023density} where the onsite potential disorder is manually put throughout the wire, one does not have a straightforward way to estimate the boundary between the weak and intermediate disorder regime, or the intermediate and strong disorder regime.
The fundamental reason is that the variance of disorder is not directly mapped to the variance of zero energy states, and therefore one has to generate a series of random disorder configurations (and simulate an ensemble of conductance spectra for each disorder realization) to study the variance of zero energies, which is inversely related to the relaxation time $\tau$, and finally approximately estimate the mean free path $\xi_{\text{MFP}}$ from a huge collection of simulated disordered data.
In contrast, in the spatially varying effective mass model, the mean free path $\xi_{\text{MFP}}$ comes out rather clearly and naturally from the underlying disorder physics, namely, it is the distance over which the mass remains uniform (or equivalently average segment length), i.e, $\xi_{\text{MFP}}=\frac{L}{n+1}$. The weak (strong) disorder regime is when $\xi_{\text{SC}}<\xi_{\text{MFP}}$ ($\xi_{\text{SC}}>\xi_{\text{MFP}}$), while the intermediate regime is when $\xi_{\text{SC}}\sim \xi_{\text{MFP}}$.
This is the huge advantage of our new approach--- it enables us to include only the minimal necessary aspect of disorder by building in the MFP physically and naturally in the effective Majorana model instead of getting overwhelmed by calculations over disorder distributions or many different disorder configurations.

In the following sections, we will first demonstrate how this spatially varying effective mass model can reproduce the same results as the previous models which include disorder explicitly in the Hamiltonian qualitatively, especially, reproducing the findings of the recent experiment from Microsoft Quantum on the InAs-Al nanowire~\cite{microsoftquantum2023inasal} in Sec.~\ref{sec:MSFT}.
We will then study how this new model behaves under the change of parameters $k$ in Sec.~\ref{sec:dependence_of_k}, and different wire lengths $L$ in Sec.~\ref{sec:dependence_of_L}.
Finally, following this criterion, we will demonstrate that we can bring a nontopological wire back to a topological wire by only increasing the proximitized SC gap which effectively decreases the SC coherence length $\xi_{\text{SC}}$ to satisfy the hierarchy of $\xi_{\text{SC}}<\xi_{\text{MFP}}$ in Sec.~\ref{sec:increase_SC_gap}.

\subsection{Simulation of experimental data}\label{sec:MSFT}
\begin{figure*}[ht]
    \centering
    \includegraphics[width=6.8in]{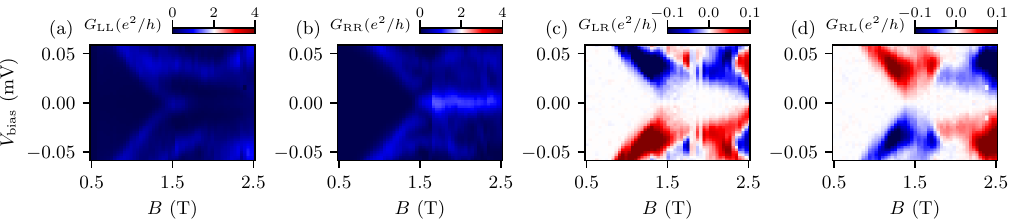}
    \caption{
    The positive experimental result, Fig 11 in Ref.~\onlinecite{microsoftquantum2023inasal}, passes the TGP.
    (a-d) Local ($G_{\text{LL}}$ and $G_{\text{RR}}$) and nonlocal ($G_{\text{LR}}$ and $G_{\text{RL}}$) conductance spectra as the function of magnetic field. 
    The false color is rendered with a new diverging color scheme from 0 to $4e^2/h$ than the original print.
    }
    \label{fig:exp_positive}
\end{figure*}

In this section, we sample different disorder configurations to reproduce the recent experiment in~\cite{microsoftquantum2023inasal}.
Our goal is to demonstrate that the disorder physics can be qualitatively captured by the spatially varying effective mass model.
Therefore, we aim to reproduce the two representative results in the experiment: 
(1) a positive result that passes the topological gap protocol (TGP), e.g., Fig. 11 in Ref.~\onlinecite{microsoftquantum2023inasal} (also shown here in Fig.~\ref{fig:exp_positive}).
(2) a negative result that fails the TGP, e.g., Fig. 20 in Ref.~\onlinecite{microsoftquantum2023inasal} (also shown here in Fig.~\ref{fig:exp_negative}).
For the details on TGP, which Ref.~\cite{microsoftquantum2023inasal} uses as the definitive tool for determining topology (``positive") or not (``negative"), we refer to the Microsoft article~\cite{microsoftquantum2023inasal}. We mention that TGP is, for all practical purposes, equivalent to our earlier work on local and nonlocal conductance for Majorana confirmation as presented in~\cite{pan2021threeterminal}.

\begin{figure*}[ht]
    \centering
    \includegraphics[width=6.8in]{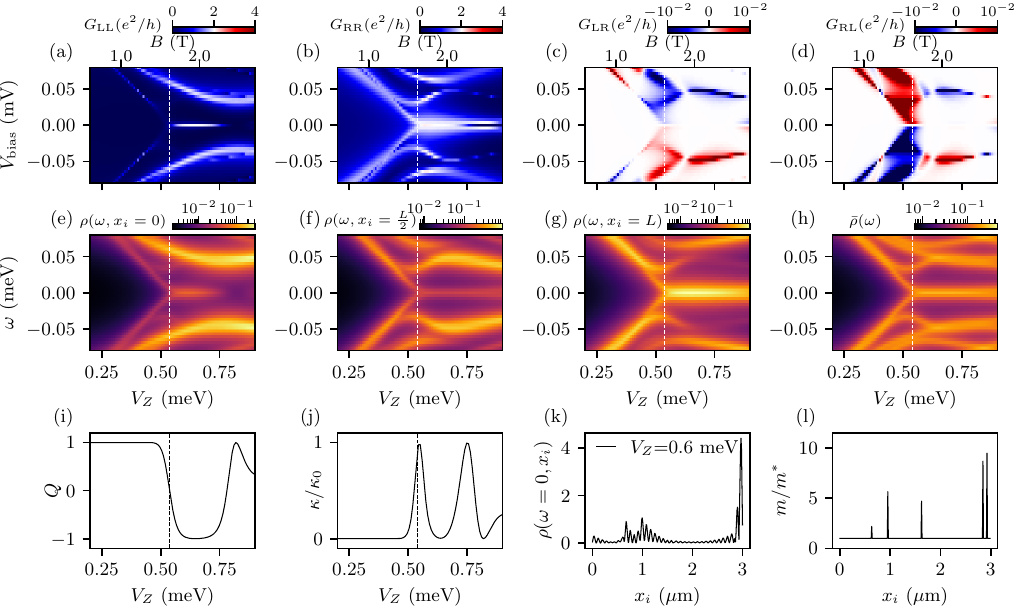}
    \caption{
    Simulation of the positive result Fig. 11 in Ref.~\cite{microsoftquantum2023inasal} (also in Fig.~\ref{fig:exp_positive}) with $n=5$, $k=9$, and $L=3~\mu$m. 
    (a-d) Local ($G_{\text{LL}}$ and $G_{\text{RR}}$) and nonlocal ($G_{\text{LR}}$ and $G_{\text{RL}}$) conductance spectra. The bottom (top) axis shows the Zeeman (magnetic) field. The vertical dashed line indicates nominal TQPT. 
    (e-g) LDOS at left, midpoint, and right end. 
    (h) Total DOS.
    (i-j) Topological visibility $Q$ and thermal conductance $\kappa$.
    (k) LDOS at zero energy for $V_Z=0.6$ meV.
    (l) Spatial profile of effective mass. 
    }
    \label{fig:thy_positive}
\end{figure*}

\begin{figure*}[ht]
    \centering
    \includegraphics[width=6.8in]{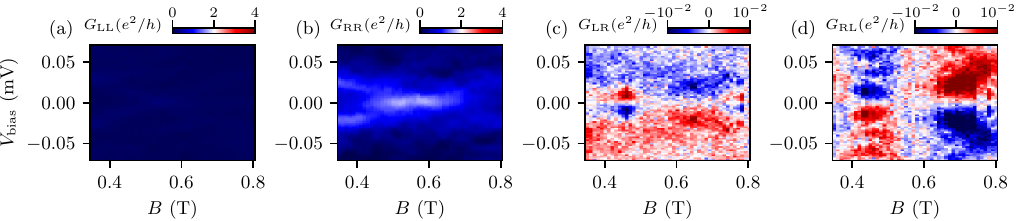}
    \caption{
    The negative experimental result, Fig 20 in Ref.~\onlinecite{microsoftquantum2023inasal}, fails to pass the TGP.
    (a-d) Local ($G_{\text{LL}}$ and $G_{\text{RR}}$) and nonlocal ($G_{\text{LR}}$ and $G_{\text{RL}}$) conductance spectra as the function of magnetic field. 
    The false color is rendered with a new diverging color scheme from 0 to $4e^2/h$ than the original print.
    }
    \label{fig:exp_negative}
\end{figure*}

Figure~\ref{fig:exp_positive} is the positive result in the sense that the local conductance spectra from both ends exhibit ZBCPs beyond the same critical field, with the left end disappearing at a smaller field than the right, and the gap closing and reopening feature can be tracked from the nonlocal conductance, which happens at the same field at the onset of ZBCP. 
Therefore, we reproduce these crucial features in Fig.~\ref{fig:thy_positive} using the new model in a 3-micron long wire (same as the experiment) by choosing the parameter of $n=5$, $k=9$ in the varying effective mass model.
From the top row of Fig.~\ref{fig:thy_positive}, we can already notice the similarity between the theoretical simulation and experimental results which manifest the correlation between local conductance spectra, and gap closing and reopening from nonlocal conductance spectra.
From the bottom row of Fig.~\ref{fig:thy_positive}, we confirm the topological regime between $V_Z\in [0.54,0.75]$ meV by computing the TV $Q=-1$ (Fig.~\ref{fig:thy_positive}(i)), and showing a peak in the thermal conductance quantized at $\kappa_0$ (Fig.~\ref{fig:thy_positive}(j)).
In Fig.~\ref{fig:thy_positive}(k), we show the LDOS at a specific Zeeman field for a zero energy state, which clearly shows that the wire hosts two pairs of MZMs (where the first pair extends between [0, 0.6$\mu$m]), and the second pair extends between [1$\mu$m, 3$\mu$m]. The LDOS at the midpoint also indicates a zero energy bulk state in Fig.~\ref{fig:thy_positive}(f).
This localization of the Majorana modes is consistent with the positions of the impurities (Fig.~\ref{fig:thy_positive}(l)), which effectively segment the wire into several quantum dots.
This positive result belongs to the weak disorder regime, which can manifest nontrivial topological visibility in Fig.~\ref{fig:pristine} over a region of $V_Z\in[0.54,0.75]$ meV.
This is also consistent with our criterion since $n=5$ corresponds to the MFP being $\xi_{\text{MFP}}=0.5~\mu$m, which is larger than $\xi_{\text{SC}}=0.36~\mu$m. 
However, the weak local conductance (Fig.~\ref{fig:thy_positive}(a)) and the absence of the peak in the zero energy LDOS at the left ends (Fig.~\ref{fig:thy_positive}(k)) imply a potential issue that the wire, despite showing a nontrivial topological invariant, may not be useful for the braiding and fusion due to the absence of isolated MZMs on both ends of the wire. We will address this issue later in Sec.~\ref{sec:results_indicators}.
Note that the presence or absence of end localized states is not included in the Microsoft TGP because this physics is not directly accessible in their transport spectroscopy.

\begin{figure*}[ht]
    \centering
    \includegraphics[width=6.8in]{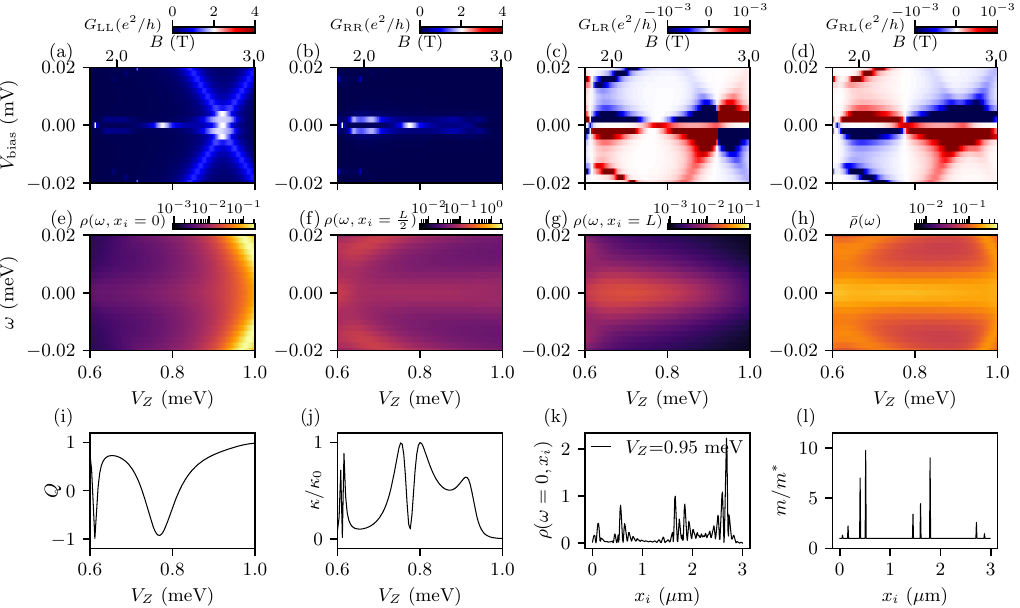}
    \caption{
    Simulation of the negative result Fig. 20 in Ref.~\cite{microsoftquantum2023inasal} (also in Fig.~\ref{fig:exp_negative}) with $n=9$, $k=9$, and $L=3~\mu$m.
    (a-d) Local ($G_{\text{LL}}$ and $G_{\text{RR}}$) and nonlocal ($G_{\text{LR}}$ and $G_{\text{RL}}$) conductance spectra. The bottom (top) axis shows the Zeeman (magnetic) field. The vertical dashed line indicates nominal TQPT. 
    (e-g) LDOS at left, midpoint, and right end. 
    (h) Total DOS.
    (i-j) Topological visibility $Q$ and thermal conductance $\kappa$.
    (k) LDOS at zero energy for $V_Z=0.95$ meV.
    (l) Spatial profile of effective mass.
    }
    \label{fig:thy_negative}
\end{figure*}

On the contrary, Figure~\ref{fig:exp_negative} is a definitive negative result because the local conductance spectra from both ends do not exhibit any correlation between the appearance of ZBCPs, and the gap closing and reopening feature cannot be inferred from the nonlocal conductance.
In this scenario, we can reproduce the similar `negative' features by increasing the number of impurities, as shown in Fig.~\ref{fig:thy_negative} in the same wire length $L=3~\mu$m with $n=9$ and $k=9$.
From the top row in Fig.~\ref{fig:thy_negative}, we simulate and reproduce the absence of the correlation and gap closing and reopening feature.
This impurity configuration destroys the topological regime as shown in Fig.~\ref{fig:thy_negative}(i,j), where the TV $Q>0$ and the thermal conductance does not show one single clear peak.
To understand the microscopic details, we show the LDOS at zero energy and a Zeeman energy near 0.95 meV as shown in Fig.~\ref{fig:thy_negative}(k). We notice multiple pairs of MZMs extended over the wire, which is consistent with the spatial distribution of impurities in Fig.~\ref{fig:thy_negative}(l).
Therefore, this negative result belongs to the intermediate disorder regime, which is consistent with the fact that it fails the TGP, and also with our simple criterion as $\xi_{\text{MFP}}=0.3~\mu$m, which is of the same order of $\xi_{\text{SC}}=0.36~\mu$m (in the pristine wire). 
(Note that this estimate of 0.36 $\mu$m serves as a lower bound--- we can roughly estimate the actual SC coherence in the presence of disorder by considering the ratio of two topological gaps with and without the disorder, which gives an actual $\xi_{\text{SC}}$ approaches infinity as the gap size approaches zero.)

The qualitative agreement between Figs.~\ref{fig:exp_positive} and \ref{fig:thy_positive}, and between Figs.~\ref{fig:exp_negative} and \ref{fig:thy_negative} is not a result of fine-tuning. Other disorder configurations using a similar value of $n$ can produce the same qualitative results, including whether the gap reopens, and whether the left and right conductance are correlated. 
These results are presented in Appendix~\ref{app:MSFT} along with the statistics for the positive and negative results summarized in Table~\ref{tab:MSFT}.
It is noteworthy that the positive simulations in general look more similar to the positive experimental data than the negative simulations look like the negative experimental data, which makes sense since the negative results are by definition an infinitely larger set than the positive results.
Of course, what matters are the positive results (and a well-defined way of ruling out the negative results).

\begin{table}[ht]
    \centering
    \caption{Statistics of the positive and negative simulations for a 3-micron wire.}
    \begin{tabular}{ccccc}
    \hline
     $k$ & $n$ & $\dfrac{\text{Positive simulations}}{\text{Ensemble size}}$ &  Examples \\
    \hline
     9 & 5 & 1/100 &  Fig.~\ref{fig:thy_positive} \\
     9 & 6 & 2/100 &  Figs.~\ref{fig:649} and~\ref{fig:658} \\
     9 & 7 & 1/100 &  Fig.~\ref{fig:703} \\
    \hline
    \hline
     $k$ &$n$  & $\dfrac{\text{Negative simulations}}{\text{Ensemble size}}$ & Examples \\
    \hline
      9 & 9 & 4/100 & Figs.~\ref{fig:thy_negative},~\ref{fig:925},~\ref{fig:941}, and~\ref{fig:971} \\
      9 & 10 & 1/50 & Fig.~\ref{fig:1001} \\
      9 & 11 & 1/50 & Fig.~\ref{fig:1113} \\
      9 & 12 & 1/50 & Fig.~\ref{fig:1201} \\
        \hline
    \end{tabular}
    \label{tab:MSFT}
\end{table}

\subsection{Dependence of $k$}\label{sec:dependence_of_k}
In the previous section, we have demonstrated that $n$ is the most relevant parameter that directly determines the topology by controlling the mean free path $\xi_{\text{MFP}}=\frac{L}{n+1}$.
In this section, we will study the effect of the other parameter $k$ (which controls the variance of the effective mass of the impurities).

Intuitively, since a larger value of $k$ will allow a larger effective mass of the impurities, which suppresses the hopping across the impurities more, the precise value of $k$ should not affect the topology as long as it is large enough to produce effective segmentation.  (In fact, even for $n$, a similar caveat applies in the sense that the value of $n$ is irrelevant except in the sense of satisfying or violating the all-important inequality of $\xi_{\text{MFP}}> \xi_{\text{SC}}$ or not.)

\begin{figure*}[ht]
    \centering
    \includegraphics[width=6.8in]{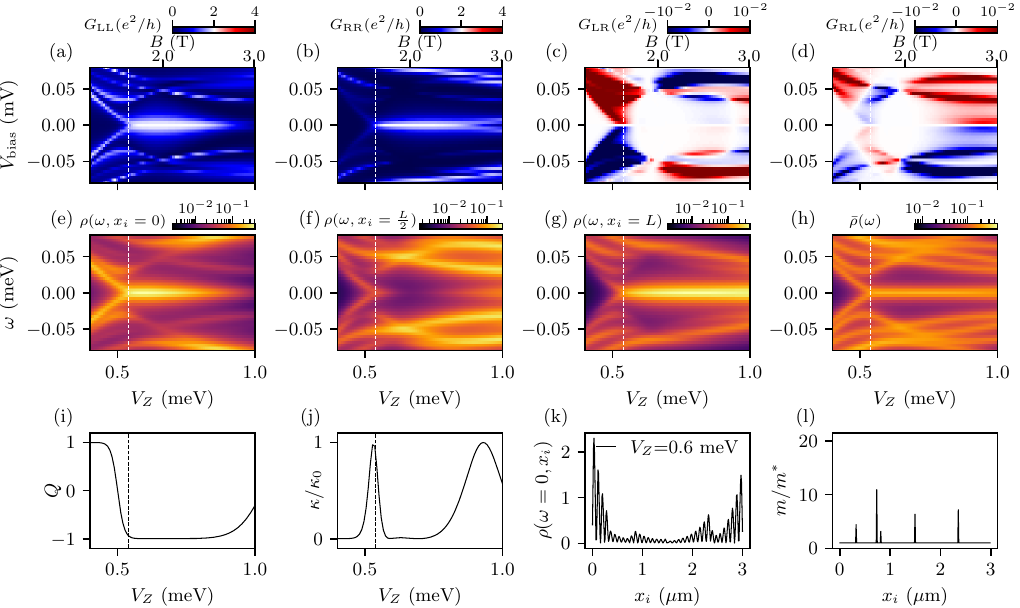}
    \caption{
    $n=5$, $k=20$, and $L=3~\mu$m. 
    (a-d) Local ($G_{\text{LL}}$ and $G_{\text{RR}}$) and nonlocal ($G_{\text{LR}}$ and $G_{\text{RL}}$) conductance spectra. The bottom (top) axis shows the Zeeman (magnetic) field. The vertical dashed line indicates nominal TQPT. 
    (e-g) LDOS at left, midpoint, and right end. 
    (h) Total DOS.
    (i-j) Topological visibility $Q$ and thermal conductance $\kappa$.
    (k) LDOS at zero energy for $V_Z=0.6$ meV.
    (l) Spatial profile of effective mass.
    }
    \label{fig:k=20}
\end{figure*}

\begin{figure*}[ht]
    \centering
    \includegraphics[width=6.8in]{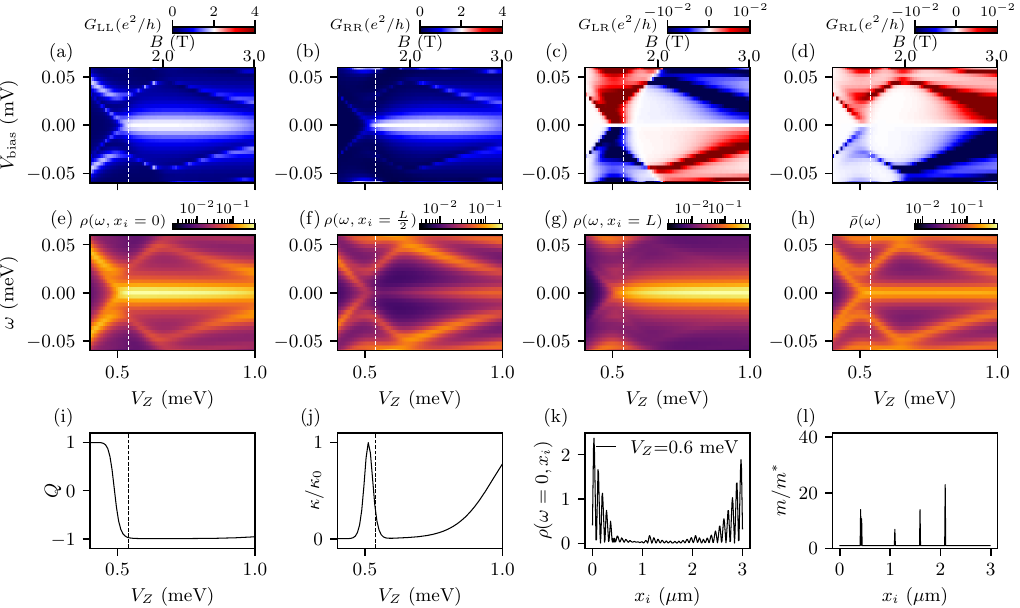}
    \caption{
    $n=5$, $k=40$, and $L=3~\mu$m. 
    (a-d) Local ($G_{\text{LL}}$ and $G_{\text{RR}}$) and nonlocal ($G_{\text{LR}}$ and $G_{\text{RL}}$) conductance spectra. The bottom (top) axis shows the Zeeman (magnetic) field. The vertical dashed line indicates nominal TQPT. 
    (e-g) LDOS at left, midpoint, and right end. 
    (h) Total DOS.
    (i-j) Topological visibility $Q$ and thermal conductance $\kappa$.
    (k) LDOS at zero energy for $V_Z=0.6$ meV.
    (l) Spatial profile of effective mass.
    }
    \label{fig:k=40}
\end{figure*}

\begin{figure*}[ht]
    \centering
    \includegraphics[width=6.8in]{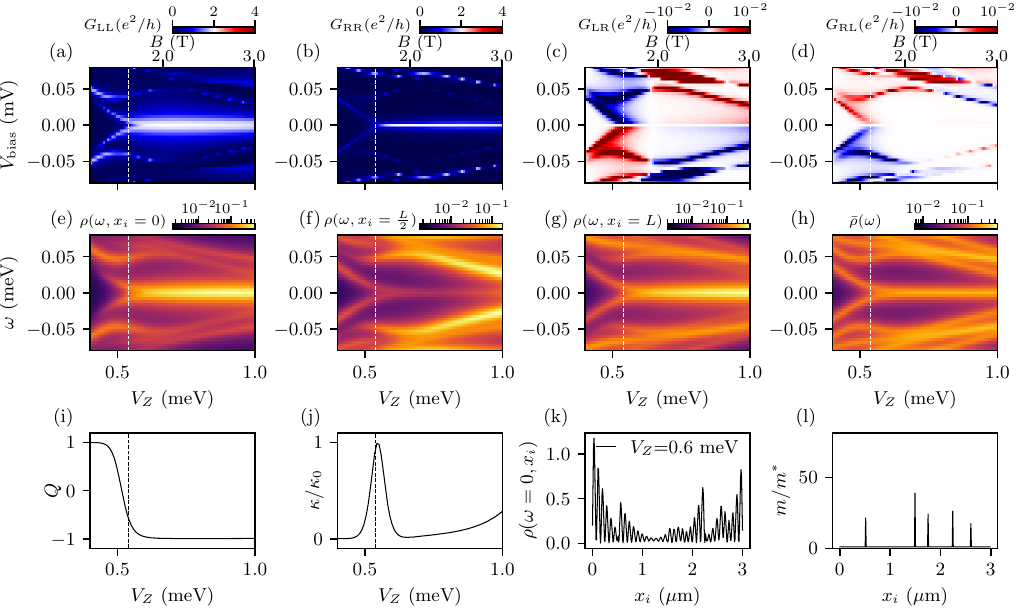}
    \caption{
    $n=5$, $k=80$, and $L=3~\mu$m. 
    (a-d) Local ($G_{\text{LL}}$ and $G_{\text{RR}}$) and nonlocal ($G_{\text{LR}}$ and $G_{\text{RL}}$) conductance spectra. The bottom (top) axis shows the Zeeman (magnetic) field. The vertical dashed line indicates nominal TQPT. 
    (e-g) LDOS at left, midpoint, and right end. 
    (h) Total DOS.
    (i-j) Topological visibility $Q$ and thermal conductance $\kappa$.
    (k) LDOS at zero energy for $V_Z=0.6$ meV.
    (l) Spatial profile of effective mass.
    }
    \label{fig:k=80}
\end{figure*}

\begin{figure*}[ht]
    \centering
    \includegraphics[width=6.8in]{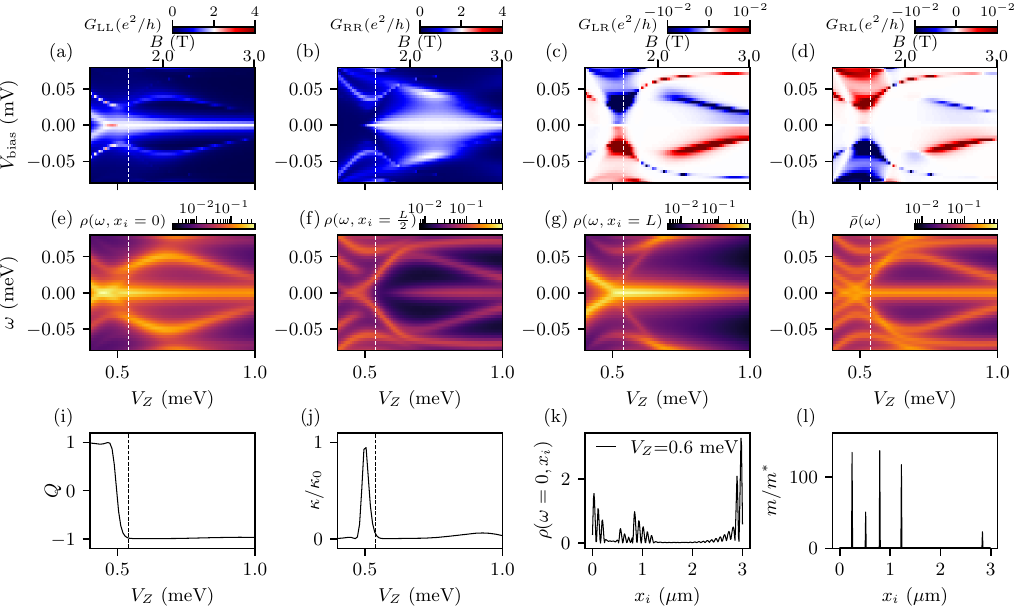}
    \caption{
    $n=5$, $k=160$, and $L=3~\mu$m. 
    (a-d) Local ($G_{\text{LL}}$ and $G_{\text{RR}}$) and nonlocal ($G_{\text{LR}}$ and $G_{\text{RL}}$) conductance spectra. The bottom (top) axis shows the Zeeman (magnetic) field. The vertical dashed line indicates nominal TQPT. 
    (e-g) LDOS at left, midpoint, and right end. 
    (h) Total DOS.
    (i-j) Topological visibility $Q$ and thermal conductance $\kappa$.
    (k) LDOS at zero energy for $V_Z=0.6$ meV.
    (l) Spatial profile of effective mass.
    }
    \label{fig:k=160}
\end{figure*}

Our numerical simulation verifies this intuition for $k$. We present typical results for larger $k$ as shown in Fig.~\ref{fig:k=20} for $k=20$, Fig.~\ref{fig:k=40} for $k=40$, Fig.~\ref{fig:k=80} for $k=80$, and Fig.~\ref{fig:k=160} for $k=160$.
All these results have the same value of $n=5$, and we find similar results as the positive simulations in Fig.~\ref{fig:thy_positive}, which belong to the weak disorder.
Besides these similar results shown from Fig.~\ref{fig:k=20} to Fig.~\ref{fig:k=160}, we present more examples in Appendix~\ref{app:dependence_of_k}. These results all confirm that the value of $k$ does not affect the topology as much as $n$ does since $k$ does not determine the $\xi_{\text{MFP}}$ as $n$ does--- $k$ just has to be large enough to produce an effective segmentation of the wire.

\subsection{Dependence of $L$}\label{sec:dependence_of_L}
One remaining parameter of experimental relevance is the wire length $L$.
Actually, $L$ is irrelevant when $L>\xi_{\text{MFP}}$, but in short enough wires, it is possible that the $L$ is the length cutoff for the MFP.
In this section, we will present the results for an extremely long wire of $L=10~\mu$m. 
We will demonstrate that changing the system size $L$ can affect the mean free path $\xi_{\text{MFP}}=\frac{L}{n+1}$, assuming a fixed number of impurities in the wire with changing length, and therefore change the topology.
Note that if the impurity density is kept fixed, then obviously increasing $L$ would increase $n$ proportionally and no new physics emerges, since it is the value of $n$ that determines the topology.
Especially, we will show how one has to scale the number of impurities $n$ proportionally to the wire length $L$, in order to keep the same disorder regime.
Basically, we show explicitly that if the impurity density is held fixed (the likely physical situation), increasing $L$ does nothing constructive for obtaining topology.  What is necessary is decreasing the impurity density, which cannot be achieved just by increasing $L$ if $n$ is not fixed.

\begin{figure*}[ht]
    \centering
    \includegraphics[width=6.8in]{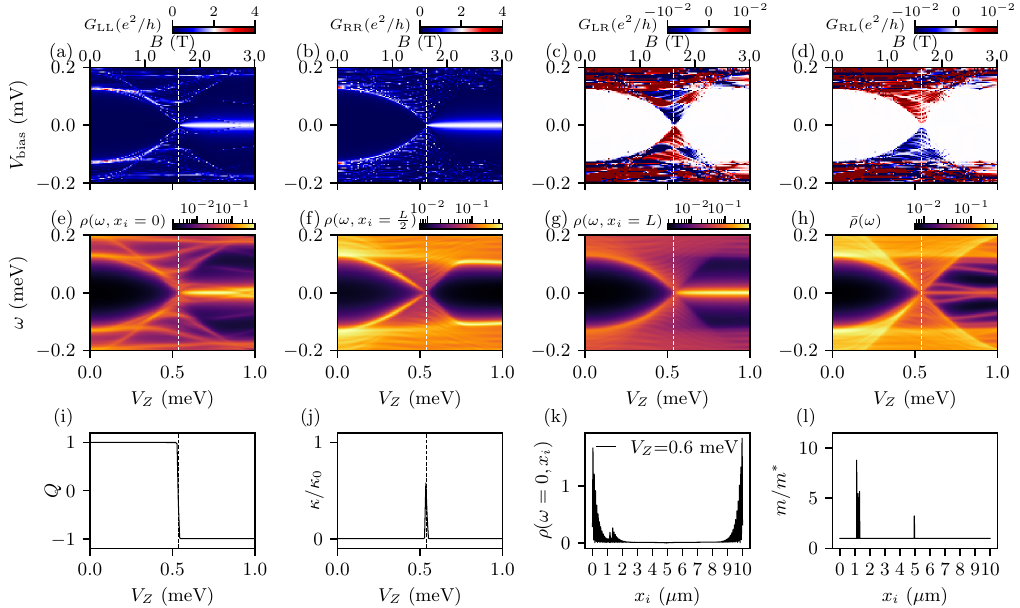}
    \caption{
    Long wire limit $L=10~\mu$m with $n$=5 and $k$=9.
    (a-d) Local ($G_{\text{LL}}$ and $G_{\text{RR}}$) and nonlocal ($G_{\text{LR}}$ and $G_{\text{RL}}$) conductance spectra. The bottom (top) axis shows the Zeeman (magnetic) field. The vertical dashed line indicates nominal TQPT. 
    (e-g) LDOS at left, midpoint, and right end. 
    (h) Total DOS.
    (i-j) Topological visibility $Q$ and thermal conductance $\kappa$.
    (k) LDOS at zero energy for $V_Z=0.6$ meV.
    (l) Spatial profile of effective mass.
    }
    \label{fig:n5L10}
\end{figure*}

We first fix the number of impurities $n=5$ and $k=9$, and present the typical result in Fig.~\ref{fig:n5L10} for a 10-micron wire.
We find that the result is very similar to the pristine wire with very low variance.
We note that in this process the effective impurity density has decreased by more than a factor of 3, thus making the situation similar to the 3-micron long wire with $n=1$! 
For almost any impurity realizations with $n=5$ in 10-micron wires, local conductance spectra consistently show the quantized ZBCP beyond TQPT at $V_{Zc}\sim 0.54$ meV and the nonlocal conductance spectra show the gap closing and reopening feature at the same critical field.
For the topological visibility and thermal conductance, we consistently find the abrupt change of TV from $+1$ to $-1$, and the peak in the thermal conductance quantized at $\kappa/\kappa_0=1$.
The zero energy LDOS at a Zeeman field $V_Z$ inside the topological regime shows one pair of MZMs localized at the wire ends. The only difference from the pristine case is that the LDOS in the bulk of the wire shows a very weak peak due to the impurity near the position at $1$ micron. 
This extreme long wire limit is clearly in the weak disorder regime, which even can be considered as the pristine case because it shows a much more stable topology than the previous 3-micron wire with the same $n=5$ as shown in Fig.~\ref{fig:thy_positive}.
This phenomenon can be understood from the same rule of thumb Eq.~\eqref{eq:hierarchy} that $\xi_{\text{MFP}}$ now increases from 0.5 $\mu$m to 1.6 $\mu$m due to the increase of $L$ from 3 $\mu$m to 10 $\mu$m, which becomes much longer than the SC coherence length $\xi_{\text{SC}}=0.36~\mu$m (since other parameters remain the same).
We emphasize that if the impurity density is kept fixed (instead of the impurity number) while increasing $L$, there would be no effect on the topology.  Thus, increasing the wire length without decreasing the impurity density does no good whatsoever.  Disorder and wire length are closely connected, and this connection is obvious in the new model, and hidden in the standard disorder (in the chemical potential) model.

\begin{figure*}[ht]
    \centering
    \includegraphics[width=6.8in]{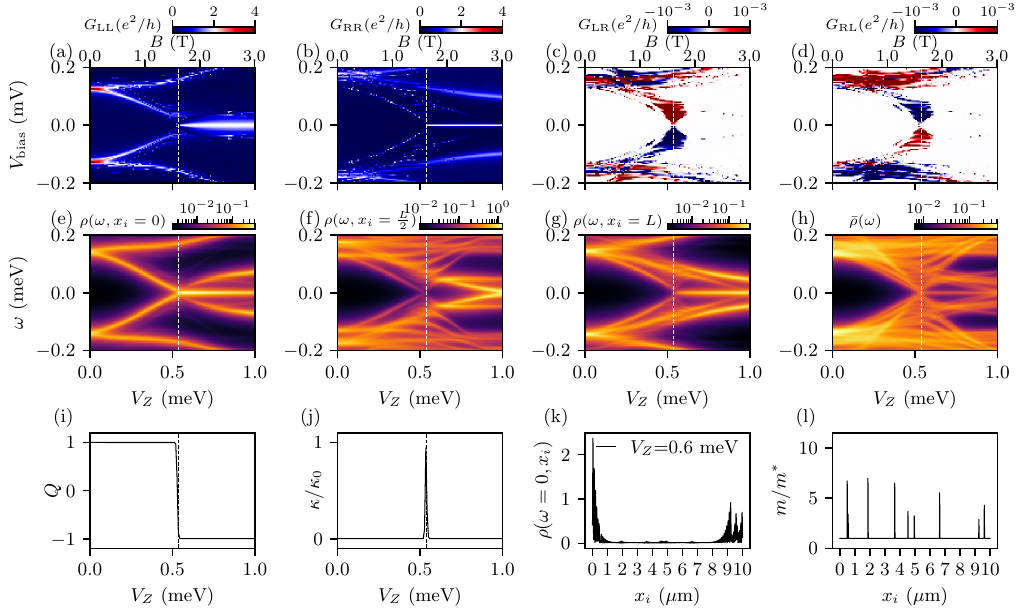}
    \caption{
    Long wire limit $L=10~\mu$m with $n$=5 and $k$=9.
    (a-d) Local ($G_{\text{LL}}$ and $G_{\text{RR}}$) and nonlocal ($G_{\text{LR}}$ and $G_{\text{RL}}$) conductance spectra. The bottom (top) axis shows the Zeeman (magnetic) field. The vertical dashed line indicates nominal TQPT. 
    (e-g) LDOS at left, midpoint, and right end. 
    (h) Total DOS.
    (i-j) Topological visibility $Q$ and thermal conductance $\kappa$.
    (k) LDOS at zero energy for $V_Z=0.6$ meV.
    (l) Spatial profile of effective mass.
    }
    \label{fig:n9L10}
\end{figure*}

Following the same logic, we find that even $n=9$ in the extremely long wire of $L=10~\mu$m can still be considered in the weak disorder regime to manifest a robust topology beyond TQPT. We present such a typical example in Fig.~\ref{fig:n9L10}.
This is in contrast to the previous case of $L=3~\mu$m, where $n=9$ is in the intermediate disorder regime as shown in Fig.~\ref{fig:thy_negative}.
This is for the same reason: $\xi_{\text{MFP}}$ now becomes 1 $\mu$m from 0.3 $\mu$m, which is much longer than the $\xi_{\text{SC}}=0.36~\mu$m.
Equivalently, $n=9$ in the $L=10$ micron case has an impurity density of $\sim 1.1$ which is smaller than the impurity density of 1.67 in the $n=5$ and $L=3$ micron case.
More examples of the extremely long wires are presented in Appendix~\ref{app:dependence_of_L}.

\begin{figure*}[ht]
    \centering
    \includegraphics[width=6.8in]{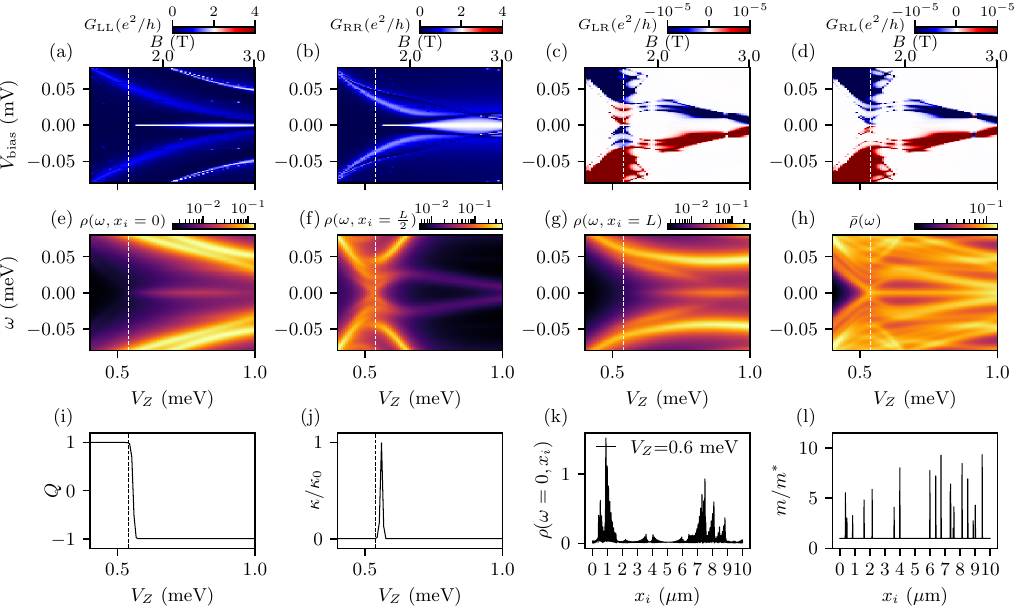}
    \caption{
        Long wire limit $L=10~\mu$m with $n$=19 and $k$=9.
    (a-d) Local ($G_{\text{LL}}$ and $G_{\text{RR}}$) and nonlocal ($G_{\text{LR}}$ and $G_{\text{RL}}$) conductance spectra. The bottom (top) axis shows the Zeeman (magnetic) field. The vertical dashed line indicates nominal TQPT. 
    (e-g) LDOS at left, midpoint, and right end. 
    (h) Total DOS.
    (i-j) Topological visibility $Q$ and thermal conductance $\kappa$.
    (k) LDOS at zero energy for $V_Z=0.6$ meV.
    (l) Spatial profile of effective mass.
    }
    \label{fig:n19L10}
\end{figure*}

The previous two examples imply that one has to choose a much larger $n$ proportionally in the extreme long wire limit to keep the wire in the same disorder regime. 
That is to say, in a 10-micron wire, to simulate what happened for $n=5$ in the 3-micron wire, we need to increase $n=19$ as shown in Fig.~\ref{fig:n19L10}. 
This is simply because $10/20$ is exactly the same as $3/6$-- both wires now have similar segmentation and comparable MFP values.
We notice similar features as Fig.~\ref{fig:thy_positive}--- though there is a gap closing and reopening feature in the nonlocal conductance, the reopened topological gap will eventually close and the maximal topological gap ($\sim0.03$ meV) is also much smaller compared to the parent SC gap at zero field ($\sim0.3$ meV), similarly to the experimental positive result in Fig.~\ref{fig:exp_positive}. 
Whether such a small gap should be called topological or not is a semantic matter.  Braiding will be difficult to accomplish in systems with such small gaps.

\begin{figure*}[ht]
    \centering
    \includegraphics[width=6.8in]{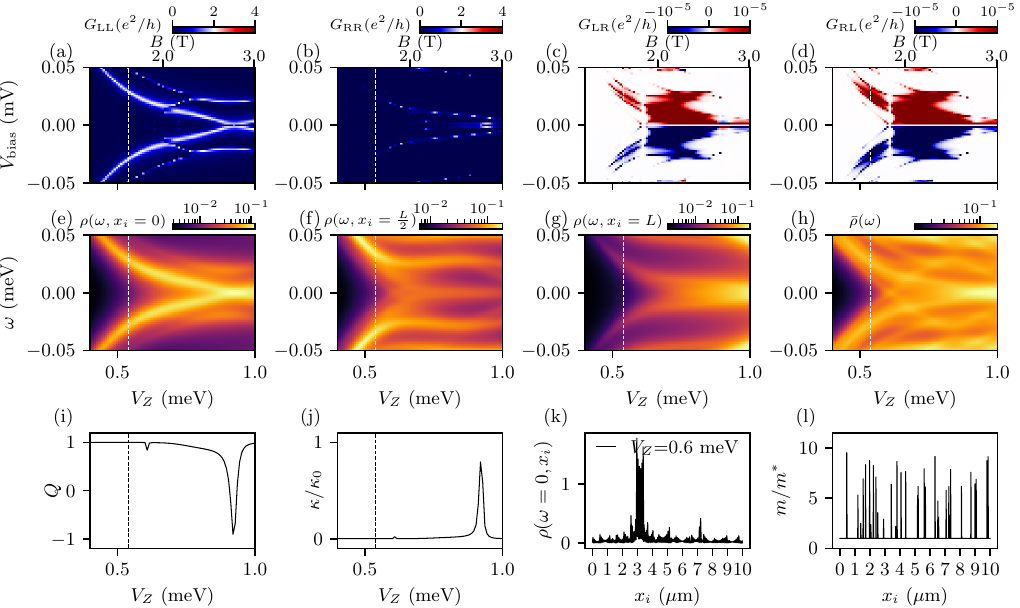}
    \caption{
        Long wire limit $L=10~\mu$m with $n$=32 and $k$=9.
        (a-d) Local ($G_{\text{LL}}$ and $G_{\text{RR}}$) and nonlocal ($G_{\text{LR}}$ and $G_{\text{RL}}$) conductance spectra. The bottom (top) axis shows the Zeeman (magnetic) field. The vertical dashed line indicates nominal TQPT. 
        (e-g) LDOS at left, midpoint, and right end. 
        (h) Total DOS.
        (i-j) Topological visibility $Q$ and thermal conductance $\kappa$.
        (k) LDOS at zero energy for $V_Z=0.6$ meV.
        (l) Spatial profile of effective mass.
    }
    \label{fig:n32L10}
\end{figure*}

Furthermore, in order to simulate the negative result (Fig.~\ref{fig:exp_negative}) using an extremely long wire, we need to further increase the number of impurities $n=32$ as shown in Fig.~\ref{fig:n32L10} so that the impurity density stays a constant.
In the first row of Fig.~\ref{fig:n32L10}, we observe a ZBCP appearing only at the left conductance spectrum but not at the right conductance spectrum. 
In addition, the gap reopening feature is also absent, and the Zeeman field where the bulk gap closes (Fig.~\ref{fig:n32L10}(c)) is not even at the same field as the onset of the ZBCP in the left local conductance spectrum (Fig.~\ref{fig:n32L10}(a)).
In the bottom row, we confirm that the system does not undergo a TQPT as the TV almost always remains positive, and the thermal conductance does not show a quantized peak. 
This wire with $n=32$ is in the intermediate disorder regime because the mean free path $\xi_{\text{MFP}}=0.3~\mu$m is already shorter than the (lower limit of) the SC coherence length $\xi_{\text{SC}}=0.36~\mu$m. 
More examples of pristine, weak disorder, and intermediate disorder in a 10-micron wire can be found in Appendix~\ref{app:dependence_of_L}.


\subsection{Increase the SC gap to make a nontopological wire topological}\label{sec:increase_SC_gap}
\begin{figure*}[ht]
    \centering
    \includegraphics[width=6.8in]{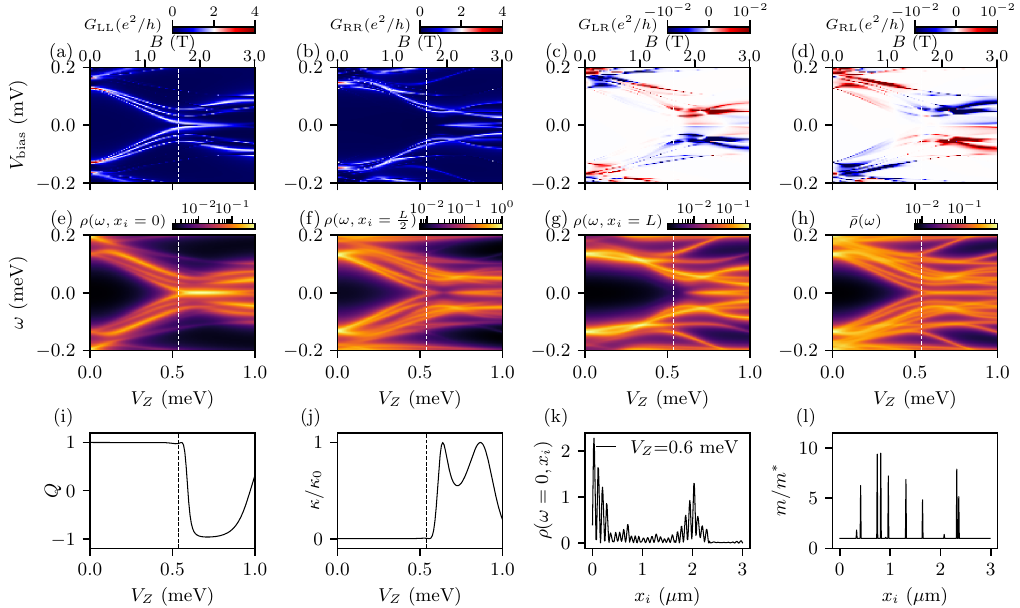}
    \caption{
    The 3-micron wire in the strong disorder regime with $n=11$ and $k=9$, and SC-SM coupling strength is 0.2 meV.
    (a-d) Local ($G_{\text{LL}}$ and $G_{\text{RR}}$) and nonlocal ($G_{\text{LR}}$ and $G_{\text{RL}}$) conductance spectra. The bottom (top) axis shows the Zeeman (magnetic) field. The vertical dashed line indicates nominal TQPT. 
    (e-g) LDOS at left, midpoint, and right end. 
    (h) Total DOS.
    (i-j) Topological visibility $Q$ and thermal conductance $\kappa$.
    (k) LDOS at zero energy for $V_Z=0.6$ meV.
    (l) Spatial profile of effective mass.
    }
    \label{fig:1113}
\end{figure*}

Finally, we demonstrate the ability to make a nontopological wire topological by only increasing the proximitized SC gap as a larger SC gap decreases the SC coherence length $\xi_{\text{SC}}$. 
We choose Fig.~\ref{fig:1113} as an example of a nontopological 3-micron wire with $n=11$ and $k=9$. 
Here, because the MFP $\xi_{\text{MFP}}=0.25~\mu$m is already smaller than the lower bound of the SC coherence length $\xi_{\text{SC}}=0.36~\mu$m, the wire was originally in the strong disorder regime. 
In the top row of Fig.~\ref{fig:1113},  the local conductance spectra show ZBCPs that only persist for a small energy interval of the Zeeman field ($V_Z\in[0.6,0.7]$ meV), and there is no correlation between the left and right conductance spectra. The nonlocal conductance spectra also do not manifest the gap reopening feature.
In the bottom row, the TV shows a negative value over a small energy interval, and the thermal conductance does not show a single sharp peak.
At a specific Zeeman field at 0.6 meV in Fig.~\ref{fig:1113}(k), the LDOS at zero energy only shows one peak at the left end. Another peak is localized in the bulk region at $x_i\sim2~\mu$m due to the spatial distribution of the impurity. 

\begin{figure*}[ht]
    \centering
    \includegraphics[width=6.8in]{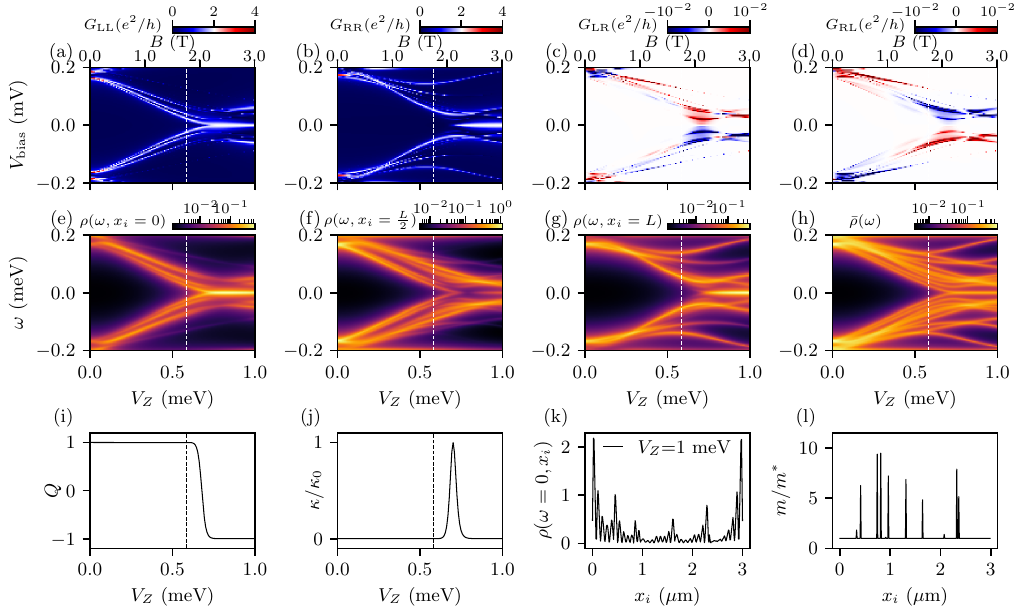}
    \caption{
    The 3-micron wire in the intermediate disorder regime with $n=11$ and $k=9$, and SC-SM coupling strength is 0.3 meV.
    (a-d) Local ($G_{\text{LL}}$ and $G_{\text{RR}}$) and nonlocal ($G_{\text{LR}}$ and $G_{\text{RL}}$) conductance spectra. The bottom (top) axis shows the Zeeman (magnetic) field. The vertical dashed line indicates nominal TQPT. 
    (e-g) LDOS at left, midpoint, and right end. 
    (h) Total DOS.
    (i-j) Topological visibility $Q$ and thermal conductance $\kappa$.
    (k) LDOS at zero energy for $V_Z=1$ meV.
    (l) Spatial profile of effective mass.
    }
    \label{fig:1113_0p3}
\end{figure*}

To make it topological without reducing the disorder, we increase the SC gap by raising the SC-SM coupling strength (0.2 meV). 
Our first attempt is to increase SC-SM coupling strength to 0.3 meV. 
Because the SC coherence length $\xi_{\text{SC}}\propto v_F/\Delta$, where $v_F$ is the Fermi velocity in the SM (assumed to be invariant under the change of SC-SM coupling), and $\Delta$ is the topological gap (due to the proximitized SC gap in the SM), a higher SC-SM coupling of 0.3 meV effectively decreases the lower limit of SC coherence length from 0.36 $\mu$m to 0.24 $\mu$m, and thus, brings the SC coherence length $\xi_{\text{SC}}$ to the similar order of the MFP $\xi_{\text{MFP}}=0.25~\mu$m.
The numerical results are shown in Fig.~\ref{fig:1113_0p3}.
Now the ZBCP in local conductance spectra (Fig.~\ref{fig:1113_0p3}(a-b)) begin to persist for a larger energy interval, although the correlation between the two ends is still absent.  
The nonlocal conductance spectra (Fig.~\ref{fig:1113_0p3}(c-d)) begin to manifest the gap reopening feature.
In the TV (Fig.~\ref{fig:1113_0p3}(i)) and the thermal conductance (Fig.~\ref{fig:1113_0p3}(j)), we find that the topological regime becomes more visible and robust.
In the LDOS at zero energy at a Zeeman field $V_Z=1$ meV (Fig.~\ref{fig:1113_0p3}(k)), where both local conductance spectra manifest ZBCP, we observe the zero energy state localized at both ends of the wire. 
The only issue is that the bulk regime hosts multiple pairs of MZMs in the bulk at the positions of impurities (Fig.~\ref{fig:1113_0p3}(l)).
This implies that the wire starts to enter an intermediate disorder regime.

\begin{figure*}[ht]
    \centering
    \includegraphics[width=6.8in]{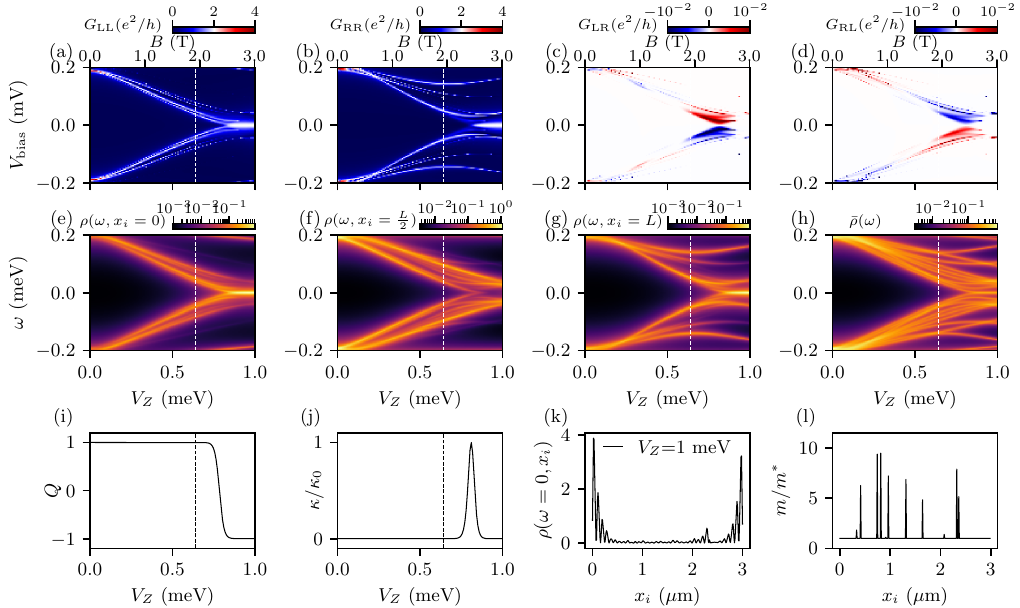}
    \caption{
    The 3-micron wire in the weak disorder regime with $n=11$ and $k=9$, and SC-SM coupling strength is 0.4 meV.
    (a-d) Local ($G_{\text{LL}}$ and $G_{\text{RR}}$) and nonlocal ($G_{\text{LR}}$ and $G_{\text{RL}}$) conductance spectra. The bottom (top) axis shows the Zeeman (magnetic) field. The vertical dashed line indicates nominal TQPT. 
    (e-g) LDOS at left, midpoint, and right end. 
    (h) Total DOS.
    (i-j) Topological visibility $Q$ and thermal conductance $\kappa$.
    (k) LDOS at zero energy for $V_Z=1$ meV.
    (l) Spatial profile of effective mass.
    }
    \label{fig:1113_0p4}
\end{figure*}

Furthermore, if we continue to increase the SC-SM coupling to 0.4 meV, which effectively brings the SC coherence length $\xi_{\text{SC}}$ further down to 0.18 $\mu$m, which is already smaller than the MFP $\xi_{\text{MFP}}=0.25~\mu$m, we find that the wire becomes even more topological, as shown in Fig.~\ref{fig:1113_0p4}.
In this case, the local conductance spectra (Fig.~\ref{fig:1113_0p4}(a-b)) show clear quantized ZBCPs correlated at both ends, and the nonlocal conductance spectra (Fig.~\ref{fig:1113_0p4}(c-d)) show the gap closing and reopening features at a larger Zeeman field. 
The larger TQPT is the result of both a larger proximitized SC gap and the disorder.
In the TV (Fig.~\ref{fig:1113_0p4}(i)) and thermal conductance (Fig.~\ref{fig:1113_0p4}(j)), we find clear change of $Q$ and peak of $\kappa$.
In the LDOS at zero energy and $V_Z=1$ meV, we find the zero energy state localized at both ends, with the bulk states being greatly suppressed compared to the previous case (Fig.~\ref{fig:1113_0p3}(k)).
This means that the wire now enters the weak disorder regime.
All of this is being achieved in our new model simply by decreasing $\xi_{\text{SC}}$  while keeping $\xi_{\text{MFP}}$ fixed.

From Fig.~\ref{fig:1113} to Fig.~\ref{fig:1113_0p4}, we improve the topology of the wire by only increasing the SC gap, which exemplifies the ability to make a nontopological wire topological by increasing the proximity-induced SC gap in the nanowire. 

\section{Topological indicators}\label{sec:results_indicators}
However, from the previous conductance spectra (e.g., Fig.~\ref{fig:thy_negative} and~\ref{fig:1113}), we  notice an issue:
The TV itself may not serve as a good indicator in the disordered short wire.
The dilemma that the topological invariant manifests a nontrivial negative value, but without the isolated MZMs localized at the ends of the wire, has been extensively discussed in Ref.~\onlinecite{dassarma2023spectral}.
This is important not just because disorder suppresses and obscures all topologies, but also because in the nonabelian anyon-based topological quantum computing~\cite{freedman2003topological,nayak2008nonabelian,alicea2011nonabelian,sarma2015majorana}, it is only when the isolated MZMs are localized at ends of the wire that they can be used to perform fusions and braiding.
Thus, the lack of TV being able to indicate topology is an extremely important problem, which must somehow be dealt with, particularly since most current experimental wires are likely to be in the intermediate disorder regime where $\xi_{\text{SC}}$ and $\xi_{\text{MFP}}$ are not that different.
Therefore, in this section, we aim to find a more reliable topological indicator by incorporating information about the spatial distribution of the MZMs.
In principle, the criteria to detect the topological MZMs should characterize both of the following features:
\begin{enumerate}
    \item Nontrivial topological invariant
    \item Isolated zero energy states localized at the ends of the wire
\end{enumerate}

The first feature is already captured by the TV, and therefore, we will focus on the indicators that can capture the second feature.
\subsection{Indicators for localized topological MZMs}\label{sec:definition_indicators}
Because the information on the spatial distribution of the MZMs can only be detected by the quantity that is defined over the entire wire, we choose to focus on the LDOS instead of the conductance spectra, which only probe the properties of the ends of the wire.
Therefore, based on the different aspects of the LDOS, we propose and discuss four distinct possible indicators.

\subsubsection{Indicator 1: LDOS end weight}
The first indicator follows the intuitive idea of directly quantifying the relative weight of the zero energy states localized at the ends of the wire. Therefore, we define it as
\begin{equation}
    I^{(1)}= \frac{\sum_{x_i\in \text{Ends}}\rho(\omega=0,x_i)}{\sum_{x_i\in [0,L]}\rho(\omega=0,x_i)},
\end{equation}
Here, the only degree of freedom is the length scale of ends.
This is because, in the actual finite-size wire, the localized MZMs are not strictly localized at the absolute (geometric) ends while extending over a finite length at the ends, where the length scale is the Majorana localization length (same as SC coherence length $\xi_{\text{SC}}$).
Therefore, we can choose to define the length scale of `ends' as $\xi_{\text{SC}}/2$. 
In all the previous results with SC-SM coupling strength of 0.2 meV and 3-micron wires, this corresponds to the end regions defined as $[0,0.18]\mu\text{m}\cup [2.82,3]\mu\text{m}$, since $\xi_{\text{SC}}=0.36~\mu$m in the pristine limit.

We can then roughly estimate this weight $I^{(1)}$ as
$\frac{\int_{x=0}^{\xi/2} \abs{\psi(x)}^2 dx}{\int_{x=0}^{\infty} \abs{\psi(x)}^2 dx}        \approx\frac{\int_{x=0}^{\xi/2} e^{-2x/\xi} dx}{\int_{x=0}^{\infty} e^{-2x/\xi} dx}        =\frac{e-1}{e}\approx 63\%$.
Here, $\psi(x)$ is the wave function in a semi-infinite pristine long wire.  
This can serve as a reference threshold for whether the second criterion of localized MZMs is satisfied.

However, this threshold tends to underestimate topology because we ignore the oscillatory part in the wave function, and the reality is always the wires with finite size where these oscillations are typically present adversely affecting topology because of the overlap between the end MZMs. 
Therefore, we can relax the threshold a bit, by choosing a smaller value of 50\%, as a working definition to define isolated localized states. Namely, $I^{(1)}>50\%$ means the localized MZMs are present at the ends of the wire.
In principle, one can impose a more stringent criterion by shortening the length scale of the end region and/or decreasing the threshold of $I^{(1)}$.

\subsubsection{Indicator 2: Total number of MZMs}
The second indicator we consider is the total number of MZMs within the wire. This quantity is originally proposed in Ref.~\onlinecite{dassarma2023spectral} to differentiate the MZMs from the trivial fermionic zero-energy Andreev bound states (ABSs).
The quantity is defined based on the LDOS as
\begin{equation}
    I^{(2)}= \pi \eta \sum_{x_i\in [0,L]}\rho(\omega=0,x_i).
\end{equation}
and, therefore, we expect the ideal pair of MZMs to show the value of 2, and any value larger than 2 means more than one pair of MZMs exists, which are likely to live in the bulk of the wire, causing potential topology problems. In contrast, any value smaller than 2 means the absence of the zero-energy state since MZMs must come in pairs.
In practice, we choose a tolerance of $\pm 10\%$, i.e., any value of $I^{(2)}$ in the range of [1.8,2.2] is considered to indicate the presence of the isolated localized MZMs at the ends.

\subsubsection{Indicator 3: Kullback-Leibler divergence}
The third indicator is motivated by statistics and information theory. 
In principle, the LDOS can be viewed as an (unnormalized) probability distribution of the position of a particle because $\rho(\omega,x)\propto\sum_{\epsilon_i} \delta(\omega-\epsilon_i)\abs{\psi_i(x)}^2$. 
Therefore, if we have a reference state that represents a standard `isolated localized MZMs' state, we can compare any arbitrary state with this reference state.
The measure of ``distance'' (rigorously speaking, it is divergence because it is not symmetric) will provide an indicator of whether these two states (two probability distributions) are similar.
Therefore, we choose the reference state to be the LDOS for a zero-energy state in the pristine limit at a Zeeman field which maximaizes the topological gap (i.e., $V_Z=0.65$ meV in Fig.~\ref{fig:pristine}(k)). Thus, this indicator is defined as
\begin{equation}
    \begin{split}
        I^{(3)}=&D_{\text{KL}}(\rho(\omega=0,x_i)_{\text{test}}||\rho(\omega=0,x_i)_{\text{ref}})\\
        =&D_{\text{KL}}(\rho(\omega=0,x_i)_{\text{disorder}}||\rho(\omega=0,x_i)_{\text{MZM}}).        
    \end{split}
\end{equation}
Here we choose Kullback-Leibler (KL) divergence as the measure of `distance' defined as 
\begin{equation}
    D_{\text{KL}}(p(x)||q(x))= \int p(x) \log(\frac{p(x)}{q(x)}) dx,
\end{equation}
where $p(x)$ and $q(x)$ are the LDOS after the normalization over the entire wire.

The KL divergence is minimized to zero if the $p(x)=q(x)$ while could diverge to infinity if $p(x)$ and $q(x)$ are very different.
Therefore, we set the threshold to be 0.9 corresponding to the KL divergence between a trivial state (e.g., $V_Z<0.54$ meV in Fig.~\ref{fig:pristine}) and the reference state (i.e., $V_Z=0.65$ meV in Fig.~\ref{fig:pristine}). Namely, we conclude an isolated localized state if $I^{(3)}<0.9$. 

\subsubsection{Indicator 4: Bulk and end LDOS ratio}
The fourth indicator is motivated directly by the fact that for any well-defined localized MZMs, we must have the LDOS vanishing in the bulk of the wire and large at the ends of the wire.
Therefore, we just impose these restrictions by having:
\begin{enumerate}
    \item The amplitude of LDOS at the midpoint of the wire is less than the LDOS peaks at the ends by a certain threshold.
    \item The amplitude of LDOS in the bulk of the wire can never exceed the average of the peaks on two ends of the wire.
\end{enumerate}

The logic of these two restrictions leads to a possible implementation to define the following set of two ratios:
\begin{enumerate}
    \item Ratio of midpoint LDOS over end LDOS, i.e.,
    \begin{equation}\label{eq:ratio1}
        I_1^{(4)}=\frac{\rho(\omega=0,x_i=\frac{L}{2})}{\frac{1}{2}\left[ \displaystyle\max_{x_i \in [0,\frac{\xi}{2}]} \rho(\omega=0,x_i) + \max_{x_i \in [L-\frac{\xi}{2},L]} \rho(\omega=0,x_i) \right]}.
    \end{equation}
    \item Ratio of maximal LDOS in the bulk over the end LDOS, i.e.,
    \begin{equation}\label{eq:ratio2}
        I_2^{(4)}=\frac{\displaystyle\max_{x_i\in \text{Bulk}}\rho(\omega=0,x_i)}{\frac{1}{2}\left[ \displaystyle\max_{x_i \in [0,\frac{\xi}{2}]} \rho(\omega=0,x_i) + \max_{x_i \in [L-\frac{\xi}{2},L]} \rho(\omega=0,x_i) \right]}.
    \end{equation}
\end{enumerate}
Here the end LDOS is take as the ``first peak'' from the end.
We choose the threshold for the first ratio as $I_1^{(4)}<4\%$, and for the second ratio as $I_2^{(4)}<1$ to declare the isolated localized MZMs.

\subsection{Benchmark of the four indicators}\label{sec:benchmark}

Given these four indicators, we need to test whether what we previously concluded visually by looking at the LDOS can now be quantified by the indicators.
Therefore, we benchmark them by applying the four indicators to the previous results ranging from pristine regime to strong disorder regime.
We find that the four indicators actually can have different predictions for topology in disordered short wires.
There is also the issue of some arbitrariness in assigning the threshold on the benchmark for each indicator, but this arbitrariness is also present in practice because true topology applies only in the thermodynamic limit, and a finite system is always subject to some threshold subjectivity. This can to some extent be adjusted by applying the indicators to the pristine system first.

\subsubsection{Pristine}\label{sec:benchmark_pristine}

\begin{figure*}[ht]
    \centering
    \includegraphics[width=6.8in]{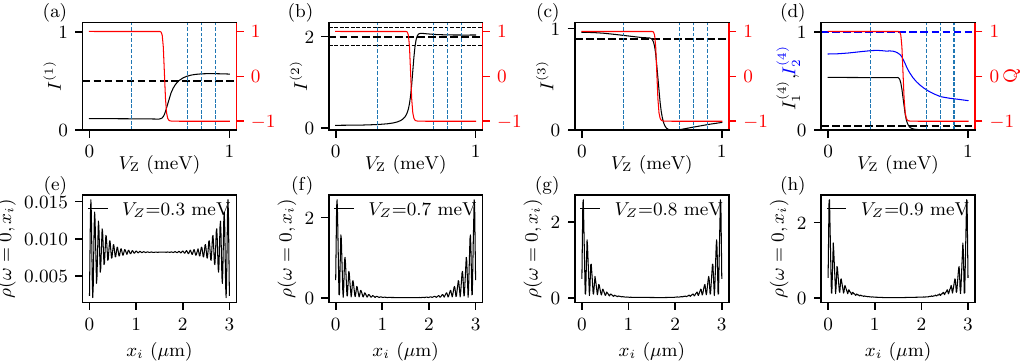}
    \caption{
        The benchmark of four indicators on the pristine 3-micron wire in Fig.~\ref{fig:pristine}. 
        (a) End weight $I^{(1)}$ (black) on the left axis, and TV $Q$ (red) on the right axis as a function of $V_Z$ (same for (a-d)). The horizontal dashed line indicates the threshold of $I^{(1)}$ (50\%). 
        (b) Total number of MZMs $I^{(2)}$ (black) on the left axis. The horizontal dashed line indicates the ideal value of $I^{(2)}$, and the dotted line indicates the tolerant range of $I^{(2)}$ around 2 ([1.8,2.2]).
        (c) KL divergence $I^{(3)}$ (black) on the left axis. The horizontal dashed line indicates the threshold of $I^{(3)}$ (0.9).
        (d) Ratio of midpoint LDOS and end LDOS $I_1^{(4)}$ (black), and the ratio of maximal in the bulk and end LDOS $I_2^{(4)}$ (blue). The horizontal dashed line indicates the threshold of $I_1^{(4)}$, 0.04 (black), and $I_2^{(4)}$, 1 (blue), respectively.
        (e-h) The LDOS at four Zeeman fields indicated by the vertical dashed line in (a-d).
        }
    \label{fig:pristine_metrics}
\end{figure*}

We first apply the four indicators to the pristine wire to provide a reference for the subsequent benchmarks as shown in Fig.~\ref{fig:pristine_metrics}(a-d). Each panel on the first row shows one indicator from $I^{(1)}$ to $I^{(4)}$ as a function of the Zeeman field $V_Z$. 
We also present the corresponding TV $Q$ on the right axis (red line) for comparison.

In Fig.~\ref{fig:pristine_metrics}(a), $I^{(1)}$ shows that the end weight is very low in the trivial regime, while it saturates around 0.6 in the topological regime, close to the ideal value of 0.63 as mentioned above.
In Fig.~\ref{fig:pristine_metrics}(b), $I^{(2)}$ shows that the total number of MZMs is zero in the trivial regime, while it is 2 in the topological regime as expected.
In Fig.~\ref{fig:pristine_metrics}(c), $I^{(3)}$ shows that the KL divergence is around 1 in the trivial regime (because the LDOS in the trivial regime is very different from that in the topological regime), and then becomes very small in the topological regime. (The minimum of zero at $V_Z=0.65$ meV is simply because the reference state is chosen there.)
In Fig.~\ref{fig:pristine_metrics}(d), $I_1^{(4)}$ (black line) shows that the ratio of the midpoint LDOS and the end LDOS becomes exponentially small beyond TQPT, and $I_2^{(4)}$ (blue line) shows that bulk LDOS can never exceed the end LDOS.
All these indicators work as expected in the pristine limit.

In the bottom panels of Fig.~\ref{fig:pristine_metrics}, we present the LDOS at four Zeeman fields, where Fig.~\ref{fig:pristine_metrics}(e) is in the trivial regime, and Figs.~\ref{fig:pristine_metrics}(f-h) are in the topological regime, corresponding to the vertical dashed lines in the top panels.
This provides a direct visualization of how the LDOS at zero energy evolves from an extended state (Fig.~\ref{fig:pristine_metrics}(e)) to a localized state (Figs.~\ref{fig:pristine_metrics}(f-h)) in the pristine limit.

All four indicates can accurately manifest the change in the distribution of LDOS for zero energy state, and their predicted criticalities converge to the change of TV.
The fact that all four indicators give consistent results is due to the perfect bulk-edge correspondence in the topological regime in the pristine limit. 
However, things become more complicated in the interplay of disorder and short wire--- the topological phase transition and the change in the LDOS from extended to localized state may not happen at the same $V_Z$. Therefore, the four indicators, which emphasize different aspects, may give different results.

\subsubsection{Weak disorder regime}\label{sec:benchmark_weak_disorder}
Thus, to provide physical insights on the short disordered wire, instead of just stating everything is complicated, we need to revisit the previous disordered results and see how the four indicators behave.

\begin{figure*}[ht]
    \centering
    \includegraphics[width=6.8in]{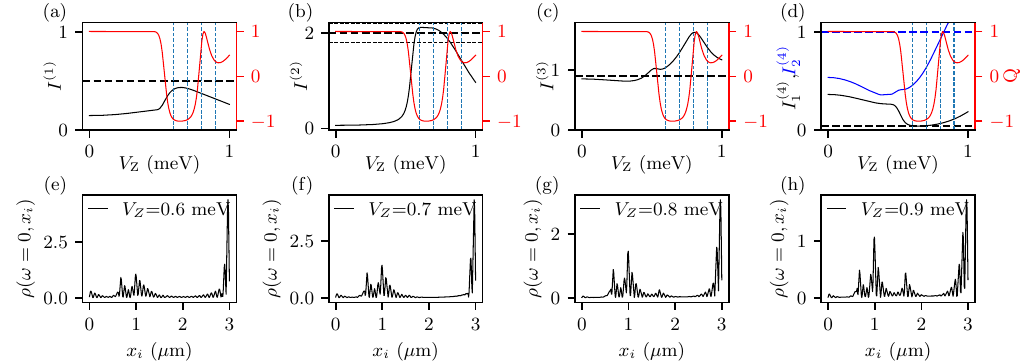}
    \caption{
    The benchmark of four indicators on the weak disorder 3-micron wire with $n=5$ and $k=9$ in Fig.~\ref{fig:thy_positive}. 
    (a) End weight $I^{(1)}$ (black) on the left axis, and TV $Q$ (red) on the right axis as a function of $V_Z$ (same for (a-d)). The horizontal dashed line indicates the threshold of $I^{(1)}$ (50\%). 
    (b) Total number of MZMs $I^{(2)}$ (black) on the left axis. The horizontal dashed line indicates the ideal value of $I^{(2)}$, and the dotted line indicates the tolerant range of $I^{(2)}$ around 2 ([1.8,2.2]).
    (c) KL divergence $I^{(3)}$ (black) on the left axis. The horizontal dashed line indicates the threshold of $I^{(3)}$ (0.9).
    (d) Ratio of midpoint LDOS and end LDOS $I_1^{(4)}$ (black), and the ratio of maximal in the bulk and end LDOS $I_2^{(4)}$ (blue). The horizontal dashed line indicates the threshold of $I_1^{(4)}$, 0.04 (black), and $I_2^{(4)}$, 1 (blue), respectively.
    (e-h) The LDOS at four Zeeman fields indicated by the vertical dashed line in (a-d).
    }
    \label{fig:thy_positive_metrics}
\end{figure*}

We first revisit the positive results with $n=5$ (Fig.~\ref{fig:thy_positive}) to see whether the four indicators can tell us that the wire is without isolated MZMs localized at both ends of the wire, despite the TV being negative within the regime of $V_Z\in[0.6,0.7]$ meV.
This is a crucial test for the efficacy of these indicators since TV by itself suggests that the system is topological.
We present the benchmark of the four metrics in Fig.~\ref{fig:thy_positive_metrics}.

In Fig.~\ref{fig:thy_positive_metrics}(a), $I^{(1)}$ never meets the threshold of end localized state ($I^{(1)}<0.5$) when the TV is -1. This is consistent with Fig.~\ref{fig:thy_positive} and Fig.~\ref{fig:thy_positive_metrics}(e-h) where the LDOS is not well localized at the left end. Thus, this indicator $I^{(1)}$ correctly reflects the absence of isolated localized MZM at the left end.

In Fig.~\ref{fig:thy_positive_metrics}(b), $I^{(2)}$ tells that the total number of MZMs is 2, which passes the criterion.
However, in the LDOS on the bottom panel, we do see another pair of MZMs close to the left end. The failure to detect another pair of peaks may be due to the arbitrariness of how zero energy is defined (i.e., a resolution problem). Namely, in finite wires, because MZMs overlap in the bulk, the total number of MZMs may change if one calculates it on a finer scale of energy. Therefore, this indicator $I^{(2)}$ fails here.

In Fig.~\ref{fig:thy_positive_metrics}(c), KL divergence $I^{(3)}$ shows a value larger than one, indicating that the distribution is very different from the reference distribution for the localized MZMs. Therefore, this indicator $I^{(3)}$ correctly supports the conclusion.

In Fig.~\ref{fig:thy_positive_metrics}(d), although the ratio of maximal LDOS over the end LDOS in the bulk remains $I_2^{(4)}$ smaller than 1 within the regime where TV is negative, the ratio of midpoint over end LDOS $I_1^{(4)}$ is always larger than the threshold 0.04, which violates the criterion. Therefore, it indicates the absence of isolated localized end states, which itself is a correct indicator.

\subsubsection{Strong disorder regime}\label{sec:benchmark_strong_disorder}

\begin{figure*}[ht]
    \centering
    \includegraphics[width=6.8in]{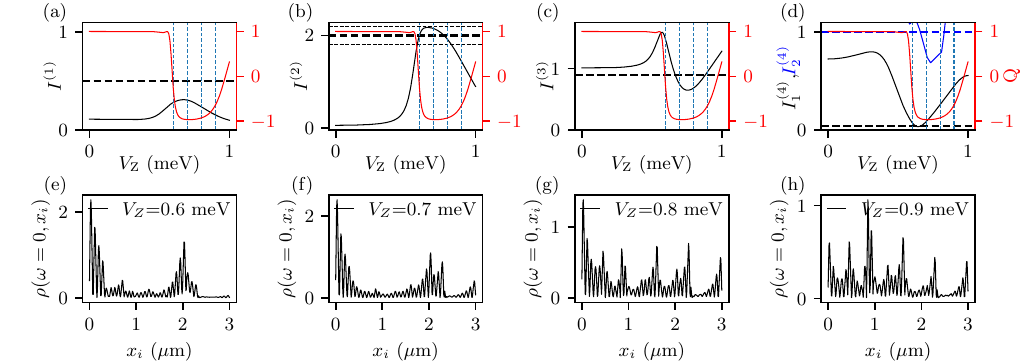}
    \caption{
    The benchmark of four indicators on the strong disorder 3-micron wire with $n=11$ and $k=9$ in Fig.~\ref{fig:1113}. 
    (a) End weight $I^{(1)}$ (black) on the left axis, and TV $Q$ (red) on the right axis as a function of $V_Z$ (same for (a-d)). The horizontal dashed line indicates the threshold of $I^{(1)}$ (50\%). 
    (b) Total number of MZMs $I^{(2)}$ (black) on the left axis. The horizontal dashed line indicates the ideal value of $I^{(2)}$, and the dotted line indicates the tolerant range of $I^{(2)}$ around 2 ([1.8,2.2]).
    (c) KL divergence $I^{(3)}$ (black) on the left axis. The horizontal dashed line indicates the threshold of $I^{(3)}$ (0.9).
    (d) Ratio of midpoint LDOS and end LDOS $I_1^{(4)}$ (black), and the ratio of maximal in the bulk and end LDOS $I_2^{(4)}$ (blue). The horizontal dashed line indicates the threshold of $I_1^{(4)}$, 0.04 (black), and $I_2^{(4)}$, 1 (blue), respectively.
    (e-h) The LDOS at four Zeeman fields indicated by the vertical dashed line in (a-d).
    }
    \label{fig:1113_metrics}
\end{figure*}

Having verified the four indicators applying to the weak disorder regime, we now consider a nontopological wire in the strong disorder regime (Fig.~\ref{fig:1113}) where the TV shows negative values over $V_Z=0.6$ to 0.9 meV.
The results are presented in Fig.~\ref{fig:1113_metrics}.

In Fig.~\ref{fig:1113_metrics}(a), the end weight $I^{(1)}$ is always much smaller than the threshold of 0.5, which correctly indicates the absence of localized MZMs at the ends of the wire.
In Fig.~\ref{fig:1113_metrics}(b), although $I^{(2)}$ shows 2 MZMs, there are more peaks in the bulk of the wire as shown in Fig.~\ref{fig:1113_metrics}(f-h). Therefore, this indicator $I^{(2)}$ fails here.
In Fig.~\ref{fig:1113_metrics}(c), the KL divergence $I^{(3)}$ shows a much larger value than 1, which is a correct indicator.
In Fig.~\ref{fig:1113_metrics}(d), the ratio of the maximal bulk LDOS over the end LDOS $I_2^{(4)}$ (blue line) is mostly larger than 1, which does not pass the threshold. Thus, it is also a correct indicator.

Therefore, our indicators, except for the total number of MZMs $I^{(2)}$, can correctly reproduce the correct conclusion of the presence or absence of isolated localized MZMs at the ends of the wire regardless of the disorder regime.
We believe that the failure of $I^{(2)}$ most likely arises from the energy resolution issues associated with defining a zero energy.

\subsubsection{Statistics of the four indicators}\label{sec:statistics_indicators}
We present more benchmark results in Appendix~\ref{app:indicator}, and summarize the success/failure statistics of the four indicators in Table~\ref{tab:indicator}.

From Table~\ref{tab:indicator}, we can conclude that the second indicator $I^{(2)}$ is the least reliable one (with accuracy less than 28\%) because the number of MZMs at zero energy is a bit arbitrary and also ill-defined if one looks at a finer scales of energy.
The first indicator $I^{(1)}$ using the end weight works well in most scenarios (with an accuracy of around 71\%). It is physical and intuitive, with the only arbitrariness being the length scale of the end and the threshold of the weight. However, once these two are fixed, the indicator is well-defined.
The third indicator $I^{(3)}$ using KL divergence is mathematically rigorous and scores even better (with an accuracy of around 85\%). However, it is difficult to connect $I^{(3)}$ directly to a physical quantity since it is a measure of the ``divergence'' between two probability distributions.
The fourth indicator, $I_1^{(4)}$ and $I_2^{(4)}$, directly uses the ratio of bulk LDOS over the edge LDOS while also ensuring the bulk LDOS does not exceed the edge LDOS. This is simple, intuitive, and also the most accurate with a perfect statistical success in our analysis.

\begin{table*}[ht]
    \centering
    \caption{Summary for four indicators}
    \label{tab:indicator}
    \begin{tabularx}{\linewidth}{XXXX}
        \hline
        Method & Accuracy & Pros & Cons\\
        \hline
        End weight, $I^{(1)}$ & {5/7 (Figs.~\ref{fig:thy_positive_metrics}, \ref{fig:649_metrics}, \ref{fig:971_metrics}, \ref{fig:1001_metrics},  \ref{fig:1113_metrics})} & Simple and intuitive & {Arbitary in the length scale of `end' and threshold}\\
        \hline
        Total number of MZMs, $I^{(2)}$ & {2/7 (Figs.~\ref{fig:658_metrics}, \ref{fig:703_metrics})} & {Simple} & Lowest accuracy and arbitrary\\
        \hline
        KL divergence, $I^{(3)}$ & {6/7 (Figs.~\ref{fig:thy_positive_metrics}, \ref{fig:649_metrics}, \ref{fig:703_metrics}, \ref{fig:971_metrics}, \ref{fig:1001_metrics}, \ref{fig:1113_metrics})} & Statistical meaning & Indirect to physics\\
        \hline
        Ratio, $I_1^{(4)}$ and $I_2^{(4)}$ & 7/7 (Figs.~\ref{fig:thy_positive_metrics}, \ref{fig:649_metrics},~\ref{fig:658_metrics}, \ref{fig:703_metrics}, \ref{fig:971_metrics},~\ref{fig:1001_metrics}, \ref{fig:1113_metrics}) & Simple, intuitive, and the most accurate &  \\
    \end{tabularx}
\end{table*}

Therefore, in the following section, we will choose the third $I^{(3)}$ and fourth indicators $I^{(4)}$ as reliable indicators to detect the localized MZMs in the disordered short wire.

\subsection{Predictive power of TV}\label{sec:predictive_power}
With the benchmark results for the four indicators, we now study the predictive power of the topological visibility $Q$.
Namely, we generate an ensemble of $L=3~\mu$m with $k=9$ for different $n$ ($n=0$, and $5\le n\le 12$), and study the correlation between the topological visibility $Q$  and the $I^{(3)}$, and $I^{(4)}$, respectively.

\begin{figure*}[ht]
    \centering
    \includegraphics[width=6.8in]{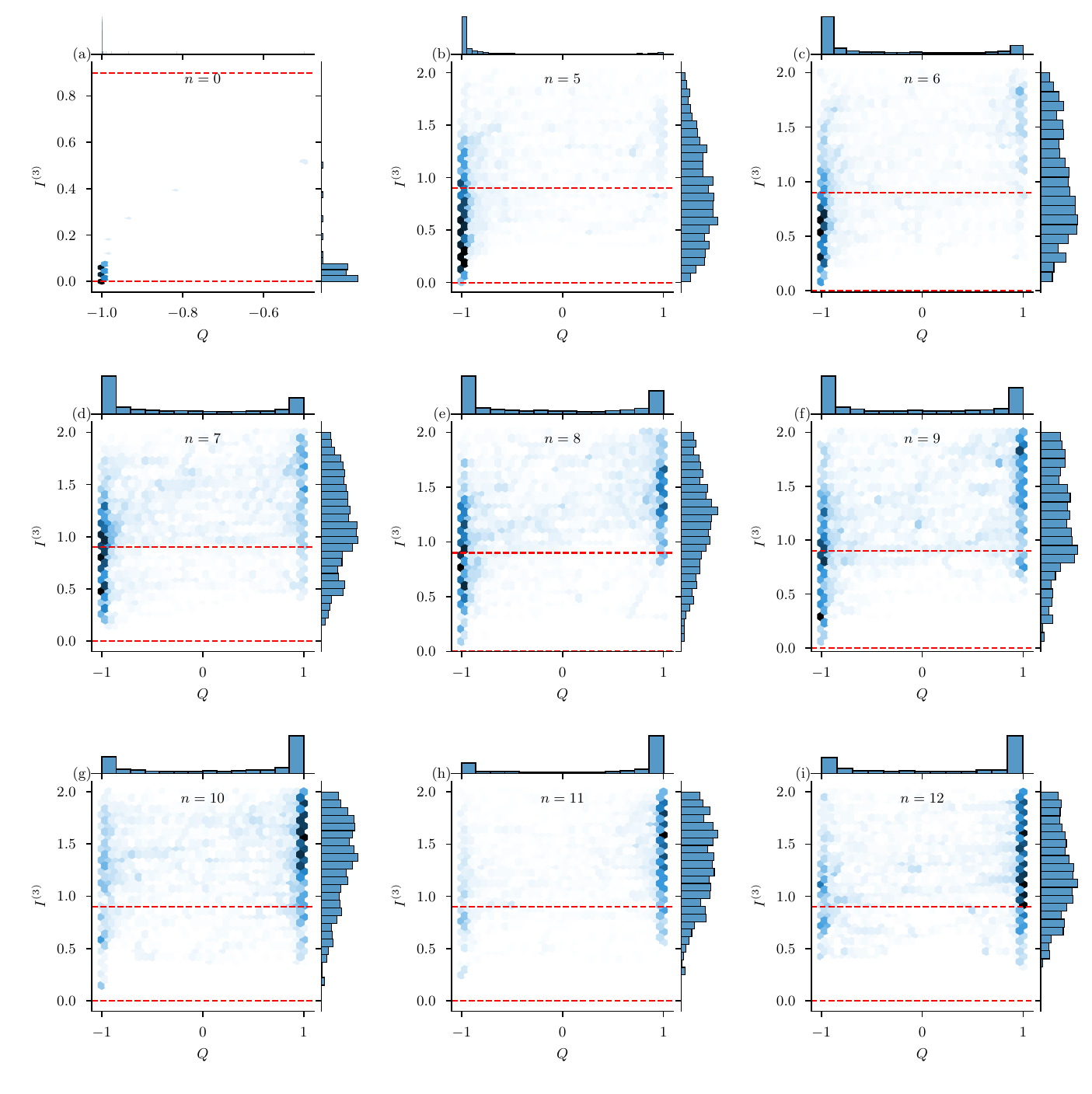}
    \caption{
    The joint distribution of TV $Q$ and KL divergence $I^{(3)}$ for 3-micron wires with $k=9$ for different $n=0$ (a), and $5\le n\le 12$ (b-i). 
    The regime between two red dashed lines indicates data points satisfying the threshold.
    Outliers ($I^{(3)}>2$) are suppressed for visibility.
    }
    \label{fig:joint_KL}
\end{figure*}

We first show the joint distribution of TV $Q$ and KL divergence $I^{(3)}$ in Fig.~\ref{fig:joint_KL} from the pristine limit $n=0$ to the strong disorder regime $n=12$. The density of the data points is interpolated using kernel density estimation and rasterized in hexagon bins.

Each panel shows all the data with a fixed $n$, where each data point in the panel corresponds to one spatial profile of zero-energy LDOS $\rho(\omega=0,x_i)$ at a specific Zeeman field $V_Z$ that is in the (nominal) topological regime (i.e., $V_Z>0.54$ meV).
Namely, we exclude the data points in the obviously trivial regime below the TQPT. This is because the LDOS at zero energy in the trivial regime is always inside the gap, and is usually very tiny (e.g., Fig.~\ref{fig:pristine_metrics}(e)) and does not have any features of interest.
The regions between two red dashed lines indicate the data points that have localized MZMs at the ends of the wire, i.e., $0<I^{(3)}<0.9$.
Around each panel, the top axis shows the marginal distribution of TV, while the right axis shows the marginal distribution of KL divergence $I^{(3)}$.

In the pristine limit at $n=0$ (Fig.~\ref{fig:joint_KL}(a)), we find that both TV and $I^{(3)}$ are correlated as all data points are localized on the lower right region of the panel, corresponding to $(Q, I^{(3)})\sim(-1,0)$
The absence of data points with $Q=+1$ is because we intentionally exclude the trivial regime.

As disorder is introduced, since $n=5$ (Fig.~\ref{fig:joint_KL}(b)), we find that the correlation between TV $Q$ and $I^{(3)}$ is relaxed. 
$I^{(3)}$ first begins to spread out, meaning that, although most of the data points still have a nontrivial TV, they begin to deviate from the localized MZMs state, i.e., bulk LDOS begins to grow.

As the disorder continues to increase, the weight of $I^{(3)}$ near zero is further reduced, and the TV also begins to spread out. For example, at $n=7$ (Fig.~\ref{fig:joint_KL}(d)), the trivial state begins to populate as indicated by the peak at $Q=+1$ on the top axis.

At $n=9$ (Fig.~\ref{fig:joint_KL}(f)), the topological states and trivial states are roughly balanced as indicated by the equal weights of $Q=+1$ and $Q=-1$.
Therefore, $n=9$ serves as a critical point, below which the topological states are statistically more likely, while above which the trivial states are more likely.
This is also consistent with our previous estimate of the intermediate disorder regime using the criterion Eq.~\eqref{eq:hierarchy}, where the SC coherence length ($\xi_{\text{SC}}\sim 0.36~\mu$m) is comparable to the MFP ($\xi_{\text{MFP}}\sim 0.3~\mu$m).

In the strong disorder regime when $n>9$ (Fig.~\ref{fig:joint_KL}(g-i)), the localized MZMs become fewer and fewer, and the trivial states dominate.
However, the trivial states ($Q\sim+1$ ) are not the only possible outcomes, there are a still few data points with $Q=-1$ but without any isolated localized MZMs ($I^{3}>0.9$) at the ends of the wire.

This is a direct demonstration of the fact that the TV itself may not serve as a sufficient indicator for the topological localized MZMs in the disordered system. 
On the other hand, since the lower right regions of all the panels are always empty, it means that the TV can only serve as a necessary condition, namely, as long as the $Q$ is positive, the existence of localized MZMs can almost be ruled out.
These results also bring out transparently the crucially important physics of disordered short wires, as in the Microsoft experiment, that the intermediate disorder regime in finite wires cannot be cast as definitively topological or definitely trivial based just on the TV values (and conductance matrix results) with the situation being highly nuanced and complex with topology and the absence of topology may coexist with small changes in the parameter values, which is a hallmark of mesoscopic physics.  We believe that braiding in this regime is likely to fail, and one should stick to the weak disorder regime where the topology is a meaningful concept even in realistic short wires.

\begin{figure*}[ht]
    \centering
    \includegraphics[width=6.8in]{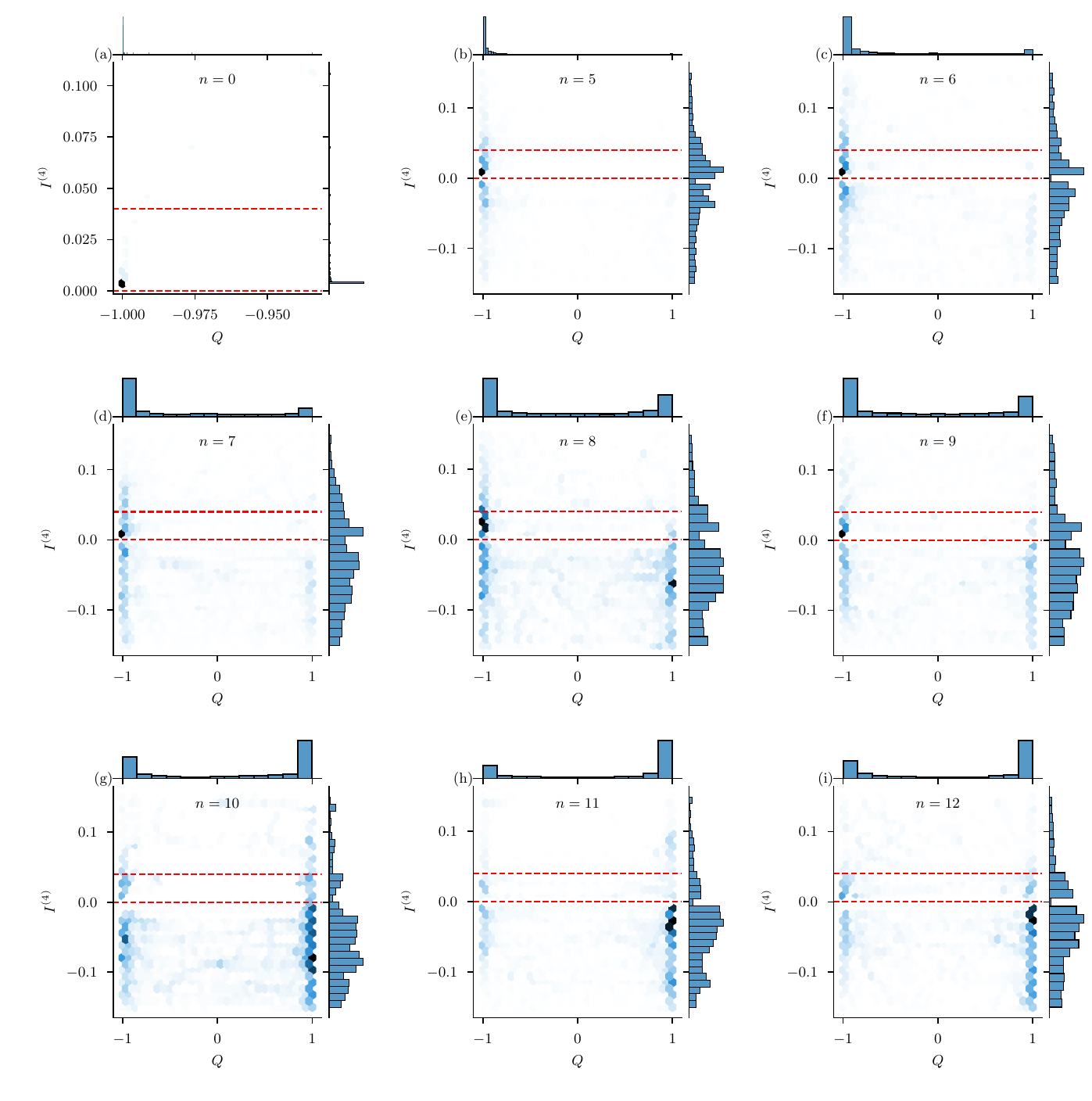}
    \caption{
    The joint distribution of TV $Q$ and ratio $I^{(4)}$ for 3-micron wires with $k=9$ for different $n=0$ (a), and $5\le n\le 12$ (b-i). 
    The regime between two red dashed lines indicates data points satisfying the threshold.
    Outliers ($\abs{I^{(4)}}>0.15$) are suppressed for visibility.
    }
    \label{fig:joint_ratio}
\end{figure*}

Similarly, we also plot the joint distribution of TV $Q$ and the fourth indicator on the ratio in Fig.~\ref{fig:joint_ratio}.
Here one technical issue is that the fourth indicator has two ratios, $I_1^{(4)}$ for the ratio between end LDOS and bulk LDOS, and $I_2^{(4)}$ for the ratio between the maximal bulk LDOS and end LDOS. 
Therefore, to avoid plotting them separated, we synthesize the two into one value $I^{(4)}$, by encoding the first ratio $I_1^{(4)}$ into the magnitude of $I^{(4)}$, and the second ratio $I_2^{(4)}$ into the sign of $I^{(4)}$--- if $I_2^{(4)}$ is larger (smaller) than 1, then the sign is negative (positive). Namely,
\begin{equation}
    I^{(4)}=\text{sign}(I_2^{(4)}-1)I_1^{(4)}.
\end{equation}
Therefore, the isolated localized MZMs are then indicated by the region of $0<I^{(4)}<0.04$, which is indicated by the regions between two red dashed lines in each panel of Fig.~\ref{fig:joint_ratio}.

From Fig.~\ref{fig:joint_ratio}, we find the same trend as what is inferred from Fig.~\ref{fig:joint_KL}. 
In the pristine limit $n=0$, all data points are concentrated at the lower left corner indicating a correlation of $Q=-1$ (nontrivial TV)and $I^{(4)}=0$ (localized state).

As disorder increases ($n$ increases), the centroid shifts gradually from the left corner above the dashed line for $I^{(4)}=0$ [$(Q, I^{(4)})\approx(-1,0^+)$] to the right corner below the dashed line for $I^{(4)}=0$ [$(Q, I^{(4)})\approx(+1,0^-)$]. This shift indicates that the state evolves from a topological localized state in the weak disorder limit ($n<9$) to a trivial bulk state in the strong disorder limit ($n>9$).

However, since there are still data points with $Q=-1$ and $I^{(4)}<0$ for any $n>0$, it means the TV is a good indicator only in the pristine wire with $n=0$, while its predictive power becomes weaker as the disorder increases ($n$ increases). This is consistent with the conclusion obtained from Fig.~\ref{fig:joint_KL}.
We think that using TV as a decisive indicator for topology is incorrect in disordered finite wires since the definition really applies in the thermodynamic limit to precisely pinpoint the TQPT only. Its use to define a topological regime may simply not work decisively in finite disordered wires.

This motivates us to directly map out the predictive power of the TV. Especially, we are concerned about the false positive (FP) rate of using TV $Q$ alone to predict the topological localized MZMs.
Here, we choose the most stringent criterion and define the true positives as the simultaneous satisfaction of all three indicators: (1) TV shows a nontrivial value; (2) $I^{(3)}<0.9$; and (3) $0<I^{(4)}<0.04$.

\begin{figure}[ht]
    \centering
    \includegraphics[width=3.4in]{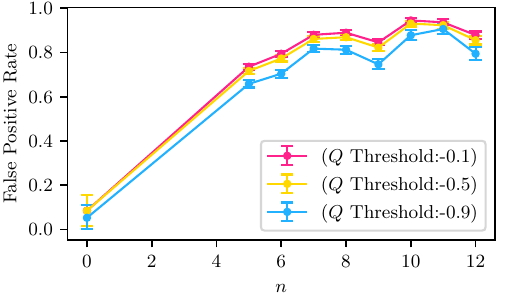}
    \caption{false-positive rate of using TV alone to predict topological localized MZM state with three choices of TV thresholds $Q<-0.1$ (red), $Q<-0.5$ (orange), and $Q<-0.9$ (cyan). The error bar indicates the 95\% binomial proportion confidence interval.}
    \label{fig:TV_threshold}
\end{figure}

The result of the FP rate is shown in Fig.~\ref{fig:TV_threshold}, where we choose three different thresholds for TV being nontrivial.
We find that the FP rate is not sensitive to the choice of TV threshold, as long as the threshold is negative. 
Therefore, in the following results, without loss of generality, we adopt the threshold of `$Q$ being negative' as $Q<-0.9$.

As disorder increases ($n$ increases), the FP rate continues to increase until around $n=9$ (intermediate disorder regime), and then it saturates at around 90\% in the strong disorder regime.
This is a relatively high FP rate, which again confirms the low predictive power of TV in the presence of disorder.
Thus, we have established that the common notion of topology being given by a binary choice (some invariant being positive or negative) does not work for finite disordered wires.

\subsection{Experimental relevance}\label{sec:exp_relevance}

\begin{figure}[ht]
    \centering
    \includegraphics[width=3.4in]{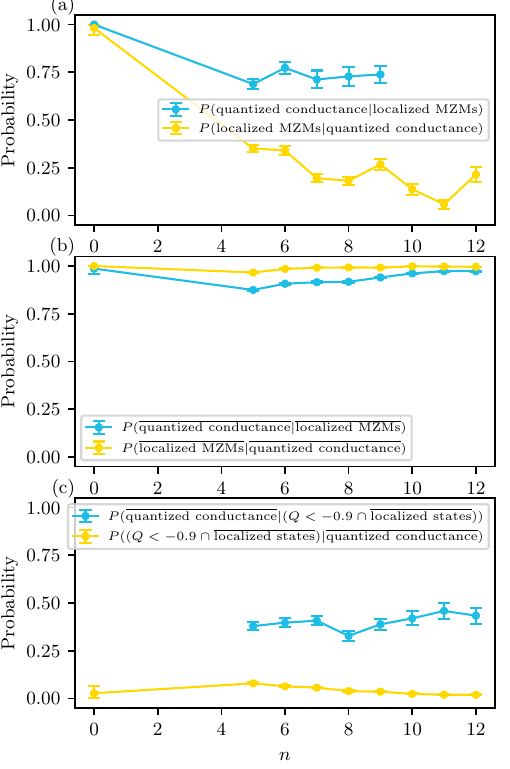}
    \caption{
        (a) The conditional probability of ``quantized conductance" given ``localized MZMs'' (blue), and ``localized MZMs'' given ``quantized conductance" (orange) for different $n$. The error bar indicates the 95\% binomial proportion confidence interval.
        (b) The conditional probability of ``nonquantized conductance'' given ``the absence of localized MZMs modes'' (blue), and ``the absence of localized MZMs modes'' given ``nonquantized conductance'' (orange) for different $n$. Here, $\overline{A}$ indicates the negation of the event $A$.
        (c) The conditional probability of ``nonquantized conductance'' given ``nontrivial topology and the absence of localized MZMs modes'' (blue), and ``nontrivial topology and the absence of localized MZMs modes'' given ``nonquantized conductance'' (orange) for different $n$. 
        }
    \label{fig:condprob}
\end{figure}

All of the previous indicators, TV $Q$, $I^{(3)}$, and $I^{(4)}$, are theoretical and can only be applied to simulations. 
Therefore, we connect our theoretical indicators to the experimental results, i.e., the differential conductance spectrum.

In this section, we aim to study whether the ZBCP strengths from both ends correlate with the existence of topological localized MZMs.
Namely, we want to answer the two questions:
(1) Does the presence of topological localized MZMs always lead to quantized ZBCP at both ends?
(2) Does quantized ZBCP at both ends always lead to the presence of topological localized MZMs?

Here, the ``localized MZMs'' are defined the same as before from the two aspects: 
(1) nontrivial topology indicated by TV; 
(2) the LDOS shows an isolated localized zero-energy state indicated by both $I^{(3)}$ and $I^{(4)}$.
The ``quantized conductance'' is defined as the ZBCP within $\pm$10\% of $2e^2/h$ (i.e., 1.8 to 2.2 $e^2/h$), namely, we require the deviation of quantized conductance as per
\begin{equation}\label{eq:G_diff}
    \Delta G\equiv\frac{1}{2}\sum_{i=\left\{ L,R \right\}}\abs{G_{ii}(V_{\text{bias}}=0)- \frac{2e^2}{h}}<0.2~\frac{e^2}{h}.
\end{equation}

To answer these two questions from a statistical point of view, we compute the Bayesian conditional probability conditioned on each other (i.e., `quantized conductance' and `localized MZMs') in Fig.~\ref{fig:condprob}(a).
Namely, we compute 
(1) the probability of measuring quantized conductance given the state is isolated localized MZMs (blue line), denoted as $P(\text{quantized conductance}|\text{localized MZM})$; 
(2) the probability of isolated localized MZMs manifesting a quantized conductance (orange line) $P(\text{localized MZM}|\text{quantized conductance})$.
Again, each Zeeman field $V_Z$ is one sample in the total ensemble (however, we did not exclude the nominal trivial regime this time). 
The missing points for large $n$ (strong disorder regime) in $P(\text{quantized conductance}|\text{localized MZM})$ arise because there is no sufficient instance of localized MZMs in the strong disorder regime in the first place (and thus it is meaningless to discuss its Bayesian conditional probability). 
Namely, $P(\text{quantized conductance}|\text{localized MZM})\equiv \frac{P(\text{quantized conductance} ~\cap~ \text{localized MZM})}{P(\text{localized MZM})}$ is not well-defined because $P(\text{localized MZM})$ itself is almost zero.

If the two events (`quantized conductance' and `localized MZMs') are equivalent, then we expect to see both conditional probabilities are 1, which is indeed the case of the pristine wire ($n=0$).
However, as disorder increases, the two conditional probabilities begin to diverge. The probability $P(\text{quantized conductance}|\text{localized MZM})$ remains close to 1 as $n>0$, however, the reciprocal case  $P(\text{localized MZM}|\text{quantized conductance})$ becomes much smaller and approach to 0 in the strong disorder regime ($n>9$).

Therefore, it would be intriguing to study those samples that fall outside of these two events. 
Namely, we would like to know the case: (1) it has localized MZMs but not quantized conductance, accounting for $P(\text{quantized conductance}|\text{localized MZM})<1$, and (2) it has quantized conductance but without the localized MZMs, accounting for $P(\text{localized MZM}|\text{quantized conductance})<1$.
We do this next.

\subsubsection{Localized MZMs without quantized ZBCP}\label{sec:localized_MZM}
\begin{figure*}[ht]
    \centering
    \includegraphics[width=6.8in]{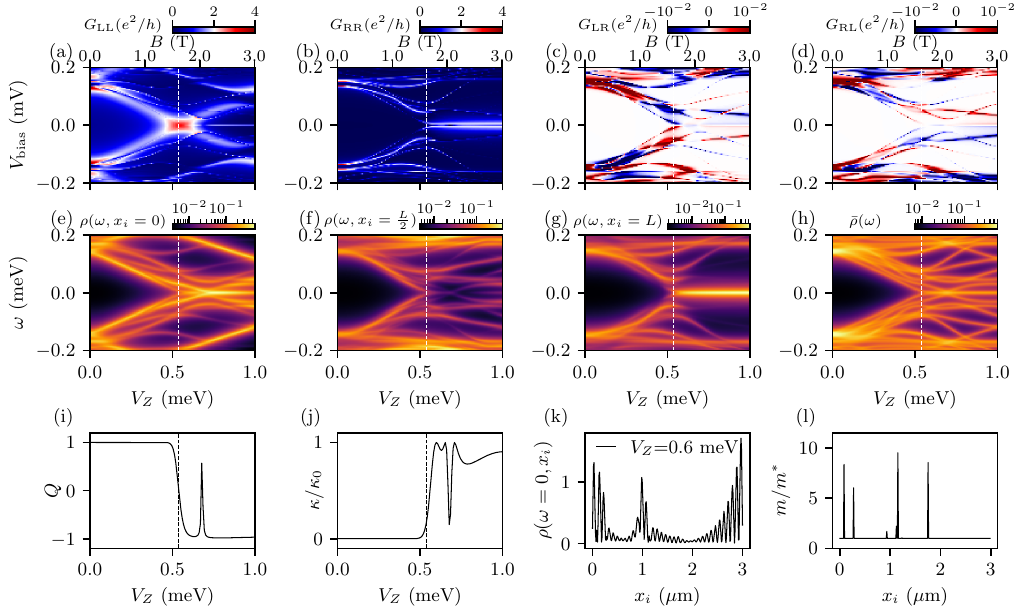}
    \caption{
    An example of localized MZMs without quantized ZBCP in a 3-micron wire with $n=7$ and $k=9$.
    (a-d) Local ($G_{\text{LL}}$ and $G_{\text{RR}}$) and nonlocal ($G_{\text{LR}}$ and $G_{\text{RL}}$) conductance spectra. The bottom (top) axis shows the Zeeman (magnetic) field. The vertical dashed line indicates nominal TQPT.
    (e-g) LDOS at left, midpoint, and right end. 
    (h) Total DOS.
    (i-j) Topological visibility $Q$ and thermal conductance $\kappa$.
    (k) LDOS at zero energy for $V_Z=0.6$ meV.
    (l) Spatial profile of effective mass. 
    }
    \label{fig:746}
\end{figure*}

\begin{figure*}[ht]
    \centering
    \includegraphics[width=6.8in]{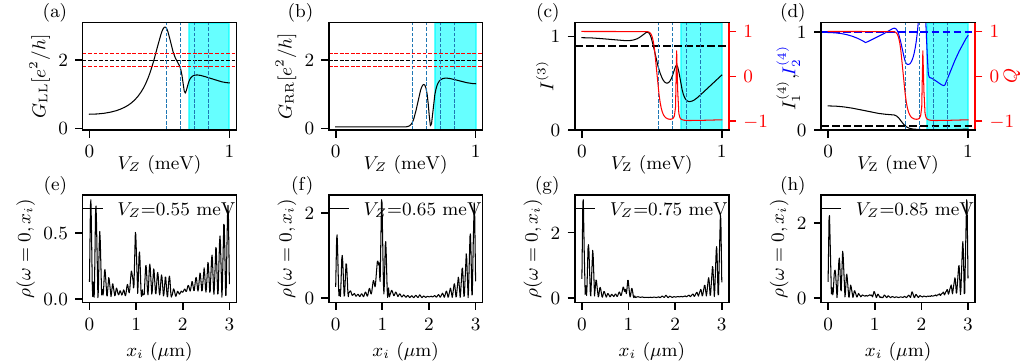}
    \caption{
    The benchmark of $I^{(3)}$ and $I_{1,2}^{(4)}$ on an example of localized MZMs without quantized ZBCP (cyan-shaded regions) in a 3-micron wire with $n=7$ and $k=9$ in Fig.~\ref{fig:746}.
    (a-b) Line cuts of the local ($G_{\text{LL}}$ and $G_{\text{RR}}$) conductance at zero bias. The red dashed line indicates the threshold of quantized conductance.
    (c) KL divergence $I^{(3)}$ (black) on the left axis, and TV $Q$ (red) on the right axis as a function of $V_Z$. The horizontal dashed line indicates the threshold of $I^{(3)}$ (0.9).
    (d) Ratio of midpoint LDOS and end LDOS $I_1^{(4)}$ (black), and the ratio of maximal in the bulk and end LDOS $I_2^{(4)}$ (blue). TV $Q$ (red) on the right axis as a function of $V_Z$. The horizontal dashed line indicates the threshold of $I_1^{(4)}$, 0.04 (black), and $I_2^{(4)}$, 1 (blue), respectively.
    (e-h) The LDOS at four Zeeman fields indicated by the vertical dashed line in (a-d).
    }
    \label{fig:746_cond_metrics}
\end{figure*}

An example of the first case is shown in Fig.~\ref{fig:746} along with Fig.~\ref{fig:746_cond_metrics}.
We find that the local conductance spectra show a weak ZBCP with a thin linewidth and small conductance over $V_Z\in[0.7,1]$ (see Fig.~\ref{fig:746}(a-b)).
In the LDOS on the bottom row, we notice that there is always a zero energy state localized in the bulk of wire at around 1 micron.
Therefore, this example should not be considered localized MZMs due to the bulk states. 
However, both indicators imply that the localized states exist from $V_Z=0.7$ to 1 meV (cyan shaded regions in Fig.~\ref{fig:746_cond_metrics}).
The reason for the indicator not capturing the bulk state is that the peak in the bulk is slightly deviated from the midpoint at 1.5 microns, and the height of the peak is also not very prominent.

We also study other examples of the same type as shown in Appendix~\ref{app:localized_MZM}, and confirm that the discrepancy of $P(\text{quantized conductance}|\text{localized MZM})$ from 1 all comes from the imperfect criterion in $I^{(3)}$ and $I^{(4)}$.
Namely, there are some data points that are not perfect localized MZMs are falsely identified as localized MZMs.
This is an artifact of the indicator $I^{(3)}$ and $I^{(4)}$ because the accuracy of these indicators is subjected to the choice of the threshold.
However, this does not mean that the indicators are not valid but should rather be understood as that they can provide a direction for improvement.
Because, ultimately, what is being proposed in this paper is the logic of using the criterion on LDOS to select the localized MZMs, and Eq.~\eqref{eq:ratio1} is just one possible implementation.
Therefore, to reduce the false-positive rate of the indicator of topological MZMs, we can impose more stringent criteria by modifying the detail of Eqs.~\eqref{eq:ratio1} and~\eqref{eq:ratio2}. For example, we can modify Eq.~\eqref{eq:ratio1} by using the average LDOS over a finite region of the bulk instead of just one point right at the midpoint, which will exclude the case shown in Fig.~\ref{fig:746}.

\subsubsection{Quantized ZBCP without localized MZMs}\label{sec:quantized_ZBCP}

The other scenario is that the ZBCP manifests the quantized conductance but without having the localized MZMs.
This case is more common because (1) it can either happen in the case of `ugly ZBCP'~\cite{pan2020physical,pan2020generic}, where the quantized ZBCP is accidentally induced by disorder without a nontrivial topological invariant (i.e., $Q>-0.9$), or (2) it can happen in the case where TV is negative but there are multiple patches of topological regime in the bulk~\cite{dassarma2023spectral} (i.e., $Q<-0.9$ but without localized states).

\begin{figure*}[ht]
    \centering
    \includegraphics[width=6.8in]{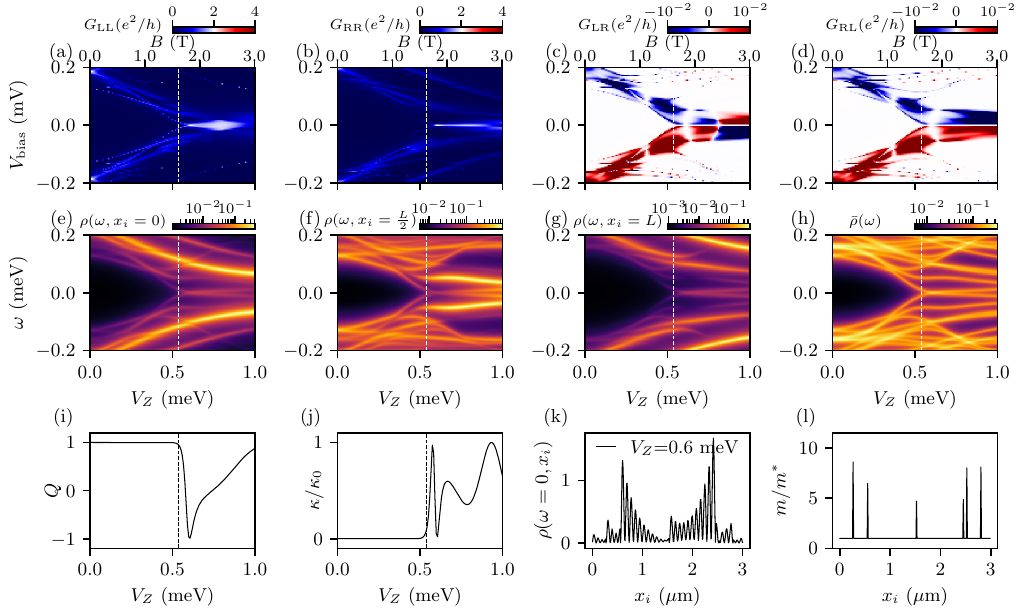}
    \caption{
    An example of quantized ZBCP without localized MZMs in a 3-micron wire with $n=6$ and $k=9$.
    (a-d) Local ($G_{\text{LL}}$ and $G_{\text{RR}}$) and nonlocal ($G_{\text{LR}}$ and $G_{\text{RL}}$) conductance spectra. The bottom (top) axis shows the Zeeman (magnetic) field. The vertical dashed line indicates nominal TQPT.
    (e-g) LDOS at left, midpoint, and right end. 
    (h) Total DOS.
    (i-j) Topological visibility $Q$ and thermal conductance $\kappa$.
    (k) LDOS at zero energy for $V_Z=0.6$ meV.
    (l) Spatial profile of effective mass. 
    }
    \label{fig:622}
\end{figure*}

\begin{figure*}[ht]
    \centering
    \includegraphics[width=6.8in]{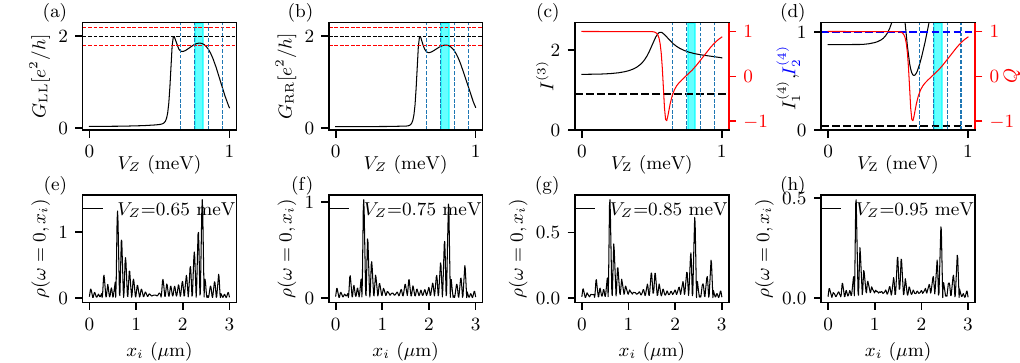}
    \caption{
    The benchmark of $I^{(3)}$ and $I_{1,2}^{(4)}$ on an example of quantized ZBCP without localized MZMs (cyan-shaded regions) in a 3-micron wire with $n=6$ and $k=9$ in Fig.~\ref{fig:622}.
    (a-b) Line cuts of the local ($G_{\text{LL}}$ and $G_{\text{RR}}$) conductance at zero bias. The red dashed line indicates the threshold of quantized conductance.
    (c) KL divergence $I^{(3)}$ (black) on the left axis, and TV $Q$ (red) on the right axis as a function of $V_Z$. The horizontal dashed line indicates the threshold of $I^{(3)}$ (0.9).
    (d) Ratio of midpoint LDOS and end LDOS $I_1^{(4)}$ (black), and the ratio of maximal in the bulk and end LDOS $I_2^{(4)}$ (blue). TV $Q$ (red) on the right axis as a function of $V_Z$. The horizontal dashed line indicates the threshold of $I_1^{(4)}$, 0.04 (black), and $I_2^{(4)}$, 1 (blue), respectively.
    (e-h) The LDOS at four Zeeman fields indicated by the vertical dashed line in (a-d).
    }
    \label{fig:622_cond_metrics}
\end{figure*}

An example of the first case is shown in Fig.~\ref{fig:622} along with Fig.~\ref{fig:622_cond_metrics}.
In Fig.~\ref{fig:622}(a-b) and Fig.~\ref{fig:622_cond_metrics}(a-b), the local conductance spectra show the almost quantized ZBCPs from both ends. However, the nonlocal conductance spectra in Fig.~\ref{fig:622}(c-d) do not manifest any feature of the gap closing and reopening.
In Fig.~\ref{fig:622}(i-j), the TV is mostly positive when ZBCPs appear at around $V_Z=0.75$ to 0.8 meV, and the thermal conductance shows multiple peaks. However, this is the region where local conductance shows quantized ZBCPs.
In Fig.~\ref{fig:622}(k-l), we notice multiple bulk states in the LDOS at zero energy, which are localized at the same positions of the impurities.
In Fig.~\ref{fig:622_cond_metrics}(c-d), the KL divergence $I^{(3)}$ is larger than 1 and the ratio $I_1^{(4)}$ and $I_2^{(4)}$ both do not satisfy the threshold, indicating the absence of localized states.
In Fig.~\ref{fig:622_cond_metrics}(e-h), we see two bulk states localized at around 0.5 microns and 2.5 microns (outside the `end' region), which explains the large value $I^{(3)}$ and $I_{1,2}^{(4)}$ in Fig.~\ref{fig:622_cond_metrics}(c-d).

\begin{figure*}[ht]
    \centering
    \includegraphics[width=6.8in]{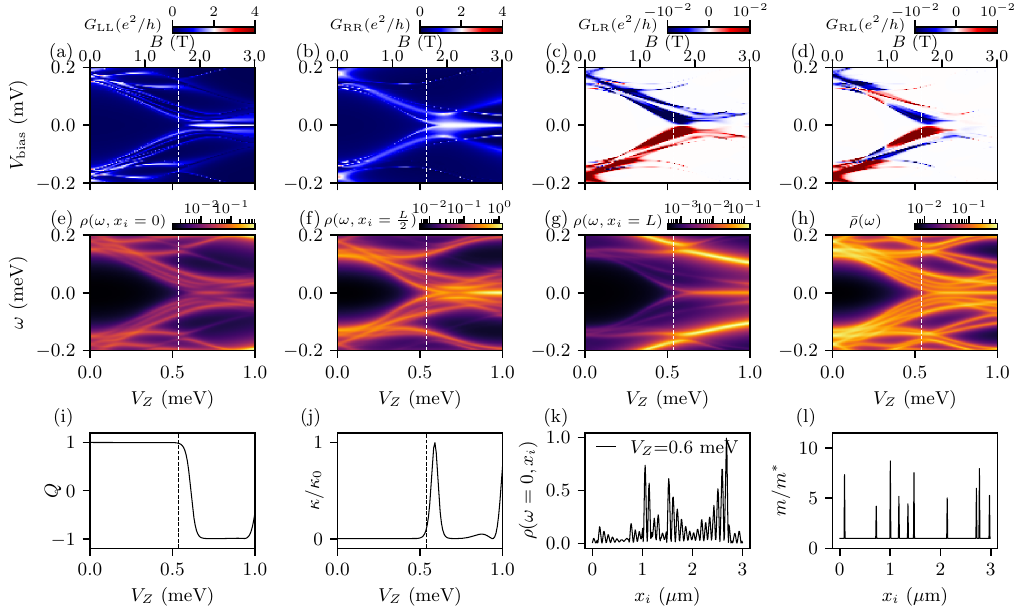}
    \caption{
    An example of quantized ZBCP without localized MZMs in a 3-micron wire with $n=10$ and $k=9$.
    (a-d) Local ($G_{\text{LL}}$ and $G_{\text{RR}}$) and nonlocal ($G_{\text{LR}}$ and $G_{\text{RL}}$) conductance spectra. The bottom (top) axis shows the Zeeman (magnetic) field. The vertical dashed line indicates nominal TQPT.
    (e-g) LDOS at left, midpoint, and right end. 
    (h) Total DOS.
    (i-j) Topological visibility $Q$ and thermal conductance $\kappa$.
    (k) LDOS at zero energy for $V_Z=0.6$ meV.
    (l) Spatial profile of effective mass. 
    }
    \label{fig:1027}
\end{figure*}

\begin{figure*}[ht]
    \centering
    \includegraphics[width=6.8in]{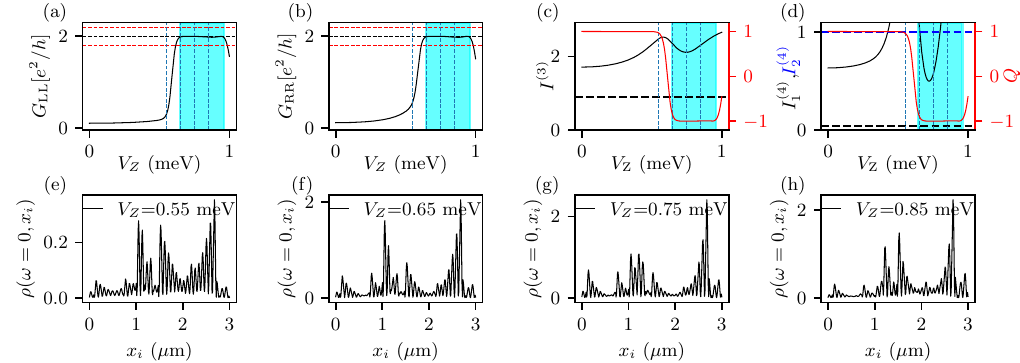}
    \caption{
    The benchmark of $I^{(3)}$ and $I_{1,2}^{(4)}$ on an example of quantized ZBCP without localized MZMs (cyan-shaded regions) in a 3-micron wire with $n=10$ and $k=9$ in Fig.~\ref{fig:1027}.  
    (a-b) Line cuts of the local ($G_{\text{LL}}$ and $G_{\text{RR}}$) conductance at zero bias. The red dashed line indicates the threshold of quantized conductance.
    (c) KL divergence $I^{(3)}$ (black) on the left axis, and TV $Q$ (red) on the right axis as a function of $V_Z$. The horizontal dashed line indicates the threshold of $I^{(3)}$ (0.9).
    (d) Ratio of midpoint LDOS and end LDOS $I_1^{(4)}$ (black), and the ratio of maximal in the bulk and end LDOS $I_2^{(4)}$ (blue). TV $Q$ (red) on the right axis as a function of $V_Z$. The horizontal dashed line indicates the threshold of $I_1^{(4)}$, 0.04 (black), and $I_2^{(4)}$, 1 (blue), respectively.
    (e-h) The LDOS at four Zeeman fields indicated by the vertical dashed line in (a-d).
    }
    \label{fig:1027_cond_metrics}
\end{figure*}

An example of the second case is shown in Fig.~\ref{fig:1027} along with Fig.~\ref{fig:1027_cond_metrics}.
In Fig.~\ref{fig:1027}(a-b), the local conductance spectra show quantized ZBCPs from both ends, and the nonlocal conductance spectra show a gap closing and reporting feature, although the reopened gap size is very tiny ($<10\%$ of the proximitized SC gap).
In this case, although the TV is negative from Fig.~\ref{fig:1027}(i), the localized state is absent because LDOS at zero energy has multiple peaks in the bulk, as shown in Fig.~\ref{fig:1027_cond_metrics}(e-h).
This is an example demonstrating the importance of isolated localized states in the LDOS. 
More examples that manifest quantized ZBCP but without localized MZMs are shown in Appendix~\ref{app:quantized_ZBCP}.

Finally, in Sec.~\ref{sec:predictive_power}, we show that if the wire manifests a trivial topological invariant, the localized states cannot exist either (see Fig.~\ref{fig:joint_KL} and Fig.~\ref{fig:joint_ratio}).
This means that these two scenarios above do not have localized states.
That is to say, we can ignore the effect of the topological invariant $Q$ and only look at whether the localized state exists or not to determine the existence of the topological localized MZMs.
Basically, in finite disordered systems the topological invariant $Q$ is simply not very meaningful, and perhaps should be avoided.

\subsubsection{Negation of localized MZMs and quantized ZBCP}\label{sec:negation}
Although the localized MZMs and quantized ZBCP are not equivalent in the general disorder case, the negation of the two events is almost equivalent.
Here, we study the case where both events, localized MZMs and quantized ZBCP, are absent.
Namely, we aim to answer the questions which correlate the negation of two events:
(1)  Does the absence of topological localized MZMs always lead to nonquantized ZBCP at both ends?
(2) Does nonquantized ZBCP at both ends always lead to the absence of topological localized MZMs?

This is important not only because these two questions are complementary to the previous two questions, but also because if any of the questions is true, then it provides a way to rule out the topological localized MZMs barely from the conductance spectrum.

Therefore, we again compute the Bayesian conditional probability conditioned in Fig.~\ref{fig:condprob}(b). Here, the overline $\overline{A}$ means the negation of the event $A$.
We find that both Bayesian conditional probabilities are close to one, and the probability of the absence of localized MZMs given nonquantized conductance (orange line in Fig.~\ref{fig:condprob}(b)) is much closer to one. 
This means that `quantized conductance' is a necessary condition for the `localized MZMs'. 
This also verifies the previous conjecture that the conditional probability of $P({\text{quantized conductance}}|\text{localized MZMs})$ should always be 1 (blue line in Fig.~\ref{fig:condprob}(a)), which was not manifested explicitly in the strong disorder due to the scarcity of `localized MZMs' itself in the strong disorder.
(Because $P(A|B)=1 \Leftrightarrow P(\overline{B}|\overline{A}=1$))

The other conditional probability `having nonquantized conductance given the absence of localized state' (blue line in Fig.~\ref{fig:condprob}(b)) is also quite close to one, but not as close as the reciprocal case.
Combined with the fact that $P({\text{localized MZMs}|\text{quantized conductance}})$ is rather small and approaches zero in the strong disorder regime, we can conclude the relative statistical occurrence of the following events:
\begin{equation}
    \begin{split}
        & P(\overline{\text{localized MZMs}} \cap \overline{\text{quantized conductance}}) \\
        \gg & P(\overline{\text{localized MZMs}} \cap {\text{quantized conductance}})\\
        \gg & P(\text{localized MZMs} \cap {\text{quantized conductance}}) \\
        \gg & P(\text{localized MZMs} \cap \overline{\text{quantized conductance}})\sim 0,        
    \end{split}
\end{equation}
The first inequality is deduced from the fact that $P(\overline{\text{quantized conductance}}|\overline{\text{localized MZMs}})$  (blue line in Fig.~\ref{fig:condprob}(b)) is close to 1. 
The second inequality is deduced from the fact that 
$P({\text{localized MZMs}|\text{quantized conductance}})$ (orange line in Fig.~\ref{fig:condprob}(a)) is very small and close to 0 in the strong disorder regime.
The third inequality is deduced simply from the fact that both $P({\text{localized MZMs}|\text{quantized conductance}})$ (blue line in Fig.~\ref{fig:condprob}(a)) and $P(\overline{\text{quantized conductance}}|\overline{\text{localized MZMs}})$ (orange line in Fig.~\ref{fig:condprob}(b)) should be 1. 

In other words, this hierarchy of the statistical occurrence means that:
(1) if the conductance spectrum does not show quantized ZBCP in experiments, then it is also unlikely to be the useful localized MZMs. 
(2) Conversely, if the state is not a localized MZM, though it can show the quantized conductance in rare cases, but mostly of the case, it will just manifest any nonquantized conductance.
We emphasize that our results are valid at $T=0$, so experimentally it applies only to very low temperatures.  Finite-temperature crossover effects are impossible to study since they would depend on other unknown parameters such as the tunnel barrier strength controlling the conductance. Topology or not can only be discussed at $T=0$ from a theoretical perspective.

\subsubsection{Topological without localized states}\label{sec:topological_without_localized}

When a state does not have localized MZMs, there could be three possibilities: 
(1) it is not topological but is localized;
(2) it is neither topological nor localized.
(3) it is topological but not localized;

The first case is actually rare because of Fig.~\ref{fig:joint_KL} and Fig.~\ref{fig:joint_ratio}, namely, as we mentioned before, the TV is a necessary condition for the localized MZMs--- if the TV is positive, then localized states is almost impossible.
For the second case, they are mostly mundane trivial states, which are not particularly interesting.
Therefore, in this section, we will focus on the third case, and study whether the ZBCP strengths from both ends correlate with whether isolated end states exist or not. Namely, we aim to answer the two questions:
(1) Does negative TV without an isolated localized end state always lead to nonquantized ZBCP at least at one end? 
(2) Does nonquantized ZBCP at least at one end always imply negative TV but without an isolated localized end state?
Note that the crucial difference between this section and the previous section Sec.~\ref{sec:negation} is whether we impose the condition of a negative TV.

\begin{figure*}[ht]
    \centering
    \includegraphics[width=6.8in]{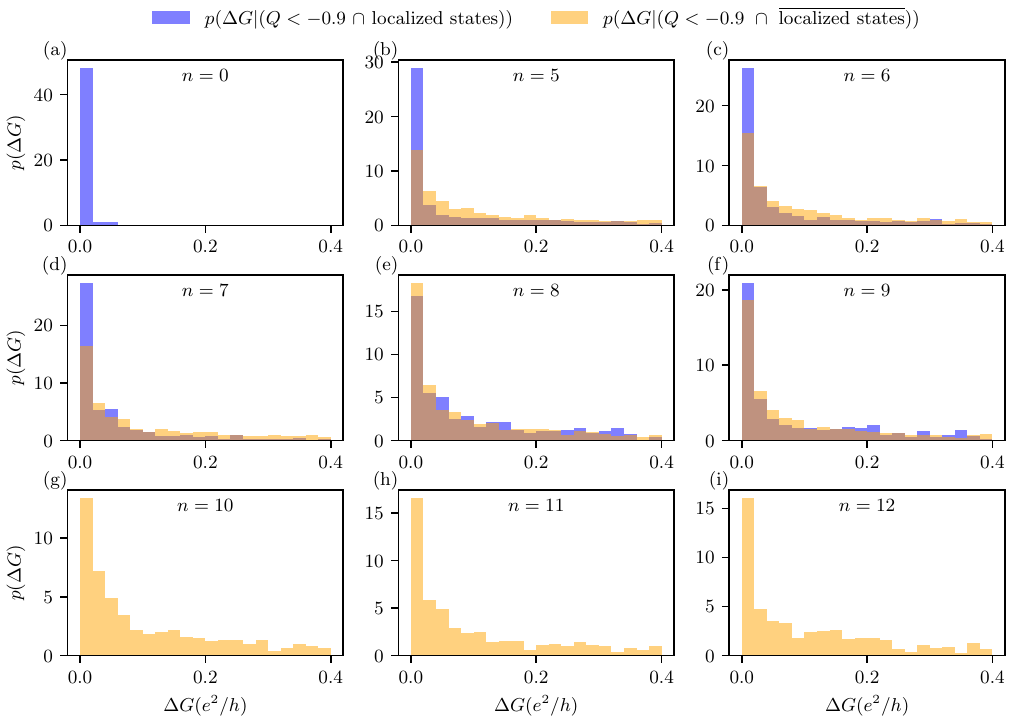}
    \caption{
    The histogram of the deviation of quantized conductance averaged over both ends, $\Delta G$, [see Eq.~\eqref{eq:G_diff}] conditioned on $Q<-0.9$ with localized states (cyan) and without localized states (orange) for different $n$. The absence of certain histograms is because of the absence of the corresponding priori probability.
    }
    \label{fig:ZBCP_distribution}
\end{figure*}

To answer the first question, we compute the probability density distribution of quantized ZBCP given $Q<-0.9$ with localized states (cyan bars) and without localized states (orange bars) for different $n$ in Fig.~\ref{fig:ZBCP_distribution}.
Namely, we use $p(\Delta G| Q<-0.9\cap \text{localized states})$ to denote the distribution of the averaged deviation from quantized conductance conditioned on the topological localized states, and similar for $p(\Delta G| Q<-0.9\cap \overline{\text{localized states}})$.

For $n=0$ in Fig.~\ref{fig:ZBCP_distribution}(a), we find a single peak at $\Delta G = 0$ for the case of topological localized states. This is consistent with the expectation that the ZBCP is always quantized if the localized MZMs are present.
The absence of the other distribution conditioned on `topological but without localized states' is because this case cannot exist in pristine wire (i.e., pristine wires always host localized MZMs in the topological regime.)

As disorder increases ($n$ increases), the distribution of the quantized ZBCP is broader for the case of `topological but without localized states' than the case of `topological with localized state'. Namely, it is more likely to have quantized ZBCP if localized MZMs are present.
However, the other distribution conditioned on `topological without localized states' also shows a peak at $\Delta G = 0$, which means that the negative TV without localized MZMs does not necessarily imply the nonquantized ZBCP. We show such probability in Fig.~\ref{fig:condprob}(c) (the blue line) where it indicates around 70\% of chances that the ZBCP is quantized even if it has a negative TV but without localized states.

In the strong disorder regime, the distribution conditioned on `topological with localized states' is missing because this is also a scenario that does not exist in the strong disorder regime (it is the same reason as in Fig.~\ref{fig:condprob}(a)).

Finally, we show the conditional probability of the reciprocal case to answer the second question in Fig.~\ref{fig:condprob}(c) (the orange line).
We find that this probability is almost always zero.
This is particularly interesting because from Fig.~\ref{fig:condprob}(b) to Fig.~\ref{fig:condprob}(c), what makes the orange line change from almost 1 to 0 is only the restriction of TV being negative. 
This indicates that, when the ZBCP is not quantized, it is not just that localized MZMs are impossible, but also that the topology itself is almost trivial. 
This is because 
\begin{equation}\label{eq:localized_MZM}
    \begin{split}
        &P(\overline{\text{localized MZMs}}|\overline{\text{quantized conductance}})\\
        =&P(\overline{(Q<-0.9) \cap \text{localized states}} |\overline{\text{quantized conductance}} )\\
        =&P((Q<-0.9) \cap \overline{\text{localized states}} |\overline{\text{quantized conductance}}  ) \\
        & +  P((Q>-0.9 )\cap \text{localized states} |\overline{\text{quantized conductance}} )\\
        & +  P((Q>-0.9) \cap \overline{\text{localized states}} |\overline{\text{quantized conductance}} )\\
        =& P((Q<-0.9) \cap \overline{\text{localized states}}|\overline{\text{quantized conductance}} )\\
        & + P((Q>-0.9) \cap \overline{\text{localized states}}|\overline{\text{quantized conductance}} )
    \end{split}
\end{equation}
The third equal sign is because $(Q>-0.9 \cap \text{localized states})$ is almost impossible as shown in Fig.~\ref{fig:joint_KL} and Fig.~\ref{fig:joint_ratio}.
Therefore, the fact that $P(\overline{\text{localized MZMs}}|\overline{\text{quantized conductance}})\sim 1$ and $P((Q<-0.9) \cap \overline{\text{localized states}}|\overline{\text{quantized conductance}} )\sim0$ indicates the large contribution from $P((Q>-0.9) \cap \overline{\text{localized states}}|\overline{\text{quantized conductance}} )$.

Note that the last equation in Eq.~\eqref{eq:localized_MZM} can be further combined to $P(\overline{\text{localized states}}|\overline{\text{quantized conductance}})$, which means that the existence of `localized MZMs' can be statistically inferred only from the existence of `localized states' because `localized states' is a more rigorous statement. 
This also demonstrates the importance of isolated localized states at the ends of the wire over the topological invariant itself.
Because whether the topological localized MZMs (tested by $Q$ plus $I^{(3)}$ and $I^{(4)}$) exist or not can be ultimately boiled down to only whether the localized states (tested by $I^{(3)}$ and $I^{(4)}$) exist or not.

\section{Discussion and conclusion}\label{sec:conclusion}

In this paper, we propose a new way to model the disorder without explicitly involving the microscopic details of disorder by using the varying effective mass in the wire.
We find that this model can capture the essential disorder physics that was previously obtained by adding an uncorrelated Gaussian random potential to the chemical potential~\cite{pan2020physical,pan2021threeterminal,dassarma2021disorderinduced}.
Besides the qualitative agreement, this method also provides a clear physical picture of the disorder effect--- segmenting the wire into quantum dots carrying different regions of topological patches~\cite{dassarma2023spectral}.
This provides us with a crystal clear and simple picture to determine the disorder regime by directly comparing the SC coherence length to the mean free path ($L/(n+1)$). 
If the SC coherence length is smaller than the mean free path, then 
the wire is in the pristine or weak disorder regime, where the topological regime is mostly protected.
If the SC coherence length is larger than the mean free path, then the wire is in the strong disorder regime, where the topological regime is almost always destroyed.
If both length scales are comparable, then the wire is in the intermediate disorder regime, where the topological regime may or may not exist because it depends on the details~\cite{motrunich2001griffiths}, and we need to look at other metrics.
(It is possible that the best current experiments are in this difficult intermediate regime whereas all the earlier experiments during 2012-2022 were in the strong disorder regime where the SC coherence length is larger than the electron mean free path, and as such, the experiments were only probing the class D anti-localization conductance peaks, conflating it with Majorana peaks.)

One success of this model is that we can reproduce both the positive (passing TGP) and negative results (failing TGP) in the recent experiment by Microsoft Quantum~\cite{microsoftquantum2023inasal} qualitatively by simply changing the number of impurities $n$ that segments the wire. 
This is not only a triumph of the theory itself but also strong evidence that the experiment is indeed in the intermediate disorder regime.
This is encouraging since the earlier experiments were in the hopeless strong disorder regime where the system is Anderson localized with no topology.

We then proceed to study the effect of other parameters in this theory. 
We find that the variance of effective mass $k$ does not affect the boundary of the disorder regime as long as it is sufficiently large ($k>10$).
On the other hand, the wire length $L$ plays a crucial role in determining the disorder regime, as it can directly control the MFP.
Therefore, to keep the same disorder regime in a longer wire, one needs to increase the number of impurities $n$ accordingly to keep the MFP invariant.
Following this logic, we demonstrate how we make a nontopological wire topological by only increasing the SC gap, which further decreases the SC coherence length until it is a smaller value than the mean free path.
Thus, future progress in the subject will come from either decreasing the disorder or increasing the SC gap (or perhaps both) as all one needs is to achieve a large value for the dimensionless ratio $\xi_{\text{MFP}}/\xi_{\text{SC}}$, which can be achieved either by increasing $\xi_{\text{MFP}}$ (by reducing disorder), or by decreasing $\xi_{\text{SC}}$ (by increasing the SC gap).
{It is important to emphasize that a pristine nanowire manifests an exponentially protected topological superconducting phase carrying non-Abelian end MZMs only in the limit of $L\gg \xi_{\text{SC}}$~\cite{kitaev2001unpaired}. Even for a pristine system, there is no topology, no protection and no isolated MZMs when the wire length $L$ is comparable to or shorter than $\xi_{\text{SC}}$ simply because the MZMs overlap strongly in this situation and simply become ordinary subgap Andreev bound states accompanied by Majorana oscillations~\cite{dassarma2012splitting}.
In the presence of disorder, the situation becomes complicated as disorder may create almost-zero energy ABS throughout the bulk of the wire, leading to a very complex scenario with no simple description and no topology. 
This is the current experimental scenario we are trying to describe using our new approach in the current work.}

In simulating the previous results, we touch on a bedrock of the topological localized MZMs. Namely, we find that the topological invariant itself is totally insufficient to predict the topological localized MZMs in the short disordered wire.
We propose the importance of the isolated localized states at the ends of the wire over the topological invariant itself as was already mentioned in Ref.~\onlinecite{dassarma2023spectral}.
This is because for any Majorana modes to be useful, it must be localized at the end of the wire.
Therefore, we propose several new indicators that characterize the spatial distribution of the LDOS to filter out the localized MZMs.
These indicators are motivated by the fact that we want the localized MZMs to be isolated at the ends, and also to suppress the bulk states.
By benchmarking these indicators with the simulation in different disorder regimes, we find two useful indicators that can predict the isolated localized states pretty well. 
The first is the KL divergence between the LDOS of a test state and a topological localized state as a reference state. 
The second one is an intuitive ratio between the end LDOS and the bulk LDOS, as well as the maximal bulk LDOS and the end LDOS, to ensure that the end LDOS is maximized while the bulk LDOS is minimized, respectively.
Combining these two indicators, we inspect the predictive power of the topological invariant (i.e., TV), and find that the TV itself is not a good indicator as the false-positive rate increases in the weak disorder regime and saturates in the strong disorder regime.
This implies that one needs to also consider the spatial distribution of the LDOS to form a reliable conclusion on the topological MZMs. 
In fact, this is essentially to say that the indicators for the localized states are more rigorous than the topological invariant itself.

Then, we use a stringent criterion by combining all three indicators and then connect the theoretical indicators to the experimentally accessible conductance.
By using the Bayesian conditional probability, we answer several crucial questions that were not seriously addressed before. For example, we find the quantized ZBCP is a necessary condition for the isolated localized MZMs to exist. Conversely, the existence of localized MZMs is almost a necessary condition for the quantized ZBCP but not absolutely. We also find the hierarchy of the statistical occurrence of the localized MZMs and quantized ZBCP in the strong disorder regime, namely, the case of `nonquantized ZBCP' without `localized MZMs' is more common than the case of `quantized ZBCP' without `localized MZMs', followed by the case of `quantized ZBCP' with `localized MZMs'(the case of `nonquantized ZBCP' with `localized MZMs' is not physically sensible).

Finally, we study the importance of the isolated localized states at the ends of the wire by asking whether the conductance is always non-quantized if the wire has nontrivial topological invariant but without localized states.
We find that the quantized ZBCP is more likely to appear if the localized MZMs are present, but the absence of localized MZMs does not necessarily imply the nonquantized ZBCP.
In addition, we find that if the conductance is not quantized, then the reason is always that the wire itself is not topological and the localized states are absent. 
Since the localized states are more rigorous criteria than the topological invariant itself, this means that if the conductance is not quantized, then the wire does not have a localized state. This suggests strongly that one should focus on whether localized states in the LDOS exist beyond the topological invariant itself to determine the existence of useful topological isolated MZMs localized at the ends of the wire.

While the Majorana nanowire problem has turned out to be much more complicated than one had thought originally~\cite{kitaev2001unpaired,lutchyn2010majorana,oreg2010helical,sau2010generic,sau2010robustness}, the important physics point is that we have learned a great deal of new physics involving the meaning of topology itself in finite disordered systems.  This knowledge transcends the specific platforms we are considering and may explain why topology has been so difficult to observe in other platforms too (e.g., quantum spin Hall quantization, edge physics in quantum hall systems). In the presence of disorder suppressing the topological energy gap, a finite system conflates topology and finite-size crossover in a highly complex manner.  Our current work shows a new way of understanding this complex interplay in Majorana platforms, which should pave the way for significant future progress.

\section*{Acknowledgements}
This work is supported by the Laboratory for Physical Sciences through the Condensed Matter Theory Center at the University of Maryland.

\bibliography{Paper_VaryingMass}

\begin{thebibliography}{54}%
\makeatletter
\providecommand \@ifxundefined [1]{%
 \@ifx{#1\undefined}
}%
\providecommand \@ifnum [1]{%
 \ifnum #1\expandafter \@firstoftwo
 \else \expandafter \@secondoftwo
 \fi
}%
\providecommand \@ifx [1]{%
 \ifx #1\expandafter \@firstoftwo
 \else \expandafter \@secondoftwo
 \fi
}%
\providecommand \natexlab [1]{#1}%
\providecommand \enquote  [1]{``#1''}%
\providecommand \bibnamefont  [1]{#1}%
\providecommand \bibfnamefont [1]{#1}%
\providecommand \citenamefont [1]{#1}%
\providecommand \href@noop [0]{\@secondoftwo}%
\providecommand \href [0]{\begingroup \@sanitize@url \@href}%
\providecommand \@href[1]{\@@startlink{#1}\@@href}%
\providecommand \@@href[1]{\endgroup#1\@@endlink}%
\providecommand \@sanitize@url [0]{\catcode `\\12\catcode `\$12\catcode `\&12\catcode `\#12\catcode `\^12\catcode `\_12\catcode `\%12\relax}%
\providecommand \@@startlink[1]{}%
\providecommand \@@endlink[0]{}%
\providecommand \url  [0]{\begingroup\@sanitize@url \@url }%
\providecommand \@url [1]{\endgroup\@href {#1}{\urlprefix }}%
\providecommand \urlprefix  [0]{URL }%
\providecommand \Eprint [0]{\href }%
\providecommand \doibase [0]{https://doi.org/}%
\providecommand \selectlanguage [0]{\@gobble}%
\providecommand \bibinfo  [0]{\@secondoftwo}%
\providecommand \bibfield  [0]{\@secondoftwo}%
\providecommand \translation [1]{[#1]}%
\providecommand \BibitemOpen [0]{}%
\providecommand \bibitemStop [0]{}%
\providecommand \bibitemNoStop [0]{.\EOS\space}%
\providecommand \EOS [0]{\spacefactor3000\relax}%
\providecommand \BibitemShut  [1]{\csname bibitem#1\endcsname}%
\let\auto@bib@innerbib\@empty
\bibitem [{\citenamefont {Lutchyn}\ \emph {et~al.}(2010)\citenamefont {Lutchyn}, \citenamefont {Sau},\ and\ \citenamefont {Das~Sarma}}]{lutchyn2010majorana}%
  \BibitemOpen
  \bibfield  {author} {\bibinfo {author} {\bibfnamefont {R.~M.}\ \bibnamefont {Lutchyn}}, \bibinfo {author} {\bibfnamefont {J.~D.}\ \bibnamefont {Sau}},\ and\ \bibinfo {author} {\bibfnamefont {S.}~\bibnamefont {Das~Sarma}},\ }\bibfield  {title} {\bibinfo {title} {Majorana {{Fermions}} and a {{Topological Phase Transition}} in {{Semiconductor-Superconductor Heterostructures}}},\ }\href {https://link.aps.org/doi/10.1103/PhysRevLett.105.077001} {\bibfield  {journal} {\bibinfo  {journal} {Physical Review Letters}\ }\textbf {\bibinfo {volume} {105}},\ \bibinfo {pages} {077001} (\bibinfo {year} {2010})}\BibitemShut {NoStop}%
\bibitem [{\citenamefont {Oreg}\ \emph {et~al.}(2010)\citenamefont {Oreg}, \citenamefont {Refael},\ and\ \citenamefont {{von Oppen}}}]{oreg2010helical}%
  \BibitemOpen
  \bibfield  {author} {\bibinfo {author} {\bibfnamefont {Y.}~\bibnamefont {Oreg}}, \bibinfo {author} {\bibfnamefont {G.}~\bibnamefont {Refael}},\ and\ \bibinfo {author} {\bibfnamefont {F.}~\bibnamefont {{von Oppen}}},\ }\bibfield  {title} {\bibinfo {title} {Helical {{Liquids}} and {{Majorana Bound States}} in {{Quantum Wires}}},\ }\href {http://link.aps.org/doi/10.1103/PhysRevLett.105.177002} {\bibfield  {journal} {\bibinfo  {journal} {Phys. Rev. Lett.}\ }\textbf {\bibinfo {volume} {105}},\ \bibinfo {pages} {177002} (\bibinfo {year} {2010})}\BibitemShut {NoStop}%
\bibitem [{\citenamefont {Sau}\ \emph {et~al.}(2010{\natexlab{a}})\citenamefont {Sau}, \citenamefont {Lutchyn}, \citenamefont {Tewari},\ and\ \citenamefont {Das~Sarma}}]{sau2010generic}%
  \BibitemOpen
  \bibfield  {author} {\bibinfo {author} {\bibfnamefont {J.~D.}\ \bibnamefont {Sau}}, \bibinfo {author} {\bibfnamefont {R.~M.}\ \bibnamefont {Lutchyn}}, \bibinfo {author} {\bibfnamefont {S.}~\bibnamefont {Tewari}},\ and\ \bibinfo {author} {\bibfnamefont {S.}~\bibnamefont {Das~Sarma}},\ }\bibfield  {title} {\bibinfo {title} {Generic {{New Platform}} for {{Topological Quantum Computation Using Semiconductor Heterostructures}}},\ }\href {https://link.aps.org/doi/10.1103/PhysRevLett.104.040502} {\bibfield  {journal} {\bibinfo  {journal} {Physical Review Letters}\ }\textbf {\bibinfo {volume} {104}},\ \bibinfo {pages} {040502} (\bibinfo {year} {2010}{\natexlab{a}})}\BibitemShut {NoStop}%
\bibitem [{\citenamefont {Sau}\ \emph {et~al.}(2010{\natexlab{b}})\citenamefont {Sau}, \citenamefont {Lutchyn}, \citenamefont {Tewari},\ and\ \citenamefont {Das~Sarma}}]{sau2010robustness}%
  \BibitemOpen
  \bibfield  {author} {\bibinfo {author} {\bibfnamefont {J.~D.}\ \bibnamefont {Sau}}, \bibinfo {author} {\bibfnamefont {R.~M.}\ \bibnamefont {Lutchyn}}, \bibinfo {author} {\bibfnamefont {S.}~\bibnamefont {Tewari}},\ and\ \bibinfo {author} {\bibfnamefont {S.}~\bibnamefont {Das~Sarma}},\ }\bibfield  {title} {\bibinfo {title} {Robustness of {{Majorana}} fermions in proximity-induced superconductors},\ }\href {https://link.aps.org/doi/10.1103/PhysRevB.82.094522} {\bibfield  {journal} {\bibinfo  {journal} {Physical Review B}\ }\textbf {\bibinfo {volume} {82}},\ \bibinfo {pages} {094522} (\bibinfo {year} {2010}{\natexlab{b}})}\BibitemShut {NoStop}%
\bibitem [{\citenamefont {Das~Sarma}(2023)}]{dassarma2023search}%
  \BibitemOpen
  \bibfield  {author} {\bibinfo {author} {\bibfnamefont {S.}~\bibnamefont {Das~Sarma}},\ }\bibfield  {title} {\bibinfo {title} {In search of {{Majorana}}},\ }\href {https://www.nature.com/articles/s41567-022-01900-9} {\bibfield  {journal} {\bibinfo  {journal} {Nature Physics}\ }\textbf {\bibinfo {volume} {19}},\ \bibinfo {pages} {165} (\bibinfo {year} {2023})}\BibitemShut {NoStop}%
\bibitem [{\citenamefont {Das}\ \emph {et~al.}(2012)\citenamefont {Das}, \citenamefont {Ronen}, \citenamefont {Most}, \citenamefont {Oreg}, \citenamefont {Heiblum},\ and\ \citenamefont {Shtrikman}}]{das2012zerobias}%
  \BibitemOpen
  \bibfield  {author} {\bibinfo {author} {\bibfnamefont {A.}~\bibnamefont {Das}}, \bibinfo {author} {\bibfnamefont {Y.}~\bibnamefont {Ronen}}, \bibinfo {author} {\bibfnamefont {Y.}~\bibnamefont {Most}}, \bibinfo {author} {\bibfnamefont {Y.}~\bibnamefont {Oreg}}, \bibinfo {author} {\bibfnamefont {M.}~\bibnamefont {Heiblum}},\ and\ \bibinfo {author} {\bibfnamefont {H.}~\bibnamefont {Shtrikman}},\ }\bibfield  {title} {\bibinfo {title} {Zero-bias peaks and splitting in an {{Al}}--{{InAs}} nanowire topological superconductor as a signature of {{Majorana}} fermions},\ }\href {https://www.nature.com/articles/nphys2479} {\bibfield  {journal} {\bibinfo  {journal} {Nature Physics}\ }\textbf {\bibinfo {volume} {8}},\ \bibinfo {pages} {887} (\bibinfo {year} {2012})}\BibitemShut {NoStop}%
\bibitem [{\citenamefont {Deng}\ \emph {et~al.}(2012)\citenamefont {Deng}, \citenamefont {Yu}, \citenamefont {Huang}, \citenamefont {Larsson}, \citenamefont {Caroff},\ and\ \citenamefont {Xu}}]{deng2012anomalous}%
  \BibitemOpen
  \bibfield  {author} {\bibinfo {author} {\bibfnamefont {M.~T.}\ \bibnamefont {Deng}}, \bibinfo {author} {\bibfnamefont {C.~L.}\ \bibnamefont {Yu}}, \bibinfo {author} {\bibfnamefont {G.~Y.}\ \bibnamefont {Huang}}, \bibinfo {author} {\bibfnamefont {M.}~\bibnamefont {Larsson}}, \bibinfo {author} {\bibfnamefont {P.}~\bibnamefont {Caroff}},\ and\ \bibinfo {author} {\bibfnamefont {H.~Q.}\ \bibnamefont {Xu}},\ }\bibfield  {title} {\bibinfo {title} {Anomalous {{Zero-Bias Conductance Peak}} in a {{Nb}}--{{InSb Nanowire}}--{{Nb Hybrid Device}}},\ }\href {http://pubs.acs.org/doi/10.1021/nl303758w} {\bibfield  {journal} {\bibinfo  {journal} {Nano Letters}\ }\textbf {\bibinfo {volume} {12}},\ \bibinfo {pages} {6414} (\bibinfo {year} {2012})}\BibitemShut {NoStop}%
\bibitem [{\citenamefont {Mourik}\ \emph {et~al.}(2012)\citenamefont {Mourik}, \citenamefont {Zuo}, \citenamefont {Frolov}, \citenamefont {Plissard}, \citenamefont {Bakkers},\ and\ \citenamefont {Kouwenhoven}}]{mourik2012signatures}%
  \BibitemOpen
  \bibfield  {author} {\bibinfo {author} {\bibfnamefont {V.}~\bibnamefont {Mourik}}, \bibinfo {author} {\bibfnamefont {K.}~\bibnamefont {Zuo}}, \bibinfo {author} {\bibfnamefont {S.~M.}\ \bibnamefont {Frolov}}, \bibinfo {author} {\bibfnamefont {S.~R.}\ \bibnamefont {Plissard}}, \bibinfo {author} {\bibfnamefont {E.~P. A.~M.}\ \bibnamefont {Bakkers}},\ and\ \bibinfo {author} {\bibfnamefont {L.~P.}\ \bibnamefont {Kouwenhoven}},\ }\bibfield  {title} {\bibinfo {title} {Signatures of {{Majorana Fermions}} in {{Hybrid Superconductor-Semiconductor Nanowire Devices}}},\ }\href {http://science.sciencemag.org/content/336/6084/1003} {\bibfield  {journal} {\bibinfo  {journal} {Science}\ }\textbf {\bibinfo {volume} {336}},\ \bibinfo {pages} {1003} (\bibinfo {year} {2012})}\BibitemShut {NoStop}%
\bibitem [{\citenamefont {Churchill}\ \emph {et~al.}(2013)\citenamefont {Churchill}, \citenamefont {Fatemi}, \citenamefont {{Grove-Rasmussen}}, \citenamefont {Deng}, \citenamefont {Caroff}, \citenamefont {Xu},\ and\ \citenamefont {Marcus}}]{churchill2013superconductornanowire}%
  \BibitemOpen
  \bibfield  {author} {\bibinfo {author} {\bibfnamefont {H.~O.~H.}\ \bibnamefont {Churchill}}, \bibinfo {author} {\bibfnamefont {V.}~\bibnamefont {Fatemi}}, \bibinfo {author} {\bibfnamefont {K.}~\bibnamefont {{Grove-Rasmussen}}}, \bibinfo {author} {\bibfnamefont {M.~T.}\ \bibnamefont {Deng}}, \bibinfo {author} {\bibfnamefont {P.}~\bibnamefont {Caroff}}, \bibinfo {author} {\bibfnamefont {H.~Q.}\ \bibnamefont {Xu}},\ and\ \bibinfo {author} {\bibfnamefont {C.~M.}\ \bibnamefont {Marcus}},\ }\bibfield  {title} {\bibinfo {title} {Superconductor-nanowire devices from tunneling to the multichannel regime: {{Zero-bias}} oscillations and magnetoconductance crossover},\ }\href {https://link.aps.org/doi/10.1103/PhysRevB.87.241401} {\bibfield  {journal} {\bibinfo  {journal} {Physical Review B}\ }\textbf {\bibinfo {volume} {87}},\ \bibinfo {pages} {241401} (\bibinfo {year} {2013})}\BibitemShut {NoStop}%
\bibitem [{\citenamefont {Finck}\ \emph {et~al.}(2013)\citenamefont {Finck}, \citenamefont {Van~Harlingen}, \citenamefont {Mohseni}, \citenamefont {Jung},\ and\ \citenamefont {Li}}]{finck2013anomalous}%
  \BibitemOpen
  \bibfield  {author} {\bibinfo {author} {\bibfnamefont {A.~D.~K.}\ \bibnamefont {Finck}}, \bibinfo {author} {\bibfnamefont {D.~J.}\ \bibnamefont {Van~Harlingen}}, \bibinfo {author} {\bibfnamefont {P.~K.}\ \bibnamefont {Mohseni}}, \bibinfo {author} {\bibfnamefont {K.}~\bibnamefont {Jung}},\ and\ \bibinfo {author} {\bibfnamefont {X.}~\bibnamefont {Li}},\ }\bibfield  {title} {\bibinfo {title} {Anomalous {{Modulation}} of a {{Zero-Bias Peak}} in a {{Hybrid Nanowire-Superconductor Device}}},\ }\href {https://link.aps.org/doi/10.1103/PhysRevLett.110.126406} {\bibfield  {journal} {\bibinfo  {journal} {Physical Review Letters}\ }\textbf {\bibinfo {volume} {110}},\ \bibinfo {pages} {126406} (\bibinfo {year} {2013})}\BibitemShut {NoStop}%
\bibitem [{\citenamefont {Nichele}\ \emph {et~al.}(2017)\citenamefont {Nichele}, \citenamefont {Drachmann}, \citenamefont {Whiticar}, \citenamefont {O'Farrell}, \citenamefont {Suominen}, \citenamefont {Fornieri}, \citenamefont {Wang}, \citenamefont {Gardner}, \citenamefont {Thomas}, \citenamefont {Hatke}, \citenamefont {Krogstrup}, \citenamefont {Manfra}, \citenamefont {Flensberg},\ and\ \citenamefont {Marcus}}]{nichele2017scaling}%
  \BibitemOpen
  \bibfield  {author} {\bibinfo {author} {\bibfnamefont {F.}~\bibnamefont {Nichele}}, \bibinfo {author} {\bibfnamefont {A.~C.~C.}\ \bibnamefont {Drachmann}}, \bibinfo {author} {\bibfnamefont {A.~M.}\ \bibnamefont {Whiticar}}, \bibinfo {author} {\bibfnamefont {E.~C.~T.}\ \bibnamefont {O'Farrell}}, \bibinfo {author} {\bibfnamefont {H.~J.}\ \bibnamefont {Suominen}}, \bibinfo {author} {\bibfnamefont {A.}~\bibnamefont {Fornieri}}, \bibinfo {author} {\bibfnamefont {T.}~\bibnamefont {Wang}}, \bibinfo {author} {\bibfnamefont {G.~C.}\ \bibnamefont {Gardner}}, \bibinfo {author} {\bibfnamefont {C.}~\bibnamefont {Thomas}}, \bibinfo {author} {\bibfnamefont {A.~T.}\ \bibnamefont {Hatke}}, \bibinfo {author} {\bibfnamefont {P.}~\bibnamefont {Krogstrup}}, \bibinfo {author} {\bibfnamefont {M.~J.}\ \bibnamefont {Manfra}}, \bibinfo {author} {\bibfnamefont {K.}~\bibnamefont {Flensberg}},\ and\ \bibinfo {author} {\bibfnamefont {C.~M.}\ \bibnamefont {Marcus}},\ }\bibfield  {title} {\bibinfo {title} {Scaling of {{Majorana Zero-Bias Conductance Peaks}}},\ }\href {https://link.aps.org/doi/10.1103/PhysRevLett.119.136803} {\bibfield  {journal} {\bibinfo  {journal} {Physical Review Letters}\ }\textbf {\bibinfo {volume} {119}},\ \bibinfo {pages} {136803} (\bibinfo {year} {2017})}\BibitemShut {NoStop}%
\bibitem [{\citenamefont {Zhang}\ \emph {et~al.}(2017)\citenamefont {Zhang}, \citenamefont {Liu}, \citenamefont {Gazibegovic}, \citenamefont {Xu}, \citenamefont {Logan}, \citenamefont {Wang}, \citenamefont {{van Loo}}, \citenamefont {Bommer}, \citenamefont {{de Moor}}, \citenamefont {Car}, \citenamefont {het Veld}, \citenamefont {{van Veldhoven}}, \citenamefont {Koelling}, \citenamefont {Verheijen}, \citenamefont {Pendharkar}, \citenamefont {Pennachio}, \citenamefont {Shojaei}, \citenamefont {Lee}, \citenamefont {Palmstrom}, \citenamefont {Bakkers}, \citenamefont {Sarma},\ and\ \citenamefont {Kouwenhoven}}]{zhang2017quantized}%
  \BibitemOpen
  \bibfield  {author} {\bibinfo {author} {\bibfnamefont {H.}~\bibnamefont {Zhang}}, \bibinfo {author} {\bibfnamefont {C.-X.}\ \bibnamefont {Liu}}, \bibinfo {author} {\bibfnamefont {S.}~\bibnamefont {Gazibegovic}}, \bibinfo {author} {\bibfnamefont {D.}~\bibnamefont {Xu}}, \bibinfo {author} {\bibfnamefont {J.~A.}\ \bibnamefont {Logan}}, \bibinfo {author} {\bibfnamefont {G.}~\bibnamefont {Wang}}, \bibinfo {author} {\bibfnamefont {N.}~\bibnamefont {{van Loo}}}, \bibinfo {author} {\bibfnamefont {J.~D.~S.}\ \bibnamefont {Bommer}}, \bibinfo {author} {\bibfnamefont {M.~W.~A.}\ \bibnamefont {{de Moor}}}, \bibinfo {author} {\bibfnamefont {D.}~\bibnamefont {Car}}, \bibinfo {author} {\bibfnamefont {R.~L. M.~O.}\ \bibnamefont {het Veld}}, \bibinfo {author} {\bibfnamefont {P.~J.}\ \bibnamefont {{van Veldhoven}}}, \bibinfo {author} {\bibfnamefont {S.}~\bibnamefont {Koelling}}, \bibinfo {author} {\bibfnamefont {M.~A.}\ \bibnamefont {Verheijen}}, \bibinfo {author} {\bibfnamefont {M.}~\bibnamefont {Pendharkar}}, \bibinfo {author} {\bibfnamefont {D.~J.}\ \bibnamefont {Pennachio}}, \bibinfo {author} {\bibfnamefont {B.}~\bibnamefont {Shojaei}}, \bibinfo {author} {\bibfnamefont {J.~S.}\ \bibnamefont {Lee}}, \bibinfo {author} {\bibfnamefont {C.~J.}\ \bibnamefont {Palmstrom}}, \bibinfo {author} {\bibfnamefont {E.~P. A.~M.}\ \bibnamefont {Bakkers}}, \bibinfo {author} {\bibfnamefont {S.~D.}\ \bibnamefont {Sarma}},\ and\ \bibinfo {author} {\bibfnamefont {L.~P.}\ \bibnamefont {Kouwenhoven}},\ }\bibfield  {title} {\bibinfo {title} {Quantized {{Majorana}} conductance},\ }\href {http://arxiv.org/abs/1710.10701} {\bibfield  {journal} {\bibinfo  {journal} {arXiv:1710.10701}\ } (\bibinfo {year} {2017})}\BibitemShut {NoStop}%
\bibitem [{\citenamefont {Zhang}\ \emph {et~al.}(2021)\citenamefont {Zhang}, \citenamefont {{de Moor}}, \citenamefont {Bommer}, \citenamefont {Xu}, \citenamefont {Wang}, \citenamefont {{van Loo}}, \citenamefont {Liu}, \citenamefont {Gazibegovic}, \citenamefont {Logan}, \citenamefont {Car}, \citenamefont {het Veld}, \citenamefont {{van Veldhoven}}, \citenamefont {Koelling}, \citenamefont {Verheijen}, \citenamefont {Pendharkar}, \citenamefont {Pennachio}, \citenamefont {Shojaei}, \citenamefont {Lee}, \citenamefont {Palmstr{\o}m}, \citenamefont {Bakkers}, \citenamefont {Sarma},\ and\ \citenamefont {Kouwenhoven}}]{zhang2021large}%
  \BibitemOpen
  \bibfield  {author} {\bibinfo {author} {\bibfnamefont {H.}~\bibnamefont {Zhang}}, \bibinfo {author} {\bibfnamefont {M.~W.~A.}\ \bibnamefont {{de Moor}}}, \bibinfo {author} {\bibfnamefont {J.~D.~S.}\ \bibnamefont {Bommer}}, \bibinfo {author} {\bibfnamefont {D.}~\bibnamefont {Xu}}, \bibinfo {author} {\bibfnamefont {G.}~\bibnamefont {Wang}}, \bibinfo {author} {\bibfnamefont {N.}~\bibnamefont {{van Loo}}}, \bibinfo {author} {\bibfnamefont {C.-X.}\ \bibnamefont {Liu}}, \bibinfo {author} {\bibfnamefont {S.}~\bibnamefont {Gazibegovic}}, \bibinfo {author} {\bibfnamefont {J.~A.}\ \bibnamefont {Logan}}, \bibinfo {author} {\bibfnamefont {D.}~\bibnamefont {Car}}, \bibinfo {author} {\bibfnamefont {R.~L. M.~O.}\ \bibnamefont {het Veld}}, \bibinfo {author} {\bibfnamefont {P.~J.}\ \bibnamefont {{van Veldhoven}}}, \bibinfo {author} {\bibfnamefont {S.}~\bibnamefont {Koelling}}, \bibinfo {author} {\bibfnamefont {M.~A.}\ \bibnamefont {Verheijen}}, \bibinfo {author} {\bibfnamefont {M.}~\bibnamefont {Pendharkar}}, \bibinfo {author} {\bibfnamefont {D.~J.}\ \bibnamefont {Pennachio}}, \bibinfo {author} {\bibfnamefont {B.}~\bibnamefont {Shojaei}}, \bibinfo {author} {\bibfnamefont {J.~S.}\ \bibnamefont {Lee}}, \bibinfo {author} {\bibfnamefont {C.~J.}\ \bibnamefont {Palmstr{\o}m}}, \bibinfo {author} {\bibfnamefont {E.~P. A.~M.}\ \bibnamefont {Bakkers}}, \bibinfo {author} {\bibfnamefont {S.~D.}\ \bibnamefont {Sarma}},\ and\ \bibinfo {author} {\bibfnamefont {L.~P.}\ \bibnamefont {Kouwenhoven}},\ }\bibfield  {title} {\bibinfo {title} {Large zero-bias peaks in {{InSb-Al}} hybrid semiconductor-superconductor nanowire devices},\ }\href {http://arxiv.org/abs/2101.11456} {\bibfield  {journal} {\bibinfo  {journal} {arXiv:2101.11456}\ } (\bibinfo {year} {2021})}\BibitemShut {NoStop}%
\bibitem [{\citenamefont {Yu}\ \emph {et~al.}(2021)\citenamefont {Yu}, \citenamefont {Chen}, \citenamefont {Gomanko}, \citenamefont {Badawy}, \citenamefont {Bakkers}, \citenamefont {Zuo}, \citenamefont {Mourik},\ and\ \citenamefont {Frolov}}]{yu2021nonmajorana}%
  \BibitemOpen
  \bibfield  {author} {\bibinfo {author} {\bibfnamefont {P.}~\bibnamefont {Yu}}, \bibinfo {author} {\bibfnamefont {J.}~\bibnamefont {Chen}}, \bibinfo {author} {\bibfnamefont {M.}~\bibnamefont {Gomanko}}, \bibinfo {author} {\bibfnamefont {G.}~\bibnamefont {Badawy}}, \bibinfo {author} {\bibfnamefont {E.~P. a.~M.}\ \bibnamefont {Bakkers}}, \bibinfo {author} {\bibfnamefont {K.}~\bibnamefont {Zuo}}, \bibinfo {author} {\bibfnamefont {V.}~\bibnamefont {Mourik}},\ and\ \bibinfo {author} {\bibfnamefont {S.~M.}\ \bibnamefont {Frolov}},\ }\bibfield  {title} {\bibinfo {title} {Non-{{Majorana}} states yield nearly quantized conductance in proximatized nanowires},\ }\href {http://www.nature.com/articles/s41567-020-01107-w} {\bibfield  {journal} {\bibinfo  {journal} {Nature Physics}\ }\textbf {\bibinfo {volume} {17}},\ \bibinfo {pages} {482} (\bibinfo {year} {2021})}\BibitemShut {NoStop}%
\bibitem [{\citenamefont {Song}\ \emph {et~al.}(2021)\citenamefont {Song}, \citenamefont {Zhang}, \citenamefont {Pan}, \citenamefont {Liu}, \citenamefont {Wang}, \citenamefont {Cao}, \citenamefont {Liu}, \citenamefont {Wen}, \citenamefont {Liao}, \citenamefont {Zhuo}, \citenamefont {Liu}, \citenamefont {Shang}, \citenamefont {Zhao},\ and\ \citenamefont {Zhang}}]{song2021large}%
  \BibitemOpen
  \bibfield  {author} {\bibinfo {author} {\bibfnamefont {H.}~\bibnamefont {Song}}, \bibinfo {author} {\bibfnamefont {Z.}~\bibnamefont {Zhang}}, \bibinfo {author} {\bibfnamefont {D.}~\bibnamefont {Pan}}, \bibinfo {author} {\bibfnamefont {D.}~\bibnamefont {Liu}}, \bibinfo {author} {\bibfnamefont {Z.}~\bibnamefont {Wang}}, \bibinfo {author} {\bibfnamefont {Z.}~\bibnamefont {Cao}}, \bibinfo {author} {\bibfnamefont {L.}~\bibnamefont {Liu}}, \bibinfo {author} {\bibfnamefont {L.}~\bibnamefont {Wen}}, \bibinfo {author} {\bibfnamefont {D.}~\bibnamefont {Liao}}, \bibinfo {author} {\bibfnamefont {R.}~\bibnamefont {Zhuo}}, \bibinfo {author} {\bibfnamefont {D.~E.}\ \bibnamefont {Liu}}, \bibinfo {author} {\bibfnamefont {R.}~\bibnamefont {Shang}}, \bibinfo {author} {\bibfnamefont {J.}~\bibnamefont {Zhao}},\ and\ \bibinfo {author} {\bibfnamefont {H.}~\bibnamefont {Zhang}},\ }\bibfield  {title} {\bibinfo {title} {Large zero bias peaks and dips in a four-terminal thin {{InAs-Al}} nanowire device},\ }\href {http://arxiv.org/abs/2107.08282} {\bibfield  {journal} {\bibinfo  {journal} {arXiv:2107.08282}\ } (\bibinfo {year} {2021})}\BibitemShut {NoStop}%
\bibitem [{\citenamefont {{Microsoft Quantum}}\ \emph {et~al.}(2023)\citenamefont {{Microsoft Quantum}}, \citenamefont {Aghaee}, \citenamefont {Akkala}, \citenamefont {Alam}, \citenamefont {Ali}, \citenamefont {Alcaraz~Ramirez}, \citenamefont {Andrzejczuk}, \citenamefont {Antipov}, \citenamefont {Aseev}, \citenamefont {Astafev}, \citenamefont {Bauer}, \citenamefont {Becker}, \citenamefont {Boddapati}, \citenamefont {Boekhout}, \citenamefont {Bommer}, \citenamefont {Bosma}, \citenamefont {Bourdet}, \citenamefont {Boutin}, \citenamefont {Caroff}, \citenamefont {Casparis}, \citenamefont {Cassidy}, \citenamefont {Chatoor}, \citenamefont {Christensen}, \citenamefont {Clay}, \citenamefont {Cole}, \citenamefont {Corsetti}, \citenamefont {Cui}, \citenamefont {Dalampiras}, \citenamefont {Dokania}, \citenamefont {{de Lange}}, \citenamefont {{de Moor}}, \citenamefont {Estrada~Salda{\~n}a}, \citenamefont {Fallahi}, \citenamefont {Fathabad}, \citenamefont {Gamble}, \citenamefont {Gardner}, \citenamefont {Govender}, \citenamefont {Griggio}, \citenamefont {Grigoryan}, \citenamefont {Gronin}, \citenamefont {Gukelberger}, \citenamefont {Hansen}, \citenamefont {Heedt}, \citenamefont {Herranz~Zamorano}, \citenamefont {Ho}, \citenamefont {Holgaard}, \citenamefont {Ingerslev}, \citenamefont {Johansson}, \citenamefont {Jones}, \citenamefont {Kallaher}, \citenamefont {Karimi}, \citenamefont {Karzig}, \citenamefont {King}, \citenamefont {Kloster}, \citenamefont {Knapp}, \citenamefont {Kocon}, \citenamefont {Koski}, \citenamefont {Kostamo}, \citenamefont {Krogstrup}, \citenamefont {Kumar}, \citenamefont {Laeven}, \citenamefont {Larsen}, \citenamefont {Li}, \citenamefont {Lindemann}, \citenamefont {Love}, \citenamefont {Lutchyn}, \citenamefont {Madsen}, \citenamefont {Manfra}, \citenamefont {Markussen}, \citenamefont {Martinez}, \citenamefont {McNeil}, \citenamefont {Memisevic}, \citenamefont {Morgan}, \citenamefont {Mullally}, \citenamefont {Nayak}, \citenamefont {Nielsen}, \citenamefont {Nielsen}, \citenamefont {Nijholt}, \citenamefont {Nurmohamed}, \citenamefont {O'Farrell}, \citenamefont {Otani}, \citenamefont {Pauka}, \citenamefont {Petersson}, \citenamefont {Petit}, \citenamefont {Pikulin}, \citenamefont {Preiss}, \citenamefont {{Quintero-Perez}}, \citenamefont {Rajpalke}, \citenamefont {Rasmussen}, \citenamefont {Razmadze}, \citenamefont {Reentila}, \citenamefont {Reilly}, \citenamefont {Rouse}, \citenamefont {Sadovskyy}, \citenamefont {Sainiemi}, \citenamefont {Schreppler}, \citenamefont {Sidorkin}, \citenamefont {Singh}, \citenamefont {Singh}, \citenamefont {Sinha}, \citenamefont {Sohr}, \citenamefont {Stankevi{\v c}}, \citenamefont {Stek}, \citenamefont {Suominen}, \citenamefont {Suter}, \citenamefont {Svidenko}, \citenamefont {Teicher}, \citenamefont {Temuerhan}, \citenamefont {Thiyagarajah}, \citenamefont {Tholapi}, \citenamefont {Thomas}, \citenamefont {Toomey}, \citenamefont {Upadhyay}, \citenamefont {Urban}, \citenamefont {Vaitiek{\.e}nas}, \citenamefont {Van~Hoogdalem}, \citenamefont {Van~Woerkom}, \citenamefont {Viazmitinov}, \citenamefont {Vogel}, \citenamefont {Waddy}, \citenamefont {Watson}, \citenamefont {Weston}, \citenamefont {Winkler}, \citenamefont {Yang}, \citenamefont {Yau}, \citenamefont {Yi}, \citenamefont {Yucelen}, \citenamefont {Webster}, \citenamefont {Zeisel},\ and\ \citenamefont {Zhao}}]{microsoftquantum2023inasal}%
  \BibitemOpen
  \bibfield  {author} {\bibinfo {author} {\bibnamefont {{Microsoft Quantum}}}, \bibinfo {author} {\bibfnamefont {M.}~\bibnamefont {Aghaee}}, \bibinfo {author} {\bibfnamefont {A.}~\bibnamefont {Akkala}}, \bibinfo {author} {\bibfnamefont {Z.}~\bibnamefont {Alam}}, \bibinfo {author} {\bibfnamefont {R.}~\bibnamefont {Ali}}, \bibinfo {author} {\bibfnamefont {A.}~\bibnamefont {Alcaraz~Ramirez}}, \bibinfo {author} {\bibfnamefont {M.}~\bibnamefont {Andrzejczuk}}, \bibinfo {author} {\bibfnamefont {A.~E.}\ \bibnamefont {Antipov}}, \bibinfo {author} {\bibfnamefont {P.}~\bibnamefont {Aseev}}, \bibinfo {author} {\bibfnamefont {M.}~\bibnamefont {Astafev}}, \bibinfo {author} {\bibfnamefont {B.}~\bibnamefont {Bauer}}, \bibinfo {author} {\bibfnamefont {J.}~\bibnamefont {Becker}}, \bibinfo {author} {\bibfnamefont {S.}~\bibnamefont {Boddapati}}, \bibinfo {author} {\bibfnamefont {F.}~\bibnamefont {Boekhout}}, \bibinfo {author} {\bibfnamefont {J.}~\bibnamefont {Bommer}}, \bibinfo {author} {\bibfnamefont {T.}~\bibnamefont {Bosma}}, \bibinfo {author} {\bibfnamefont {L.}~\bibnamefont {Bourdet}}, \bibinfo {author} {\bibfnamefont {S.}~\bibnamefont {Boutin}}, \bibinfo {author} {\bibfnamefont {P.}~\bibnamefont {Caroff}}, \bibinfo {author} {\bibfnamefont {L.}~\bibnamefont {Casparis}}, \bibinfo {author} {\bibfnamefont {M.}~\bibnamefont {Cassidy}}, \bibinfo {author} {\bibfnamefont {S.}~\bibnamefont {Chatoor}}, \bibinfo {author} {\bibfnamefont {A.~W.}\ \bibnamefont {Christensen}}, \bibinfo {author} {\bibfnamefont {N.}~\bibnamefont {Clay}}, \bibinfo {author} {\bibfnamefont {W.~S.}\ \bibnamefont {Cole}}, \bibinfo {author} {\bibfnamefont {F.}~\bibnamefont {Corsetti}}, \bibinfo {author} {\bibfnamefont {A.}~\bibnamefont {Cui}}, \bibinfo {author} {\bibfnamefont {P.}~\bibnamefont {Dalampiras}}, \bibinfo {author} {\bibfnamefont {A.}~\bibnamefont {Dokania}}, \bibinfo {author} {\bibfnamefont {G.}~\bibnamefont {{de Lange}}}, \bibinfo {author} {\bibfnamefont {M.}~\bibnamefont {{de Moor}}}, \bibinfo {author} {\bibfnamefont {J.~C.}\ \bibnamefont {Estrada~Salda{\~n}a}}, \bibinfo {author} {\bibfnamefont {S.}~\bibnamefont {Fallahi}}, \bibinfo {author} {\bibfnamefont {Z.~H.}\ \bibnamefont {Fathabad}}, \bibinfo {author} {\bibfnamefont {J.}~\bibnamefont {Gamble}}, \bibinfo {author} {\bibfnamefont {G.}~\bibnamefont {Gardner}}, \bibinfo {author} {\bibfnamefont {D.}~\bibnamefont {Govender}}, \bibinfo {author} {\bibfnamefont {F.}~\bibnamefont {Griggio}}, \bibinfo {author} {\bibfnamefont {R.}~\bibnamefont {Grigoryan}}, \bibinfo {author} {\bibfnamefont {S.}~\bibnamefont {Gronin}}, \bibinfo {author} {\bibfnamefont {J.}~\bibnamefont {Gukelberger}}, \bibinfo {author} {\bibfnamefont {E.~B.}\ \bibnamefont {Hansen}}, \bibinfo {author} {\bibfnamefont {S.}~\bibnamefont {Heedt}}, \bibinfo {author} {\bibfnamefont {J.}~\bibnamefont {Herranz~Zamorano}}, \bibinfo {author} {\bibfnamefont {S.}~\bibnamefont {Ho}}, \bibinfo {author} {\bibfnamefont {U.~L.}\ \bibnamefont {Holgaard}}, \bibinfo {author} {\bibfnamefont {H.}~\bibnamefont {Ingerslev}}, \bibinfo {author} {\bibfnamefont {L.}~\bibnamefont {Johansson}}, \bibinfo {author} {\bibfnamefont {J.}~\bibnamefont {Jones}}, \bibinfo {author} {\bibfnamefont {R.}~\bibnamefont {Kallaher}}, \bibinfo {author} {\bibfnamefont {F.}~\bibnamefont {Karimi}}, \bibinfo {author} {\bibfnamefont {T.}~\bibnamefont {Karzig}}, \bibinfo {author} {\bibfnamefont {C.}~\bibnamefont {King}}, \bibinfo {author} {\bibfnamefont {M.~E.}\ \bibnamefont {Kloster}}, \bibinfo {author} {\bibfnamefont {C.}~\bibnamefont {Knapp}}, \bibinfo {author} {\bibfnamefont {D.}~\bibnamefont {Kocon}}, \bibinfo {author} {\bibfnamefont {J.}~\bibnamefont {Koski}}, \bibinfo {author} {\bibfnamefont {P.}~\bibnamefont {Kostamo}}, \bibinfo {author} {\bibfnamefont {P.}~\bibnamefont {Krogstrup}}, \bibinfo {author} {\bibfnamefont {M.}~\bibnamefont {Kumar}}, \bibinfo {author} {\bibfnamefont {T.}~\bibnamefont {Laeven}}, \bibinfo {author} {\bibfnamefont {T.}~\bibnamefont {Larsen}}, \bibinfo {author} {\bibfnamefont {K.}~\bibnamefont {Li}}, \bibinfo {author} {\bibfnamefont {T.}~\bibnamefont {Lindemann}}, \bibinfo {author} {\bibfnamefont {J.}~\bibnamefont {Love}}, \bibinfo {author} {\bibfnamefont {R.}~\bibnamefont {Lutchyn}}, \bibinfo {author} {\bibfnamefont {M.~H.}\ \bibnamefont {Madsen}}, \bibinfo {author} {\bibfnamefont {M.}~\bibnamefont {Manfra}}, \bibinfo {author} {\bibfnamefont {S.}~\bibnamefont {Markussen}}, \bibinfo {author} {\bibfnamefont {E.}~\bibnamefont {Martinez}}, \bibinfo {author} {\bibfnamefont {R.}~\bibnamefont {McNeil}}, \bibinfo {author} {\bibfnamefont {E.}~\bibnamefont {Memisevic}}, \bibinfo {author} {\bibfnamefont {T.}~\bibnamefont {Morgan}}, \bibinfo {author} {\bibfnamefont {A.}~\bibnamefont {Mullally}}, \bibinfo {author} {\bibfnamefont {C.}~\bibnamefont {Nayak}}, \bibinfo {author} {\bibfnamefont {J.}~\bibnamefont {Nielsen}}, \bibinfo {author} {\bibfnamefont {W.~H.~P.}\ \bibnamefont {Nielsen}}, \bibinfo {author} {\bibfnamefont {B.}~\bibnamefont {Nijholt}}, \bibinfo {author} {\bibfnamefont {A.}~\bibnamefont {Nurmohamed}}, \bibinfo {author} {\bibfnamefont {E.}~\bibnamefont {O'Farrell}}, \bibinfo {author} {\bibfnamefont {K.}~\bibnamefont {Otani}}, \bibinfo {author} {\bibfnamefont {S.}~\bibnamefont {Pauka}}, \bibinfo {author} {\bibfnamefont {K.}~\bibnamefont {Petersson}}, \bibinfo {author} {\bibfnamefont {L.}~\bibnamefont {Petit}}, \bibinfo {author} {\bibfnamefont {D.~I.}\ \bibnamefont {Pikulin}}, \bibinfo {author} {\bibfnamefont {F.}~\bibnamefont {Preiss}}, \bibinfo {author} {\bibfnamefont {M.}~\bibnamefont {{Quintero-Perez}}}, \bibinfo {author} {\bibfnamefont {M.}~\bibnamefont {Rajpalke}}, \bibinfo {author} {\bibfnamefont {K.}~\bibnamefont {Rasmussen}}, \bibinfo {author} {\bibfnamefont {D.}~\bibnamefont {Razmadze}}, \bibinfo {author} {\bibfnamefont {O.}~\bibnamefont {Reentila}}, \bibinfo {author} {\bibfnamefont {D.}~\bibnamefont {Reilly}}, \bibinfo {author} {\bibfnamefont {R.}~\bibnamefont {Rouse}}, \bibinfo {author} {\bibfnamefont {I.}~\bibnamefont {Sadovskyy}}, \bibinfo {author} {\bibfnamefont {L.}~\bibnamefont {Sainiemi}}, \bibinfo {author} {\bibfnamefont {S.}~\bibnamefont {Schreppler}}, \bibinfo {author} {\bibfnamefont {V.}~\bibnamefont {Sidorkin}}, \bibinfo {author} {\bibfnamefont {A.}~\bibnamefont {Singh}}, \bibinfo {author} {\bibfnamefont {S.}~\bibnamefont {Singh}}, \bibinfo {author} {\bibfnamefont {S.}~\bibnamefont {Sinha}}, \bibinfo {author} {\bibfnamefont {P.}~\bibnamefont {Sohr}}, \bibinfo {author} {\bibfnamefont {T.}~\bibnamefont {Stankevi{\v c}}}, \bibinfo {author} {\bibfnamefont {L.}~\bibnamefont {Stek}}, \bibinfo {author} {\bibfnamefont {H.}~\bibnamefont {Suominen}}, \bibinfo {author} {\bibfnamefont {J.}~\bibnamefont {Suter}}, \bibinfo {author} {\bibfnamefont {V.}~\bibnamefont {Svidenko}}, \bibinfo {author} {\bibfnamefont {S.}~\bibnamefont {Teicher}}, \bibinfo {author} {\bibfnamefont {M.}~\bibnamefont {Temuerhan}}, \bibinfo {author} {\bibfnamefont {N.}~\bibnamefont {Thiyagarajah}}, \bibinfo {author} {\bibfnamefont {R.}~\bibnamefont {Tholapi}}, \bibinfo {author} {\bibfnamefont {M.}~\bibnamefont {Thomas}}, \bibinfo {author} {\bibfnamefont {E.}~\bibnamefont {Toomey}}, \bibinfo {author} {\bibfnamefont {S.}~\bibnamefont {Upadhyay}}, \bibinfo {author} {\bibfnamefont {I.}~\bibnamefont {Urban}}, \bibinfo {author} {\bibfnamefont {S.}~\bibnamefont {Vaitiek{\.e}nas}}, \bibinfo {author} {\bibfnamefont {K.}~\bibnamefont {Van~Hoogdalem}}, \bibinfo {author} {\bibfnamefont {D.}~\bibnamefont {Van~Woerkom}}, \bibinfo {author} {\bibfnamefont {D.~V.}\ \bibnamefont {Viazmitinov}}, \bibinfo {author} {\bibfnamefont {D.}~\bibnamefont {Vogel}}, \bibinfo {author} {\bibfnamefont {S.}~\bibnamefont {Waddy}}, \bibinfo {author} {\bibfnamefont {J.}~\bibnamefont {Watson}}, \bibinfo {author} {\bibfnamefont {J.}~\bibnamefont {Weston}}, \bibinfo {author} {\bibfnamefont {G.~W.}\ \bibnamefont {Winkler}}, \bibinfo {author} {\bibfnamefont {C.~K.}\ \bibnamefont {Yang}}, \bibinfo {author} {\bibfnamefont {S.}~\bibnamefont {Yau}}, \bibinfo {author} {\bibfnamefont {D.}~\bibnamefont {Yi}}, \bibinfo {author} {\bibfnamefont {E.}~\bibnamefont {Yucelen}}, \bibinfo {author} {\bibfnamefont {A.}~\bibnamefont {Webster}}, \bibinfo {author} {\bibfnamefont {R.}~\bibnamefont {Zeisel}},\ and\ \bibinfo {author} {\bibfnamefont {R.}~\bibnamefont {Zhao}},\ }\bibfield  {title} {\bibinfo {title} {{{InAs-Al}} hybrid devices passing the topological gap protocol},\ }\href {https://link.aps.org/doi/10.1103/PhysRevB.107.245423} {\bibfield  {journal} {\bibinfo  {journal} {Physical Review B}\ }\textbf {\bibinfo {volume} {107}},\ \bibinfo {pages} {245423} (\bibinfo {year} {2023})}\BibitemShut {NoStop}%
\bibitem [{\citenamefont {Kells}\ \emph {et~al.}(2012)\citenamefont {Kells}, \citenamefont {Meidan},\ and\ \citenamefont {Brouwer}}]{kells2012nearzeroenergy}%
  \BibitemOpen
  \bibfield  {author} {\bibinfo {author} {\bibfnamefont {G.}~\bibnamefont {Kells}}, \bibinfo {author} {\bibfnamefont {D.}~\bibnamefont {Meidan}},\ and\ \bibinfo {author} {\bibfnamefont {P.~W.}\ \bibnamefont {Brouwer}},\ }\bibfield  {title} {\bibinfo {title} {Near-zero-energy end states in topologically trivial spin-orbit coupled superconducting nanowires with a smooth confinement},\ }\href {https://link.aps.org/doi/10.1103/PhysRevB.86.100503} {\bibfield  {journal} {\bibinfo  {journal} {Physical Review B}\ }\textbf {\bibinfo {volume} {86}},\ \bibinfo {pages} {100503} (\bibinfo {year} {2012})}\BibitemShut {NoStop}%
\bibitem [{\citenamefont {Liu}\ \emph {et~al.}(2017)\citenamefont {Liu}, \citenamefont {Sau}, \citenamefont {Stanescu},\ and\ \citenamefont {Das~Sarma}}]{liu2017andreev}%
  \BibitemOpen
  \bibfield  {author} {\bibinfo {author} {\bibfnamefont {C.-X.}\ \bibnamefont {Liu}}, \bibinfo {author} {\bibfnamefont {J.~D.}\ \bibnamefont {Sau}}, \bibinfo {author} {\bibfnamefont {T.~D.}\ \bibnamefont {Stanescu}},\ and\ \bibinfo {author} {\bibfnamefont {S.}~\bibnamefont {Das~Sarma}},\ }\bibfield  {title} {\bibinfo {title} {Andreev bound states versus {{Majorana}} bound states in quantum dot-nanowire-superconductor hybrid structures: {{Trivial}} versus topological zero-bias conductance peaks},\ }\href {https://link.aps.org/doi/10.1103/PhysRevB.96.075161} {\bibfield  {journal} {\bibinfo  {journal} {Physical Review B}\ }\textbf {\bibinfo {volume} {96}},\ \bibinfo {pages} {075161} (\bibinfo {year} {2017})}\BibitemShut {NoStop}%
\bibitem [{\citenamefont {Pan}\ and\ \citenamefont {Das~Sarma}(2020)}]{pan2020physical}%
  \BibitemOpen
  \bibfield  {author} {\bibinfo {author} {\bibfnamefont {H.}~\bibnamefont {Pan}}\ and\ \bibinfo {author} {\bibfnamefont {S.}~\bibnamefont {Das~Sarma}},\ }\bibfield  {title} {\bibinfo {title} {Physical mechanisms for zero-bias conductance peaks in {{Majorana}} nanowires},\ }\href {https://link.aps.org/doi/10.1103/PhysRevResearch.2.013377} {\bibfield  {journal} {\bibinfo  {journal} {Physical Review Research}\ }\textbf {\bibinfo {volume} {2}},\ \bibinfo {pages} {013377} (\bibinfo {year} {2020})}\BibitemShut {NoStop}%
\bibitem [{\citenamefont {Brouwer}\ \emph {et~al.}(2000)\citenamefont {Brouwer}, \citenamefont {Furusaki}, \citenamefont {Gruzberg},\ and\ \citenamefont {Mudry}}]{brouwer2000localization}%
  \BibitemOpen
  \bibfield  {author} {\bibinfo {author} {\bibfnamefont {P.~W.}\ \bibnamefont {Brouwer}}, \bibinfo {author} {\bibfnamefont {A.}~\bibnamefont {Furusaki}}, \bibinfo {author} {\bibfnamefont {I.~A.}\ \bibnamefont {Gruzberg}},\ and\ \bibinfo {author} {\bibfnamefont {C.}~\bibnamefont {Mudry}},\ }\bibfield  {title} {\bibinfo {title} {Localization and {{Delocalization}} in {{Dirty Superconducting Wires}}},\ }\href {https://link.aps.org/doi/10.1103/PhysRevLett.85.1064} {\bibfield  {journal} {\bibinfo  {journal} {Physical Review Letters}\ }\textbf {\bibinfo {volume} {85}},\ \bibinfo {pages} {1064} (\bibinfo {year} {2000})}\BibitemShut {NoStop}%
\bibitem [{\citenamefont {Brouwer}\ \emph {et~al.}(2011{\natexlab{a}})\citenamefont {Brouwer}, \citenamefont {Duckheim}, \citenamefont {Romito},\ and\ \citenamefont {{von Oppen}}}]{brouwer2011topological}%
  \BibitemOpen
  \bibfield  {author} {\bibinfo {author} {\bibfnamefont {P.~W.}\ \bibnamefont {Brouwer}}, \bibinfo {author} {\bibfnamefont {M.}~\bibnamefont {Duckheim}}, \bibinfo {author} {\bibfnamefont {A.}~\bibnamefont {Romito}},\ and\ \bibinfo {author} {\bibfnamefont {F.}~\bibnamefont {{von Oppen}}},\ }\bibfield  {title} {\bibinfo {title} {Topological superconducting phases in disordered quantum wires with strong spin-orbit coupling},\ }\href {https://link.aps.org/doi/10.1103/PhysRevB.84.144526} {\bibfield  {journal} {\bibinfo  {journal} {Physical Review B}\ }\textbf {\bibinfo {volume} {84}},\ \bibinfo {pages} {144526} (\bibinfo {year} {2011}{\natexlab{a}})}\BibitemShut {NoStop}%
\bibitem [{\citenamefont {Brouwer}\ \emph {et~al.}(2011{\natexlab{b}})\citenamefont {Brouwer}, \citenamefont {Duckheim}, \citenamefont {Romito},\ and\ \citenamefont {{von Oppen}}}]{brouwer2011probability}%
  \BibitemOpen
  \bibfield  {author} {\bibinfo {author} {\bibfnamefont {P.~W.}\ \bibnamefont {Brouwer}}, \bibinfo {author} {\bibfnamefont {M.}~\bibnamefont {Duckheim}}, \bibinfo {author} {\bibfnamefont {A.}~\bibnamefont {Romito}},\ and\ \bibinfo {author} {\bibfnamefont {F.}~\bibnamefont {{von Oppen}}},\ }\bibfield  {title} {\bibinfo {title} {Probability {{Distribution}} of {{Majorana End-State Energies}} in {{Disordered Wires}}},\ }\href {https://link.aps.org/doi/10.1103/PhysRevLett.107.196804} {\bibfield  {journal} {\bibinfo  {journal} {Physical Review Letters}\ }\textbf {\bibinfo {volume} {107}},\ \bibinfo {pages} {196804} (\bibinfo {year} {2011}{\natexlab{b}})}\BibitemShut {NoStop}%
\bibitem [{\citenamefont {Liu}\ \emph {et~al.}(2012)\citenamefont {Liu}, \citenamefont {Potter}, \citenamefont {Law},\ and\ \citenamefont {Lee}}]{liu2012zerobias}%
  \BibitemOpen
  \bibfield  {author} {\bibinfo {author} {\bibfnamefont {J.}~\bibnamefont {Liu}}, \bibinfo {author} {\bibfnamefont {A.~C.}\ \bibnamefont {Potter}}, \bibinfo {author} {\bibfnamefont {K.~T.}\ \bibnamefont {Law}},\ and\ \bibinfo {author} {\bibfnamefont {P.~A.}\ \bibnamefont {Lee}},\ }\bibfield  {title} {\bibinfo {title} {Zero-{{Bias Peaks}} in the {{Tunneling Conductance}} of {{Spin-Orbit-Coupled Superconducting Wires}} with and without {{Majorana End-States}}},\ }\href {https://link.aps.org/doi/10.1103/PhysRevLett.109.267002} {\bibfield  {journal} {\bibinfo  {journal} {Physical Review Letters}\ }\textbf {\bibinfo {volume} {109}},\ \bibinfo {pages} {267002} (\bibinfo {year} {2012})}\BibitemShut {NoStop}%
\bibitem [{\citenamefont {Bagrets}\ and\ \citenamefont {Altland}(2012)}]{bagrets2012class}%
  \BibitemOpen
  \bibfield  {author} {\bibinfo {author} {\bibfnamefont {D.}~\bibnamefont {Bagrets}}\ and\ \bibinfo {author} {\bibfnamefont {A.}~\bibnamefont {Altland}},\ }\bibfield  {title} {\bibinfo {title} {Class {$D$} {{Spectral Peak}} in {{Majorana Quantum Wires}}},\ }\href {https://link.aps.org/doi/10.1103/PhysRevLett.109.227005} {\bibfield  {journal} {\bibinfo  {journal} {Physical Review Letters}\ }\textbf {\bibinfo {volume} {109}},\ \bibinfo {pages} {227005} (\bibinfo {year} {2012})}\BibitemShut {NoStop}%
\bibitem [{\citenamefont {Akhmerov}\ \emph {et~al.}(2011)\citenamefont {Akhmerov}, \citenamefont {Dahlhaus}, \citenamefont {Hassler}, \citenamefont {Wimmer},\ and\ \citenamefont {Beenakker}}]{akhmerov2011quantized}%
  \BibitemOpen
  \bibfield  {author} {\bibinfo {author} {\bibfnamefont {A.~R.}\ \bibnamefont {Akhmerov}}, \bibinfo {author} {\bibfnamefont {J.~P.}\ \bibnamefont {Dahlhaus}}, \bibinfo {author} {\bibfnamefont {F.}~\bibnamefont {Hassler}}, \bibinfo {author} {\bibfnamefont {M.}~\bibnamefont {Wimmer}},\ and\ \bibinfo {author} {\bibfnamefont {C.~W.~J.}\ \bibnamefont {Beenakker}},\ }\bibfield  {title} {\bibinfo {title} {Quantized {{Conductance}} at the {{Majorana Phase Transition}} in a {{Disordered Superconducting Wire}}},\ }\href {https://link.aps.org/doi/10.1103/PhysRevLett.106.057001} {\bibfield  {journal} {\bibinfo  {journal} {Physical Review Letters}\ }\textbf {\bibinfo {volume} {106}},\ \bibinfo {pages} {057001} (\bibinfo {year} {2011})}\BibitemShut {NoStop}%
\bibitem [{\citenamefont {Sau}\ and\ \citenamefont {Das~Sarma}(2013)}]{sau2013density}%
  \BibitemOpen
  \bibfield  {author} {\bibinfo {author} {\bibfnamefont {J.~D.}\ \bibnamefont {Sau}}\ and\ \bibinfo {author} {\bibfnamefont {S.}~\bibnamefont {Das~Sarma}},\ }\bibfield  {title} {\bibinfo {title} {Density of states of disordered topological superconductor-semiconductor hybrid nanowires},\ }\href {https://link.aps.org/doi/10.1103/PhysRevB.88.064506} {\bibfield  {journal} {\bibinfo  {journal} {Physical Review B}\ }\textbf {\bibinfo {volume} {88}},\ \bibinfo {pages} {064506} (\bibinfo {year} {2013})}\BibitemShut {NoStop}%
\bibitem [{\citenamefont {Takei}\ \emph {et~al.}(2013)\citenamefont {Takei}, \citenamefont {Fregoso}, \citenamefont {Hui}, \citenamefont {Lobos},\ and\ \citenamefont {Das~Sarma}}]{takei2013soft}%
  \BibitemOpen
  \bibfield  {author} {\bibinfo {author} {\bibfnamefont {S.}~\bibnamefont {Takei}}, \bibinfo {author} {\bibfnamefont {B.~M.}\ \bibnamefont {Fregoso}}, \bibinfo {author} {\bibfnamefont {H.-Y.}\ \bibnamefont {Hui}}, \bibinfo {author} {\bibfnamefont {A.~M.}\ \bibnamefont {Lobos}},\ and\ \bibinfo {author} {\bibfnamefont {S.}~\bibnamefont {Das~Sarma}},\ }\bibfield  {title} {\bibinfo {title} {Soft {{Superconducting Gap}} in {{Semiconductor Majorana Nanowires}}},\ }\href {https://link.aps.org/doi/10.1103/PhysRevLett.110.186803} {\bibfield  {journal} {\bibinfo  {journal} {Physical Review Letters}\ }\textbf {\bibinfo {volume} {110}},\ \bibinfo {pages} {186803} (\bibinfo {year} {2013})}\BibitemShut {NoStop}%
\bibitem [{\citenamefont {Adagideli}\ \emph {et~al.}(2014)\citenamefont {Adagideli}, \citenamefont {Wimmer},\ and\ \citenamefont {Teker}}]{adagideli2014effects}%
  \BibitemOpen
  \bibfield  {author} {\bibinfo {author} {\bibfnamefont {{\.I}.}~\bibnamefont {Adagideli}}, \bibinfo {author} {\bibfnamefont {M.}~\bibnamefont {Wimmer}},\ and\ \bibinfo {author} {\bibfnamefont {A.}~\bibnamefont {Teker}},\ }\bibfield  {title} {\bibinfo {title} {Effects of electron scattering on the topological properties of nanowires: {{Majorana}} fermions from disorder and superlattices},\ }\href {https://link.aps.org/doi/10.1103/PhysRevB.89.144506} {\bibfield  {journal} {\bibinfo  {journal} {Physical Review B}\ }\textbf {\bibinfo {volume} {89}},\ \bibinfo {pages} {144506} (\bibinfo {year} {2014})}\BibitemShut {NoStop}%
\bibitem [{\citenamefont {Pekerten}\ \emph {et~al.}(2017)\citenamefont {Pekerten}, \citenamefont {Teker}, \citenamefont {Bozat}, \citenamefont {Wimmer},\ and\ \citenamefont {Adagideli}}]{pekerten2017disorderinduced}%
  \BibitemOpen
  \bibfield  {author} {\bibinfo {author} {\bibfnamefont {B.}~\bibnamefont {Pekerten}}, \bibinfo {author} {\bibfnamefont {A.}~\bibnamefont {Teker}}, \bibinfo {author} {\bibfnamefont {{\"O}.}~\bibnamefont {Bozat}}, \bibinfo {author} {\bibfnamefont {M.}~\bibnamefont {Wimmer}},\ and\ \bibinfo {author} {\bibfnamefont {{\.I}.}~\bibnamefont {Adagideli}},\ }\bibfield  {title} {\bibinfo {title} {Disorder-induced topological transitions in multichannel {{Majorana}} wires},\ }\href {https://link.aps.org/doi/10.1103/PhysRevB.95.064507} {\bibfield  {journal} {\bibinfo  {journal} {Physical Review B}\ }\textbf {\bibinfo {volume} {95}},\ \bibinfo {pages} {064507} (\bibinfo {year} {2017})}\BibitemShut {NoStop}%
\bibitem [{\citenamefont {Ahn}\ \emph {et~al.}(2021)\citenamefont {Ahn}, \citenamefont {Pan}, \citenamefont {Woods}, \citenamefont {Stanescu},\ and\ \citenamefont {Das~Sarma}}]{ahn2021estimating}%
  \BibitemOpen
  \bibfield  {author} {\bibinfo {author} {\bibfnamefont {S.}~\bibnamefont {Ahn}}, \bibinfo {author} {\bibfnamefont {H.}~\bibnamefont {Pan}}, \bibinfo {author} {\bibfnamefont {B.}~\bibnamefont {Woods}}, \bibinfo {author} {\bibfnamefont {T.~D.}\ \bibnamefont {Stanescu}},\ and\ \bibinfo {author} {\bibfnamefont {S.}~\bibnamefont {Das~Sarma}},\ }\bibfield  {title} {\bibinfo {title} {Estimating disorder and its adverse effects in semiconductor {{Majorana}} nanowires},\ }\href {https://link.aps.org/doi/10.1103/PhysRevMaterials.5.124602} {\bibfield  {journal} {\bibinfo  {journal} {Physical Review Materials}\ }\textbf {\bibinfo {volume} {5}},\ \bibinfo {pages} {124602} (\bibinfo {year} {2021})}\BibitemShut {NoStop}%
\bibitem [{\citenamefont {Woods}\ \emph {et~al.}(2021)\citenamefont {Woods}, \citenamefont {Das~Sarma},\ and\ \citenamefont {Stanescu}}]{woods2021chargeimpurity}%
  \BibitemOpen
  \bibfield  {author} {\bibinfo {author} {\bibfnamefont {B.~D.}\ \bibnamefont {Woods}}, \bibinfo {author} {\bibfnamefont {S.}~\bibnamefont {Das~Sarma}},\ and\ \bibinfo {author} {\bibfnamefont {T.~D.}\ \bibnamefont {Stanescu}},\ }\bibfield  {title} {\bibinfo {title} {Charge-{{Impurity Effects}} in {{Hybrid Majorana Nanowires}}},\ }\href {https://link.aps.org/doi/10.1103/PhysRevApplied.16.054053} {\bibfield  {journal} {\bibinfo  {journal} {Physical Review Applied}\ }\textbf {\bibinfo {volume} {16}},\ \bibinfo {pages} {054053} (\bibinfo {year} {2021})}\BibitemShut {NoStop}%
\bibitem [{\citenamefont {Das~Sarma}\ and\ \citenamefont {Pan}(2021)}]{dassarma2021disorderinduced}%
  \BibitemOpen
  \bibfield  {author} {\bibinfo {author} {\bibfnamefont {S.}~\bibnamefont {Das~Sarma}}\ and\ \bibinfo {author} {\bibfnamefont {H.}~\bibnamefont {Pan}},\ }\bibfield  {title} {\bibinfo {title} {Disorder-induced zero-bias peaks in {{Majorana}} nanowires},\ }\href {https://link.aps.org/doi/10.1103/PhysRevB.103.195158} {\bibfield  {journal} {\bibinfo  {journal} {Physical Review B}\ }\textbf {\bibinfo {volume} {103}},\ \bibinfo {pages} {195158} (\bibinfo {year} {2021})}\BibitemShut {NoStop}%
\bibitem [{\citenamefont {Das~Sarma}\ and\ \citenamefont {Pan}(2023)}]{dassarma2023density}%
  \BibitemOpen
  \bibfield  {author} {\bibinfo {author} {\bibfnamefont {S.}~\bibnamefont {Das~Sarma}}\ and\ \bibinfo {author} {\bibfnamefont {H.}~\bibnamefont {Pan}},\ }\bibfield  {title} {\bibinfo {title} {Density of states, transport, and topology in disordered {{Majorana}} nanowires},\ }\href {https://link.aps.org/doi/10.1103/PhysRevB.108.085415} {\bibfield  {journal} {\bibinfo  {journal} {Physical Review B}\ }\textbf {\bibinfo {volume} {108}},\ \bibinfo {pages} {085415} (\bibinfo {year} {2023})}\BibitemShut {NoStop}%
\bibitem [{\citenamefont {Das~Sarma}\ \emph {et~al.}(2023)\citenamefont {Das~Sarma}, \citenamefont {Sau},\ and\ \citenamefont {Stanescu}}]{dassarma2023spectral}%
  \BibitemOpen
  \bibfield  {author} {\bibinfo {author} {\bibfnamefont {S.}~\bibnamefont {Das~Sarma}}, \bibinfo {author} {\bibfnamefont {J.~D.}\ \bibnamefont {Sau}},\ and\ \bibinfo {author} {\bibfnamefont {T.~D.}\ \bibnamefont {Stanescu}},\ }\bibfield  {title} {\bibinfo {title} {Spectral properties, topological patches, and effective phase diagrams of finite disordered {{Majorana}} nanowires},\ }\href {https://link.aps.org/doi/10.1103/PhysRevB.108.085416} {\bibfield  {journal} {\bibinfo  {journal} {Physical Review B}\ }\textbf {\bibinfo {volume} {108}},\ \bibinfo {pages} {085416} (\bibinfo {year} {2023})}\BibitemShut {NoStop}%
\bibitem [{\citenamefont {Taylor}\ \emph {et~al.}(2023)\citenamefont {Taylor}, \citenamefont {Sau},\ and\ \citenamefont {Sarma}}]{taylor2023machine}%
  \BibitemOpen
  \bibfield  {author} {\bibinfo {author} {\bibfnamefont {J.~R.}\ \bibnamefont {Taylor}}, \bibinfo {author} {\bibfnamefont {J.~D.}\ \bibnamefont {Sau}},\ and\ \bibinfo {author} {\bibfnamefont {S.~D.}\ \bibnamefont {Sarma}},\ }\bibfield  {title} {\bibinfo {title} {Machine learning {{Majorana}} nanowire disorder landscape},\ }\href {http://arxiv.org/abs/2307.11068} {\bibfield  {journal} {\bibinfo  {journal} {arXiv:2307.11068}\ } (\bibinfo {year} {2023})}\BibitemShut {NoStop}%
\bibitem [{\citenamefont {Brzezicki}\ \emph {et~al.}(2017)\citenamefont {Brzezicki}, \citenamefont {Ole{\'s}},\ and\ \citenamefont {Cuoco}}]{brzezicki2017driving}%
  \BibitemOpen
  \bibfield  {author} {\bibinfo {author} {\bibfnamefont {W.}~\bibnamefont {Brzezicki}}, \bibinfo {author} {\bibfnamefont {A.~M.}\ \bibnamefont {Ole{\'s}}},\ and\ \bibinfo {author} {\bibfnamefont {M.}~\bibnamefont {Cuoco}},\ }\bibfield  {title} {\bibinfo {title} {Driving topological phases by spatially inhomogeneous pairing centers},\ }\href {https://link.aps.org/doi/10.1103/PhysRevB.95.140506} {\bibfield  {journal} {\bibinfo  {journal} {Physical Review B}\ }\textbf {\bibinfo {volume} {95}},\ \bibinfo {pages} {140506} (\bibinfo {year} {2017})}\BibitemShut {NoStop}%
\bibitem [{\citenamefont {Barman~Ray}\ \emph {et~al.}(2021)\citenamefont {Barman~Ray}, \citenamefont {Sau},\ and\ \citenamefont {Mandal}}]{barmanray2021symmetrybreaking}%
  \BibitemOpen
  \bibfield  {author} {\bibinfo {author} {\bibfnamefont {A.}~\bibnamefont {Barman~Ray}}, \bibinfo {author} {\bibfnamefont {J.~D.}\ \bibnamefont {Sau}},\ and\ \bibinfo {author} {\bibfnamefont {I.}~\bibnamefont {Mandal}},\ }\bibfield  {title} {\bibinfo {title} {Symmetry-breaking signatures of multiple {{Majorana}} zero modes in one-dimensional spin-triplet superconductors},\ }\href {https://link.aps.org/doi/10.1103/PhysRevB.104.104513} {\bibfield  {journal} {\bibinfo  {journal} {Physical Review B}\ }\textbf {\bibinfo {volume} {104}},\ \bibinfo {pages} {104513} (\bibinfo {year} {2021})}\BibitemShut {NoStop}%
\bibitem [{\citenamefont {Pan}\ \emph {et~al.}(2021)\citenamefont {Pan}, \citenamefont {Sau},\ and\ \citenamefont {Das~Sarma}}]{pan2021threeterminal}%
  \BibitemOpen
  \bibfield  {author} {\bibinfo {author} {\bibfnamefont {H.}~\bibnamefont {Pan}}, \bibinfo {author} {\bibfnamefont {J.~D.}\ \bibnamefont {Sau}},\ and\ \bibinfo {author} {\bibfnamefont {S.}~\bibnamefont {Das~Sarma}},\ }\bibfield  {title} {\bibinfo {title} {Three-terminal nonlocal conductance in {{Majorana}} nanowires: {{Distinguishing}} topological and trivial in realistic systems with disorder and inhomogeneous potential},\ }\href {https://link.aps.org/doi/10.1103/PhysRevB.103.014513} {\bibfield  {journal} {\bibinfo  {journal} {Physical Review B}\ }\textbf {\bibinfo {volume} {103}},\ \bibinfo {pages} {014513} (\bibinfo {year} {2021})}\BibitemShut {NoStop}%
\bibitem [{\citenamefont {Motrunich}\ \emph {et~al.}(2001)\citenamefont {Motrunich}, \citenamefont {Damle},\ and\ \citenamefont {Huse}}]{motrunich2001griffiths}%
  \BibitemOpen
  \bibfield  {author} {\bibinfo {author} {\bibfnamefont {O.}~\bibnamefont {Motrunich}}, \bibinfo {author} {\bibfnamefont {K.}~\bibnamefont {Damle}},\ and\ \bibinfo {author} {\bibfnamefont {D.~A.}\ \bibnamefont {Huse}},\ }\bibfield  {title} {\bibinfo {title} {Griffiths effects and quantum critical points in dirty superconductors without spin-rotation invariance: {{One-dimensional}} examples},\ }\href {https://link.aps.org/doi/10.1103/PhysRevB.63.224204} {\bibfield  {journal} {\bibinfo  {journal} {Physical Review B}\ }\textbf {\bibinfo {volume} {63}},\ \bibinfo {pages} {224204} (\bibinfo {year} {2001})}\BibitemShut {NoStop}%
\bibitem [{\citenamefont {Kitaev}(2001)}]{kitaev2001unpaired}%
  \BibitemOpen
  \bibfield  {author} {\bibinfo {author} {\bibfnamefont {A.~Y.}\ \bibnamefont {Kitaev}},\ }\bibfield  {title} {\bibinfo {title} {Unpaired {{Majorana}} fermions in quantum wires},\ }\href {https://doi.org/10.1070%2F1063-7869%2F44%2F10s%2Fs29} {\bibfield  {journal} {\bibinfo  {journal} {Physics-Uspekhi}\ }\textbf {\bibinfo {volume} {44}},\ \bibinfo {pages} {131} (\bibinfo {year} {2001})}\BibitemShut {NoStop}%
\bibitem [{\citenamefont {Stanescu}\ \emph {et~al.}(2010)\citenamefont {Stanescu}, \citenamefont {Sau}, \citenamefont {Lutchyn},\ and\ \citenamefont {Das~Sarma}}]{stanescu2010proximity}%
  \BibitemOpen
  \bibfield  {author} {\bibinfo {author} {\bibfnamefont {T.~D.}\ \bibnamefont {Stanescu}}, \bibinfo {author} {\bibfnamefont {J.~D.}\ \bibnamefont {Sau}}, \bibinfo {author} {\bibfnamefont {R.~M.}\ \bibnamefont {Lutchyn}},\ and\ \bibinfo {author} {\bibfnamefont {S.}~\bibnamefont {Das~Sarma}},\ }\bibfield  {title} {\bibinfo {title} {Proximity effect at the superconductor--topological insulator interface},\ }\href {https://link.aps.org/doi/10.1103/PhysRevB.81.241310} {\bibfield  {journal} {\bibinfo  {journal} {Physical Review B}\ }\textbf {\bibinfo {volume} {81}},\ \bibinfo {pages} {241310} (\bibinfo {year} {2010})}\BibitemShut {NoStop}%
\bibitem [{\citenamefont {Cheng}\ \emph {et~al.}(2009)\citenamefont {Cheng}, \citenamefont {Lutchyn}, \citenamefont {Galitski},\ and\ \citenamefont {Das~Sarma}}]{cheng2009splitting}%
  \BibitemOpen
  \bibfield  {author} {\bibinfo {author} {\bibfnamefont {M.}~\bibnamefont {Cheng}}, \bibinfo {author} {\bibfnamefont {R.~M.}\ \bibnamefont {Lutchyn}}, \bibinfo {author} {\bibfnamefont {V.}~\bibnamefont {Galitski}},\ and\ \bibinfo {author} {\bibfnamefont {S.}~\bibnamefont {Das~Sarma}},\ }\bibfield  {title} {\bibinfo {title} {Splitting of {{Majorana-Fermion Modes}} due to {{Intervortex Tunneling}} in a p{\textsubscript{x}}+ip{\textsubscript{y}} {{Superconductor}}},\ }\href {https://link.aps.org/doi/10.1103/PhysRevLett.103.107001} {\bibfield  {journal} {\bibinfo  {journal} {Physical Review Letters}\ }\textbf {\bibinfo {volume} {103}},\ \bibinfo {pages} {107001} (\bibinfo {year} {2009})}\BibitemShut {NoStop}%
\bibitem [{\citenamefont {Groth}\ \emph {et~al.}(2014)\citenamefont {Groth}, \citenamefont {Wimmer}, \citenamefont {Akhmerov},\ and\ \citenamefont {Waintal}}]{groth2014kwant}%
  \BibitemOpen
  \bibfield  {author} {\bibinfo {author} {\bibfnamefont {C.~W.}\ \bibnamefont {Groth}}, \bibinfo {author} {\bibfnamefont {M.}~\bibnamefont {Wimmer}}, \bibinfo {author} {\bibfnamefont {A.~R.}\ \bibnamefont {Akhmerov}},\ and\ \bibinfo {author} {\bibfnamefont {X.}~\bibnamefont {Waintal}},\ }\bibfield  {title} {\bibinfo {title} {Kwant: A software package for quantum transport},\ }\href {http://iopscience.iop.org/article/10.1088/1367-2630/16/6/063065/meta} {\bibfield  {journal} {\bibinfo  {journal} {New Journal of Physics}\ }\textbf {\bibinfo {volume} {16}},\ \bibinfo {pages} {063065} (\bibinfo {year} {2014})}\BibitemShut {NoStop}%
\bibitem [{\citenamefont {Blonder}\ \emph {et~al.}(1982)\citenamefont {Blonder}, \citenamefont {Tinkham},\ and\ \citenamefont {Klapwijk}}]{blonder1982transition}%
  \BibitemOpen
  \bibfield  {author} {\bibinfo {author} {\bibfnamefont {G.~E.}\ \bibnamefont {Blonder}}, \bibinfo {author} {\bibfnamefont {M.}~\bibnamefont {Tinkham}},\ and\ \bibinfo {author} {\bibfnamefont {T.~M.}\ \bibnamefont {Klapwijk}},\ }\bibfield  {title} {\bibinfo {title} {Transition from metallic to tunneling regimes in superconducting microconstrictions: {{Excess}} current, charge imbalance, and supercurrent conversion},\ }\href {https://link.aps.org/doi/10.1103/PhysRevB.25.4515} {\bibfield  {journal} {\bibinfo  {journal} {Physical Review B}\ }\textbf {\bibinfo {volume} {25}},\ \bibinfo {pages} {4515} (\bibinfo {year} {1982})}\BibitemShut {NoStop}%
\bibitem [{\citenamefont {Setiawan}\ \emph {et~al.}(2015)\citenamefont {Setiawan}, \citenamefont {Brydon}, \citenamefont {Sau},\ and\ \citenamefont {Das~Sarma}}]{setiawan2015conductance}%
  \BibitemOpen
  \bibfield  {author} {\bibinfo {author} {\bibfnamefont {F.}~\bibnamefont {Setiawan}}, \bibinfo {author} {\bibfnamefont {P.~M.~R.}\ \bibnamefont {Brydon}}, \bibinfo {author} {\bibfnamefont {J.~D.}\ \bibnamefont {Sau}},\ and\ \bibinfo {author} {\bibfnamefont {S.}~\bibnamefont {Das~Sarma}},\ }\bibfield  {title} {\bibinfo {title} {Conductance spectroscopy of topological superconductor wire junctions},\ }\href {https://link.aps.org/doi/10.1103/PhysRevB.91.214513} {\bibfield  {journal} {\bibinfo  {journal} {Physical Review B}\ }\textbf {\bibinfo {volume} {91}},\ \bibinfo {pages} {214513} (\bibinfo {year} {2015})}\BibitemShut {NoStop}%
\bibitem [{\citenamefont {Rosdahl}\ \emph {et~al.}(2018)\citenamefont {Rosdahl}, \citenamefont {Vuik}, \citenamefont {Kjaergaard},\ and\ \citenamefont {Akhmerov}}]{rosdahl2018andreev}%
  \BibitemOpen
  \bibfield  {author} {\bibinfo {author} {\bibfnamefont {T.~{\"O}.}\ \bibnamefont {Rosdahl}}, \bibinfo {author} {\bibfnamefont {A.}~\bibnamefont {Vuik}}, \bibinfo {author} {\bibfnamefont {M.}~\bibnamefont {Kjaergaard}},\ and\ \bibinfo {author} {\bibfnamefont {A.~R.}\ \bibnamefont {Akhmerov}},\ }\bibfield  {title} {\bibinfo {title} {Andreev rectifier: {{A}} nonlocal conductance signature of topological phase transitions},\ }\href {https://link.aps.org/doi/10.1103/PhysRevB.97.045421} {\bibfield  {journal} {\bibinfo  {journal} {Physical Review B}\ }\textbf {\bibinfo {volume} {97}},\ \bibinfo {pages} {045421} (\bibinfo {year} {2018})}\BibitemShut {NoStop}%
\bibitem [{\citenamefont {Fulga}\ \emph {et~al.}(2011)\citenamefont {Fulga}, \citenamefont {Hassler}, \citenamefont {Akhmerov},\ and\ \citenamefont {Beenakker}}]{fulga2011scattering}%
  \BibitemOpen
  \bibfield  {author} {\bibinfo {author} {\bibfnamefont {I.~C.}\ \bibnamefont {Fulga}}, \bibinfo {author} {\bibfnamefont {F.}~\bibnamefont {Hassler}}, \bibinfo {author} {\bibfnamefont {A.~R.}\ \bibnamefont {Akhmerov}},\ and\ \bibinfo {author} {\bibfnamefont {C.~W.~J.}\ \bibnamefont {Beenakker}},\ }\bibfield  {title} {\bibinfo {title} {Scattering formula for the topological quantum number of a disordered multimode wire},\ }\href {https://link.aps.org/doi/10.1103/PhysRevB.83.155429} {\bibfield  {journal} {\bibinfo  {journal} {Physical Review B}\ }\textbf {\bibinfo {volume} {83}},\ \bibinfo {pages} {155429} (\bibinfo {year} {2011})}\BibitemShut {NoStop}%
\bibitem [{\citenamefont {Das~Sarma}\ \emph {et~al.}(2016)\citenamefont {Das~Sarma}, \citenamefont {Nag},\ and\ \citenamefont {Sau}}]{dassarma2016how}%
  \BibitemOpen
  \bibfield  {author} {\bibinfo {author} {\bibfnamefont {S.}~\bibnamefont {Das~Sarma}}, \bibinfo {author} {\bibfnamefont {A.}~\bibnamefont {Nag}},\ and\ \bibinfo {author} {\bibfnamefont {J.~D.}\ \bibnamefont {Sau}},\ }\bibfield  {title} {\bibinfo {title} {How to infer non-{{Abelian}} statistics and topological visibility from tunneling conductance properties of realistic {{Majorana}} nanowires},\ }\href {https://link.aps.org/doi/10.1103/PhysRevB.94.035143} {\bibfield  {journal} {\bibinfo  {journal} {Physical Review B}\ }\textbf {\bibinfo {volume} {94}},\ \bibinfo {pages} {035143} (\bibinfo {year} {2016})}\BibitemShut {NoStop}%
\bibitem [{\citenamefont {Freedman}\ \emph {et~al.}(2003)\citenamefont {Freedman}, \citenamefont {Kitaev}, \citenamefont {Larsen},\ and\ \citenamefont {Wang}}]{freedman2003topological}%
  \BibitemOpen
  \bibfield  {author} {\bibinfo {author} {\bibfnamefont {M.}~\bibnamefont {Freedman}}, \bibinfo {author} {\bibfnamefont {A.}~\bibnamefont {Kitaev}}, \bibinfo {author} {\bibfnamefont {M.}~\bibnamefont {Larsen}},\ and\ \bibinfo {author} {\bibfnamefont {Z.}~\bibnamefont {Wang}},\ }\bibfield  {title} {\bibinfo {title} {Topological quantum computation},\ }\href {https://www.ams.org/home/page/} {\bibfield  {journal} {\bibinfo  {journal} {Bulletin of the American Mathematical Society}\ }\textbf {\bibinfo {volume} {40}},\ \bibinfo {pages} {31} (\bibinfo {year} {2003})}\BibitemShut {NoStop}%
\bibitem [{\citenamefont {Nayak}\ \emph {et~al.}(2008)\citenamefont {Nayak}, \citenamefont {Simon}, \citenamefont {Stern}, \citenamefont {Freedman},\ and\ \citenamefont {Das~Sarma}}]{nayak2008nonabelian}%
  \BibitemOpen
  \bibfield  {author} {\bibinfo {author} {\bibfnamefont {C.}~\bibnamefont {Nayak}}, \bibinfo {author} {\bibfnamefont {S.~H.}\ \bibnamefont {Simon}}, \bibinfo {author} {\bibfnamefont {A.}~\bibnamefont {Stern}}, \bibinfo {author} {\bibfnamefont {M.}~\bibnamefont {Freedman}},\ and\ \bibinfo {author} {\bibfnamefont {S.}~\bibnamefont {Das~Sarma}},\ }\bibfield  {title} {\bibinfo {title} {Non-{{Abelian}} anyons and topological quantum computation},\ }\href {https://link.aps.org/doi/10.1103/RevModPhys.80.1083} {\bibfield  {journal} {\bibinfo  {journal} {Reviews of Modern Physics}\ }\textbf {\bibinfo {volume} {80}},\ \bibinfo {pages} {1083} (\bibinfo {year} {2008})}\BibitemShut {NoStop}%
\bibitem [{\citenamefont {Alicea}\ \emph {et~al.}(2011)\citenamefont {Alicea}, \citenamefont {Oreg}, \citenamefont {Refael}, \citenamefont {{von Oppen}},\ and\ \citenamefont {Fisher}}]{alicea2011nonabelian}%
  \BibitemOpen
  \bibfield  {author} {\bibinfo {author} {\bibfnamefont {J.}~\bibnamefont {Alicea}}, \bibinfo {author} {\bibfnamefont {Y.}~\bibnamefont {Oreg}}, \bibinfo {author} {\bibfnamefont {G.}~\bibnamefont {Refael}}, \bibinfo {author} {\bibfnamefont {F.}~\bibnamefont {{von Oppen}}},\ and\ \bibinfo {author} {\bibfnamefont {M.~P.~A.}\ \bibnamefont {Fisher}},\ }\bibfield  {title} {\bibinfo {title} {Non-{{Abelian}} statistics and topological quantum information processing in {{1D}} wire networks},\ }\href {https://www.nature.com/articles/nphys1915} {\bibfield  {journal} {\bibinfo  {journal} {Nature Physics}\ }\textbf {\bibinfo {volume} {7}},\ \bibinfo {pages} {412} (\bibinfo {year} {2011})}\BibitemShut {NoStop}%
\bibitem [{\citenamefont {Sarma}\ \emph {et~al.}(2015)\citenamefont {Sarma}, \citenamefont {Freedman},\ and\ \citenamefont {Nayak}}]{sarma2015majorana}%
  \BibitemOpen
  \bibfield  {author} {\bibinfo {author} {\bibfnamefont {S.~D.}\ \bibnamefont {Sarma}}, \bibinfo {author} {\bibfnamefont {M.}~\bibnamefont {Freedman}},\ and\ \bibinfo {author} {\bibfnamefont {C.}~\bibnamefont {Nayak}},\ }\bibfield  {title} {\bibinfo {title} {Majorana zero modes and topological quantum computation},\ }\href {https://www.nature.com/articles/npjqi20151} {\bibfield  {journal} {\bibinfo  {journal} {npj Quantum Information}\ }\textbf {\bibinfo {volume} {1}},\ \bibinfo {pages} {15001} (\bibinfo {year} {2015})}\BibitemShut {NoStop}%
\bibitem [{\citenamefont {Pan}\ \emph {et~al.}(2020)\citenamefont {Pan}, \citenamefont {Cole}, \citenamefont {Sau},\ and\ \citenamefont {Das~Sarma}}]{pan2020generic}%
  \BibitemOpen
  \bibfield  {author} {\bibinfo {author} {\bibfnamefont {H.}~\bibnamefont {Pan}}, \bibinfo {author} {\bibfnamefont {W.~S.}\ \bibnamefont {Cole}}, \bibinfo {author} {\bibfnamefont {J.~D.}\ \bibnamefont {Sau}},\ and\ \bibinfo {author} {\bibfnamefont {S.}~\bibnamefont {Das~Sarma}},\ }\bibfield  {title} {\bibinfo {title} {Generic quantized zero-bias conductance peaks in superconductor-semiconductor hybrid structures},\ }\href {https://link.aps.org/doi/10.1103/PhysRevB.101.024506} {\bibfield  {journal} {\bibinfo  {journal} {Physical Review B}\ }\textbf {\bibinfo {volume} {101}},\ \bibinfo {pages} {024506} (\bibinfo {year} {2020})}\BibitemShut {NoStop}%
\bibitem [{\citenamefont {Das~Sarma}\ \emph {et~al.}(2012)\citenamefont {Das~Sarma}, \citenamefont {Sau},\ and\ \citenamefont {Stanescu}}]{dassarma2012splitting}%
  \BibitemOpen
  \bibfield  {author} {\bibinfo {author} {\bibfnamefont {S.}~\bibnamefont {Das~Sarma}}, \bibinfo {author} {\bibfnamefont {J.~D.}\ \bibnamefont {Sau}},\ and\ \bibinfo {author} {\bibfnamefont {T.~D.}\ \bibnamefont {Stanescu}},\ }\bibfield  {title} {\bibinfo {title} {Splitting of the zero-bias conductance peak as smoking gun evidence for the existence of the {{Majorana}} mode in a superconductor-semiconductor nanowire},\ }\href {https://link.aps.org/doi/10.1103/PhysRevB.86.220506} {\bibfield  {journal} {\bibinfo  {journal} {Physical Review B}\ }\textbf {\bibinfo {volume} {86}},\ \bibinfo {pages} {220506} (\bibinfo {year} {2012})}\BibitemShut {NoStop}%
\end{thebibliography}%

\clearpage
\appendix
\onecolumngrid

\renewcommand{\thefigure}{\thesection\arabic{figure*}}
\counterwithin{figure}{section}

\section{More simulations of the experimental data}\label{app:MSFT}
\begin{figure*}[ht]
    \centering
    \includegraphics[width=6.8in]{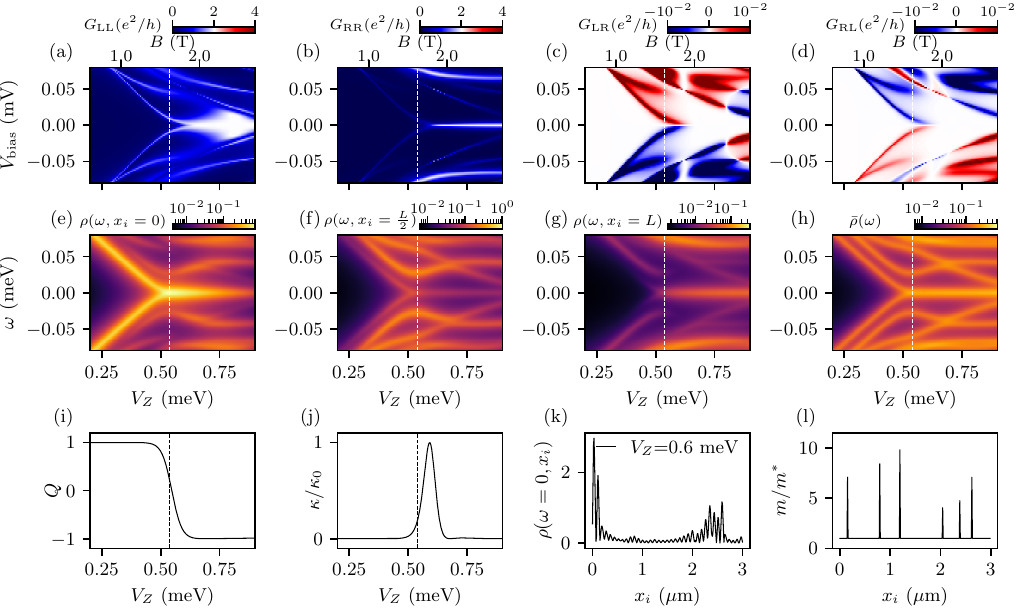}
    \caption{Simulation of the positive result Fig. 11 in Ref.~\cite{microsoftquantum2023inasal} (also in Fig.~\ref{fig:exp_positive}) with $n=6$, $k=9$, and $L=3~\mu$m. 
    Same captions as Fig.~\ref{fig:thy_positive}.
    }
    \label{fig:649}
\end{figure*}

\begin{figure*}[ht]
    \centering
    \includegraphics[width=6.8in]{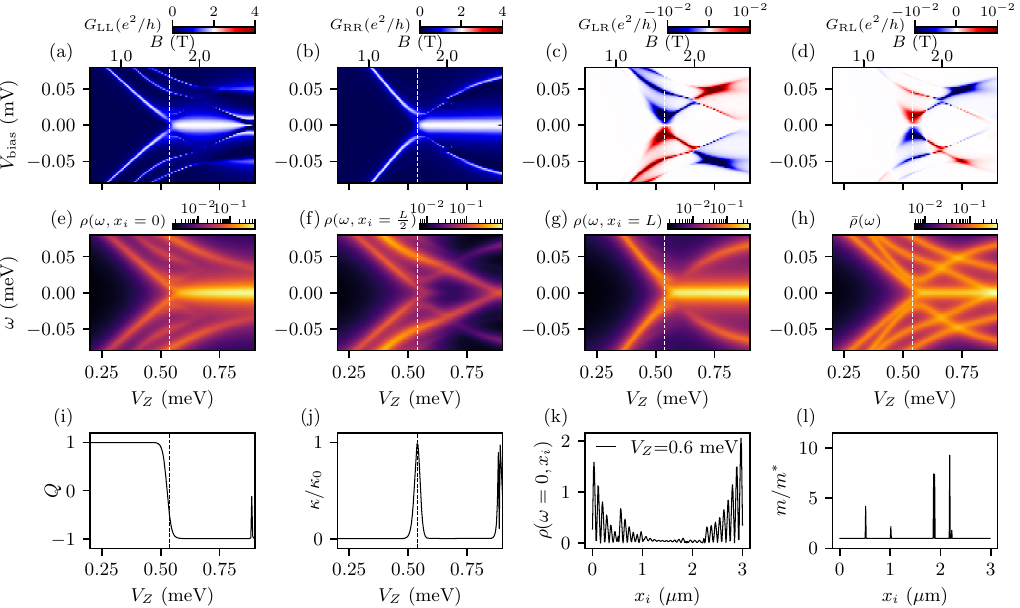}
    \caption{Simulation of the positive result Fig. 11 in Ref.~\cite{microsoftquantum2023inasal} (also in Fig.~\ref{fig:exp_positive}) with $n=6$, $k=9$, and $L=3~\mu$m. 
    Same captions as Fig.~\ref{fig:thy_positive}.
    }
    \label{fig:658}
\end{figure*}

\begin{figure*}[ht]
    \centering
    \includegraphics[width=6.8in]{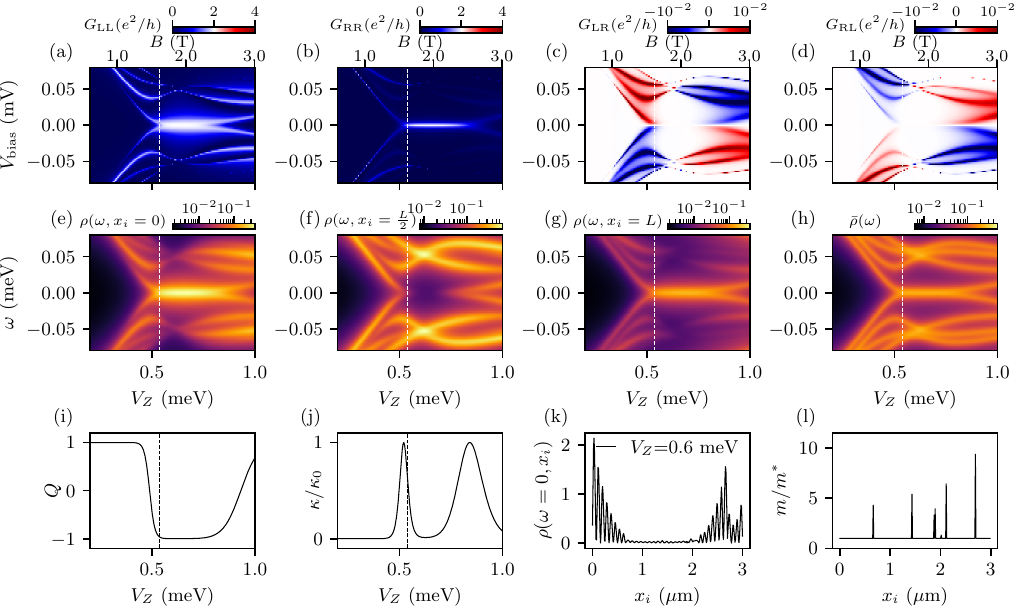}
    \caption{Simulation of the positive result Fig. 11 in Ref.~\cite{microsoftquantum2023inasal} (also in Fig.~\ref{fig:exp_positive}) with $n=7$, $k=9$, and $L=3~\mu$m. 
    Same captions as Fig.~\ref{fig:thy_positive}.
    }
    \label{fig:703}
\end{figure*}

\begin{figure*}[ht]
    \centering
    \includegraphics[width=6.8in]{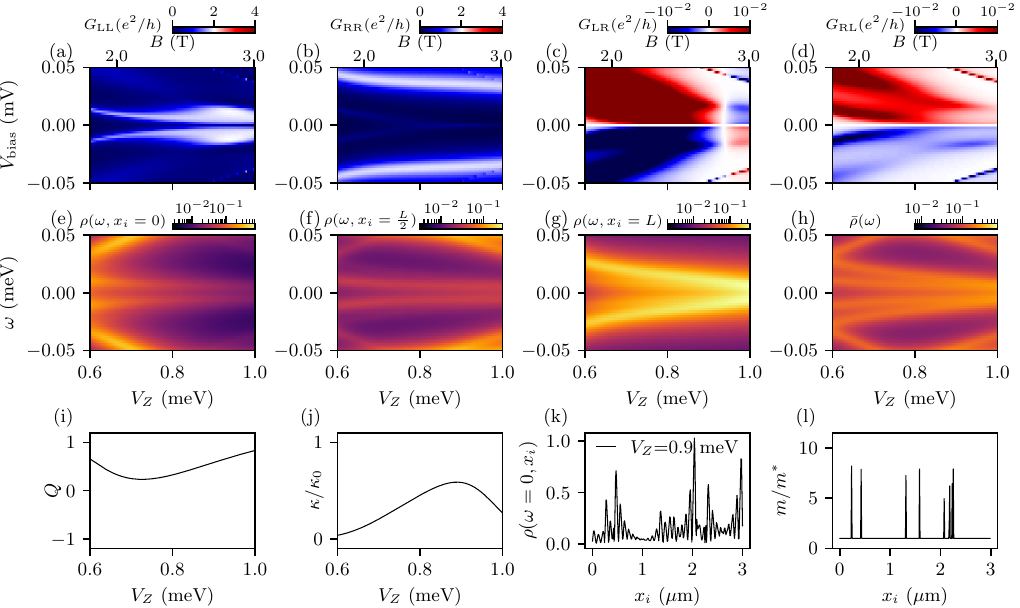}
    \caption{Simulation of the negative result Fig. 20 in Ref.~\cite{microsoftquantum2023inasal} (also in Fig.~\ref{fig:exp_negative}) with $n=9$, $k=9$, and $L=3~\mu$m.
    Same captions as Fig.~\ref{fig:thy_negative}.
    }
    \label{fig:925}
\end{figure*}

\begin{figure*}[ht]
    \centering
    \includegraphics[width=6.8in]{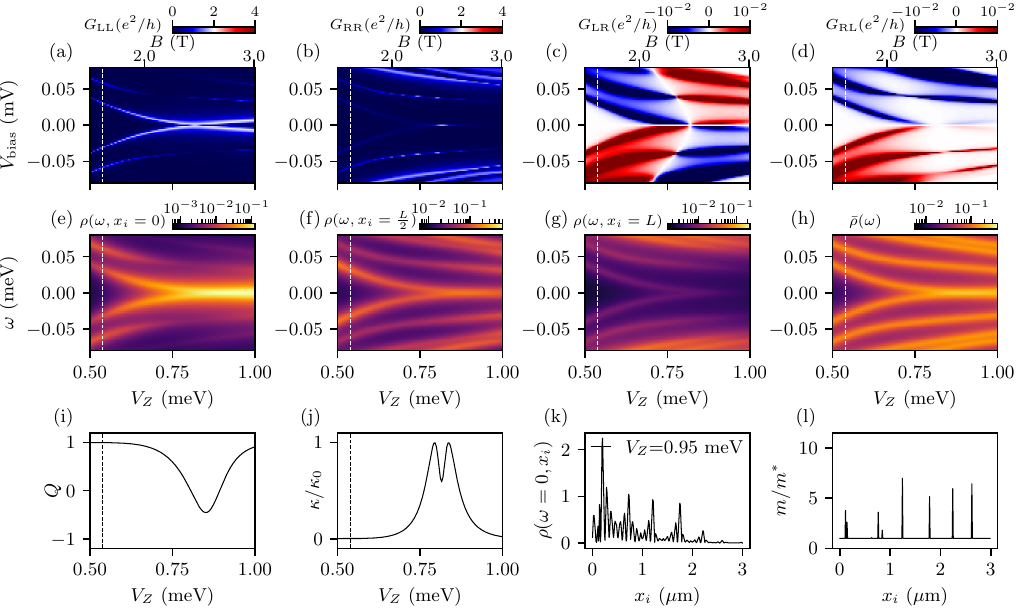}
    \caption{Simulation of the negative result Fig. 20 in Ref.~\cite{microsoftquantum2023inasal} (also in Fig.~\ref{fig:exp_negative}) with $n=9$, $k=9$, and $L=3~\mu$m.
    Same captions as Fig.~\ref{fig:thy_negative}.
    }
    \label{fig:941}
\end{figure*}

\begin{figure*}[ht]
    \centering
    \includegraphics[width=6.8in]{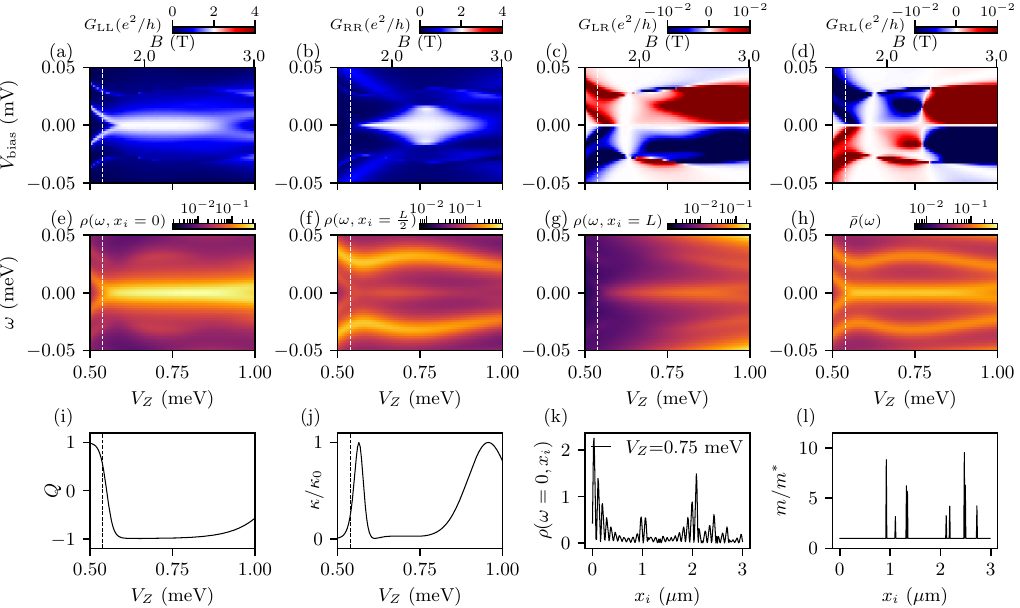}
    \caption{Simulation of the negative result Fig. 20 in Ref.~\cite{microsoftquantum2023inasal} (also in Fig.~\ref{fig:exp_negative}) with $n=9$, $k=9$, and $L=3~\mu$m.
    Same captions as Fig.~\ref{fig:thy_negative}.
    }
    \label{fig:971}
\end{figure*}

\begin{figure*}[ht]
    \centering
    \includegraphics[width=6.8in]{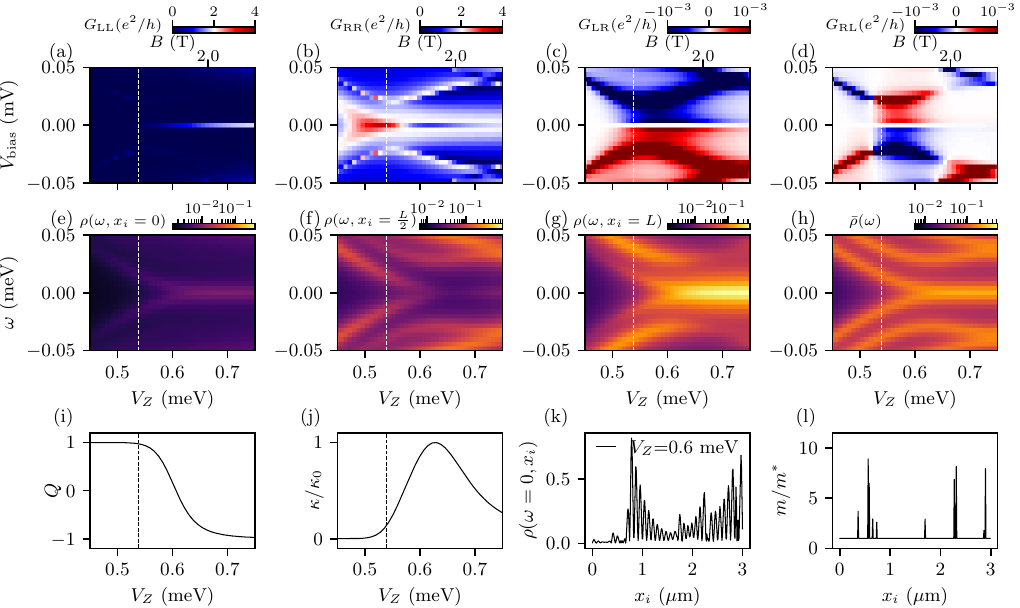}
    \caption{Simulation of the negative result Fig. 20 in Ref.~\cite{microsoftquantum2023inasal} (also in Fig.~\ref{fig:exp_negative}) with $n=10$, $k=9$, and $L=3~\mu$m.
    Same captions as Fig.~\ref{fig:thy_negative}.
    }
    \label{fig:1001}
\end{figure*}

\begin{figure*}[ht]
    \centering
    \includegraphics[width=6.8in]{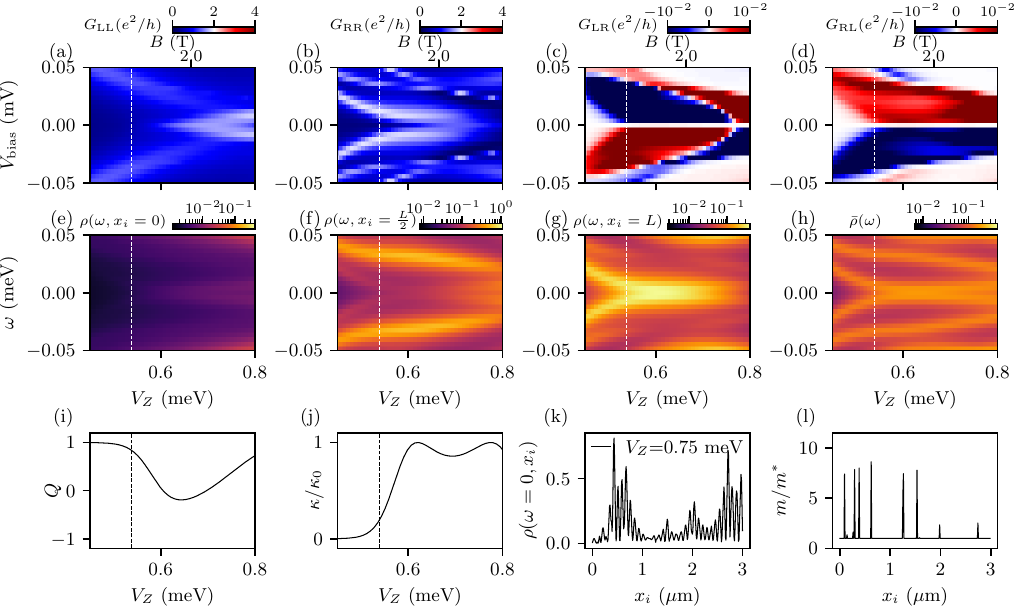}
    \caption{Simulation of the negative result Fig. 20 in Ref.~\cite{microsoftquantum2023inasal} (also in Fig.~\ref{fig:exp_negative}) with $n=12$, $k=9$, and $L=3~\mu$m.
    Same captions as Fig.~\ref{fig:thy_negative}.
    }
    \label{fig:1201}
\end{figure*}

In this section, we present more simulations of the positive results (Fig.~\ref{fig:exp_positive}) from Fig.~\ref{fig:649} to Fig.~\ref{fig:703}, and negative results (Fig.~\ref{fig:exp_negative}) from Fig.~\ref{fig:925} to Fig.~\ref{fig:1201}.

\clearpage
\section{More results showing the dependence of $k$}\label{app:dependence_of_k}
\begin{figure*}[ht]
    \centering
    \includegraphics[width=6.8in]{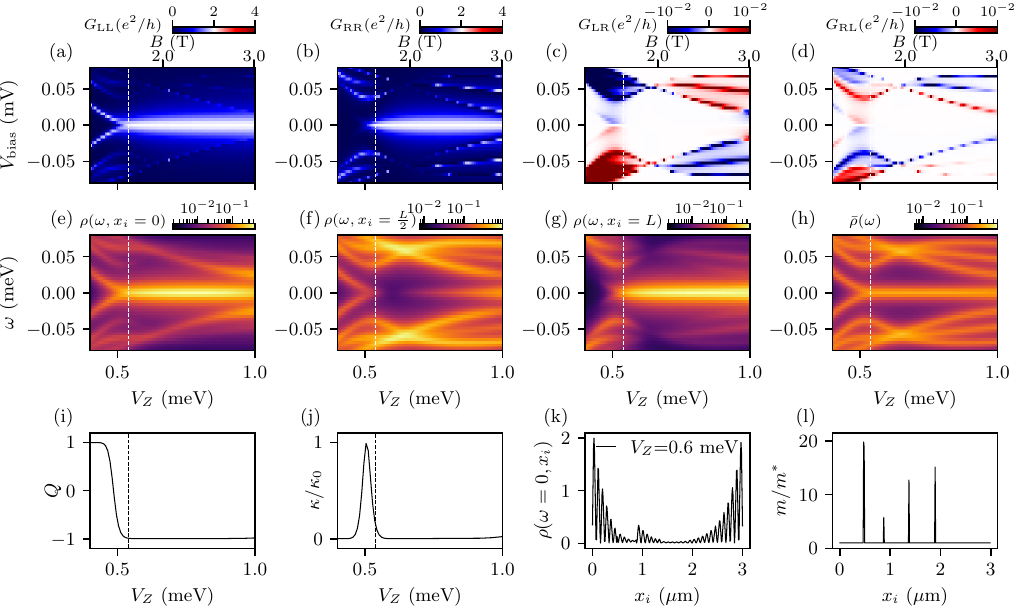}
    \caption{$n=5$, $k=20$, and $L=3~\mu$m. 
    Same captions as Fig.~\ref{fig:thy_positive}.
    }
    \label{fig:2015}
\end{figure*}

\begin{figure*}[ht]
    \centering
    \includegraphics[width=6.8in]{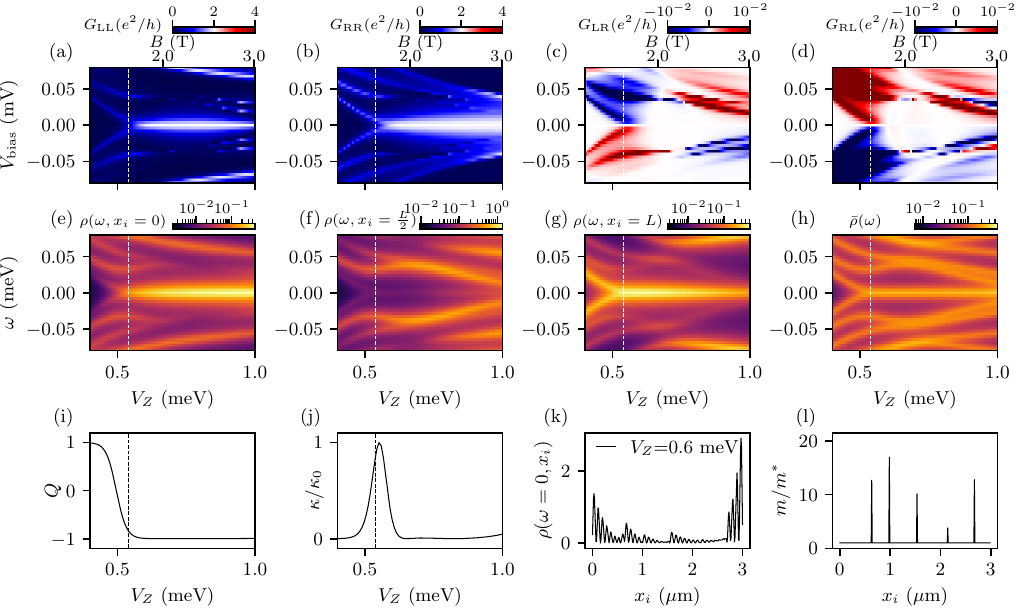}
    \caption{$n=5$, $k=20$, and $L=3~\mu$m. 
    Same captions as Fig.~\ref{fig:thy_positive}.
    }
    \label{fig:2043}
\end{figure*}

\begin{figure*}[ht]
    \centering
    \includegraphics[width=6.8in]{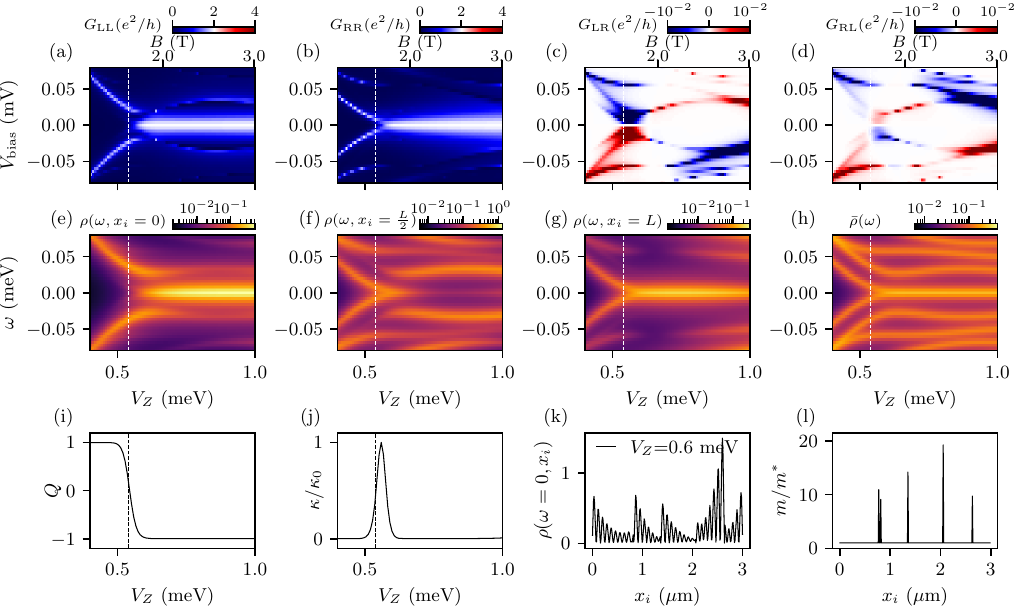}
    \caption{$n=5$, $k=20$, and $L=3~\mu$m. 
    Same captions as Fig.~\ref{fig:thy_positive}.
    }
    \label{fig:2047}
\end{figure*}

\begin{figure*}[ht]
    \centering
    \includegraphics[width=6.8in]{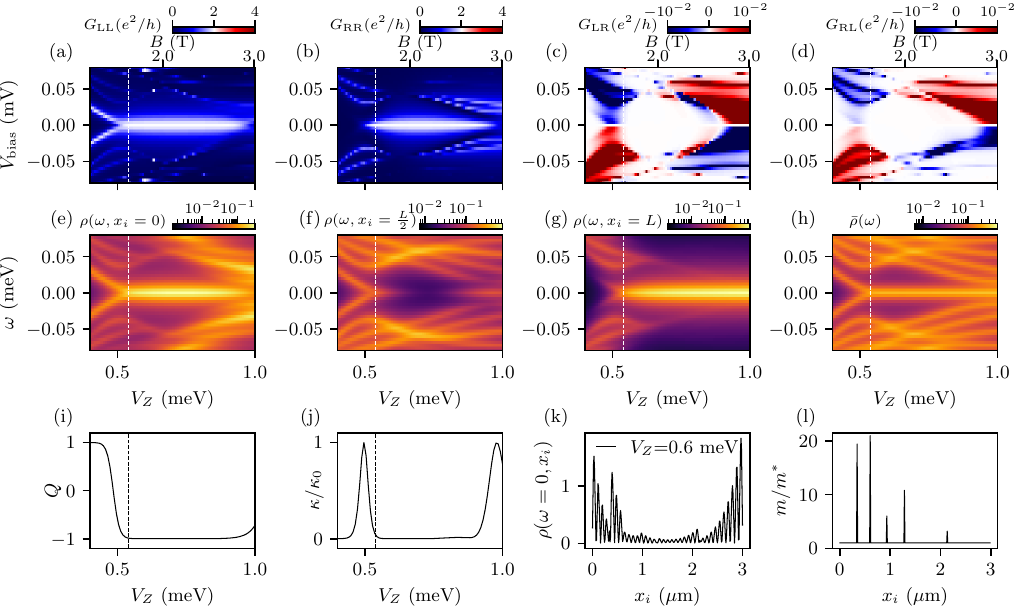}
    \caption{$n=5$, $k=20$, and $L=3~\mu$m. 
    Same captions as Fig.~\ref{fig:thy_positive}.
    }
    \label{fig:2063}
\end{figure*}

\begin{figure*}[ht]
    \centering
    \includegraphics[width=6.8in]{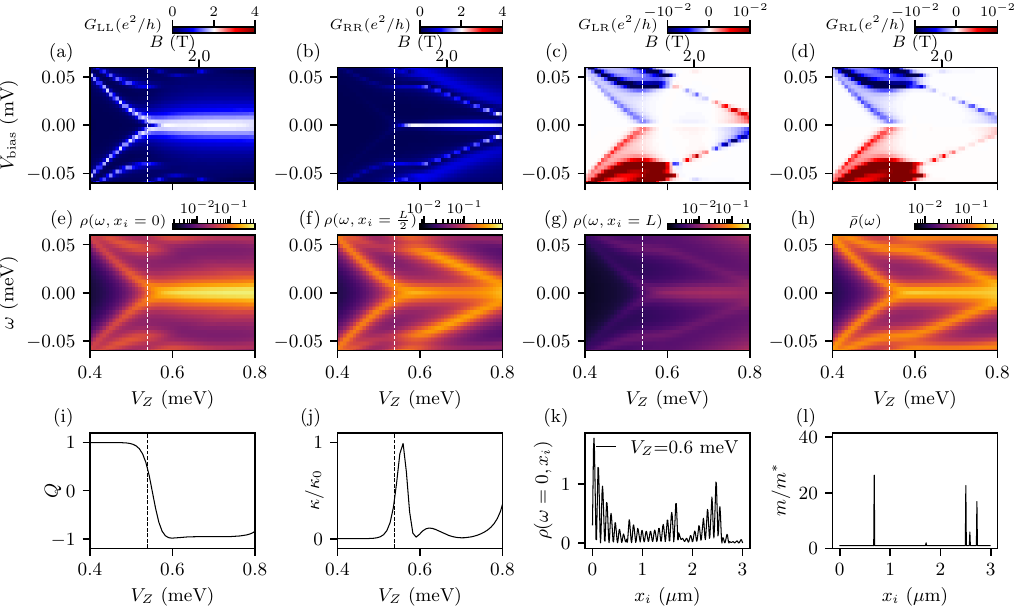}
    \caption{$n=5$, $k=40$, and $L=3~\mu$m. 
    Same captions as Fig.~\ref{fig:thy_positive}.
    }
    \label{fig:4001}
\end{figure*}

\begin{figure*}[ht]
    \centering
    \includegraphics[width=6.8in]{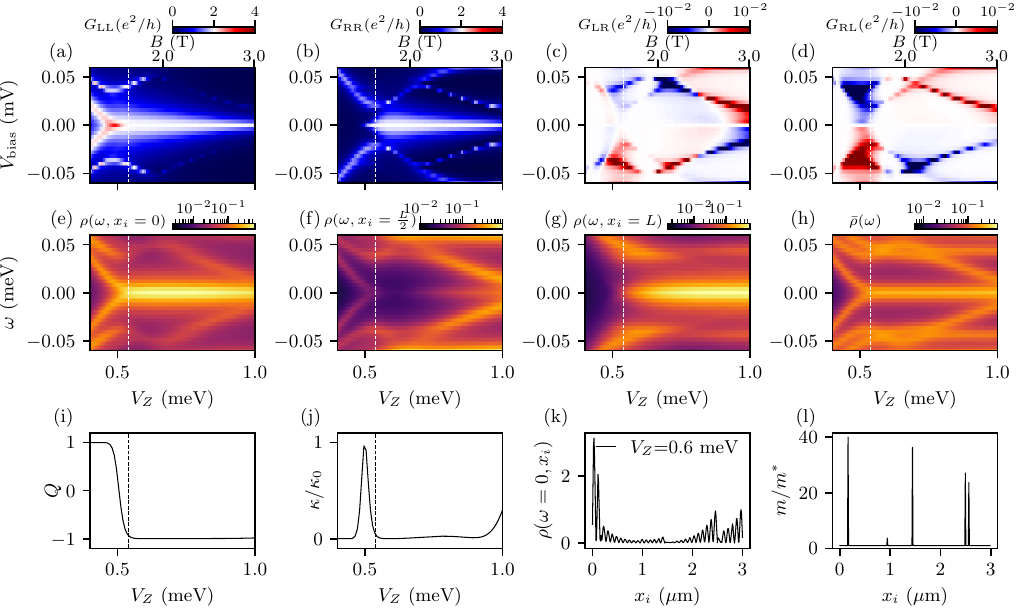}
    \caption{$n=5$, $k=40$, and $L=3~\mu$m. 
    Same captions as Fig.~\ref{fig:thy_positive}.
    }
    \label{fig:4005}
\end{figure*}

\begin{figure*}[ht]
    \centering
    \includegraphics[width=6.8in]{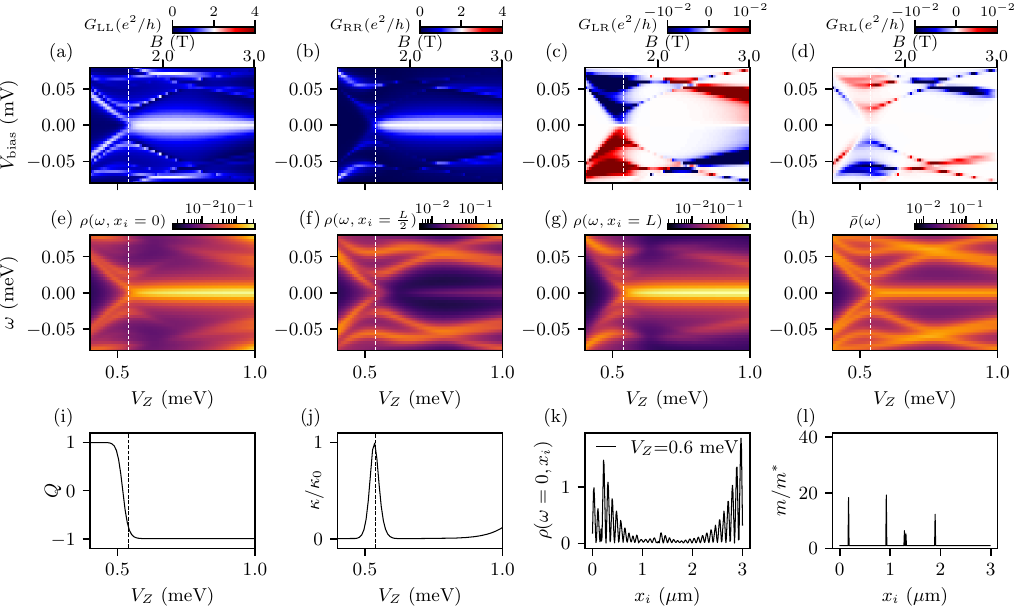}
    \caption{$n=5$, $k=40$, and $L=3~\mu$m. 
    Same captions as Fig.~\ref{fig:thy_positive}.
    }
    \label{fig:4017}
\end{figure*}

\begin{figure*}[ht]
    \centering
    \includegraphics[width=6.8in]{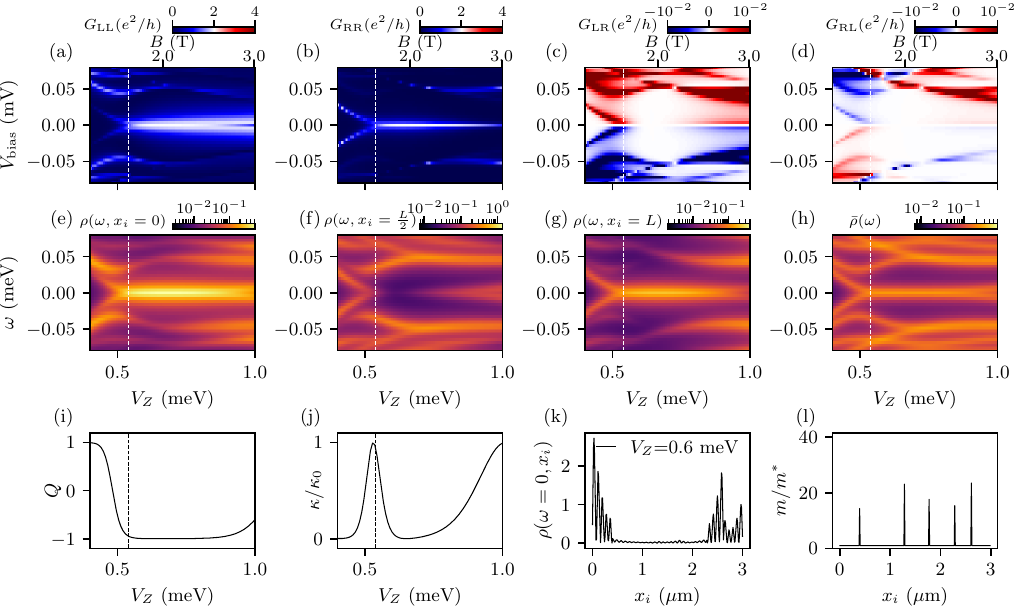}
    \caption{$n=5$, $k=40$, and $L=3~\mu$m. 
    Same captions as Fig.~\ref{fig:thy_positive}.
    }
    \label{fig:4018}
\end{figure*}

\begin{figure*}[ht]
    \centering
    \includegraphics[width=6.8in]{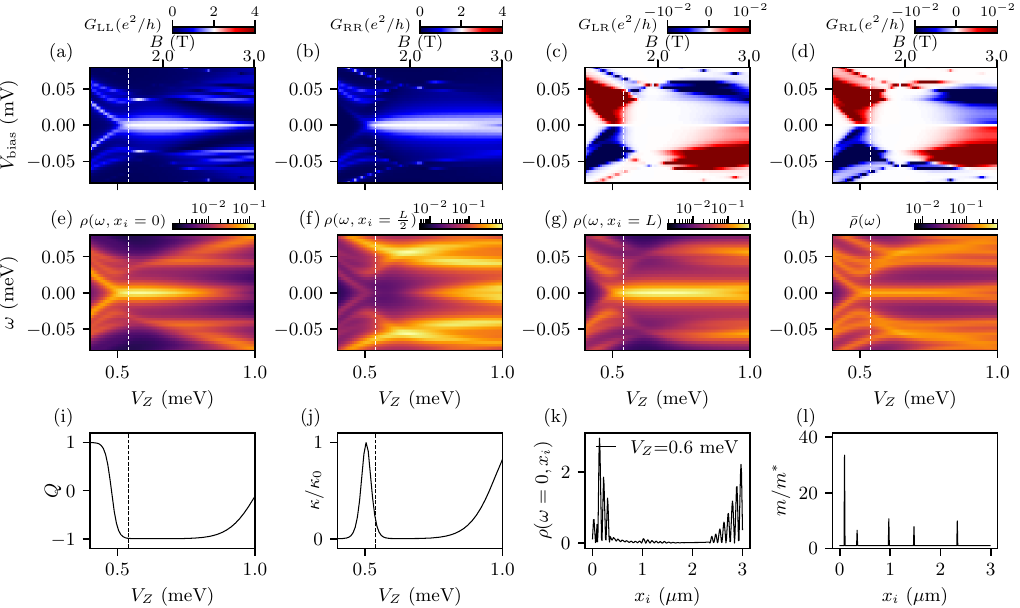}
    \caption{$n=5$, $k=40$, and $L=3~\mu$m. 
    Same captions as Fig.~\ref{fig:thy_positive}.
    }
    \label{fig:4066}
\end{figure*}

\begin{figure*}[ht]
    \centering
    \includegraphics[width=6.8in]{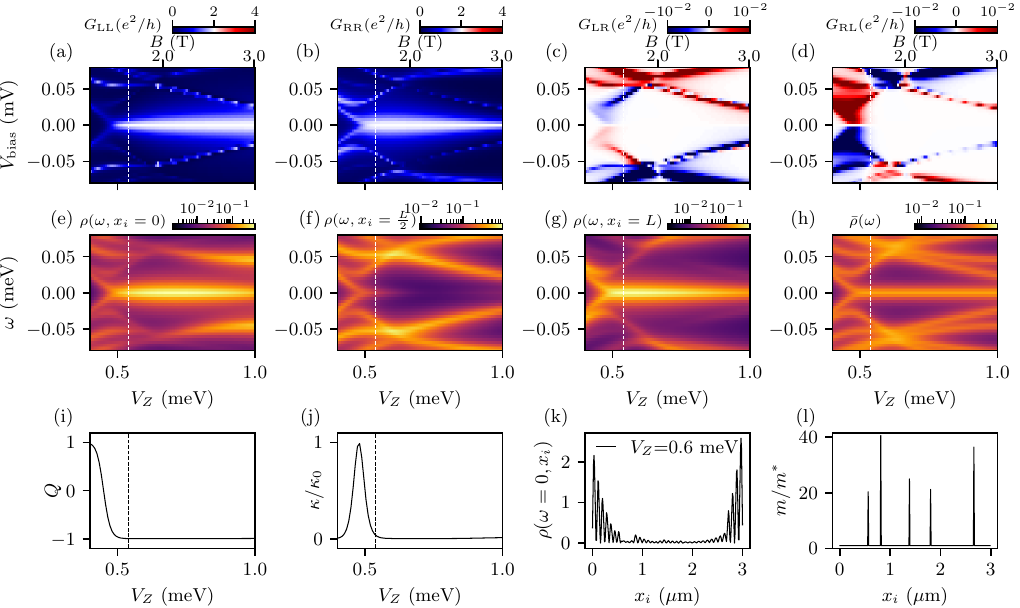}
    \caption{$n=5$, $k=40$, and $L=3~\mu$m. 
    Same captions as Fig.~\ref{fig:thy_positive}.
    }
    \label{fig:4068}
\end{figure*}

\begin{figure*}[ht]
    \centering
    \includegraphics[width=6.8in]{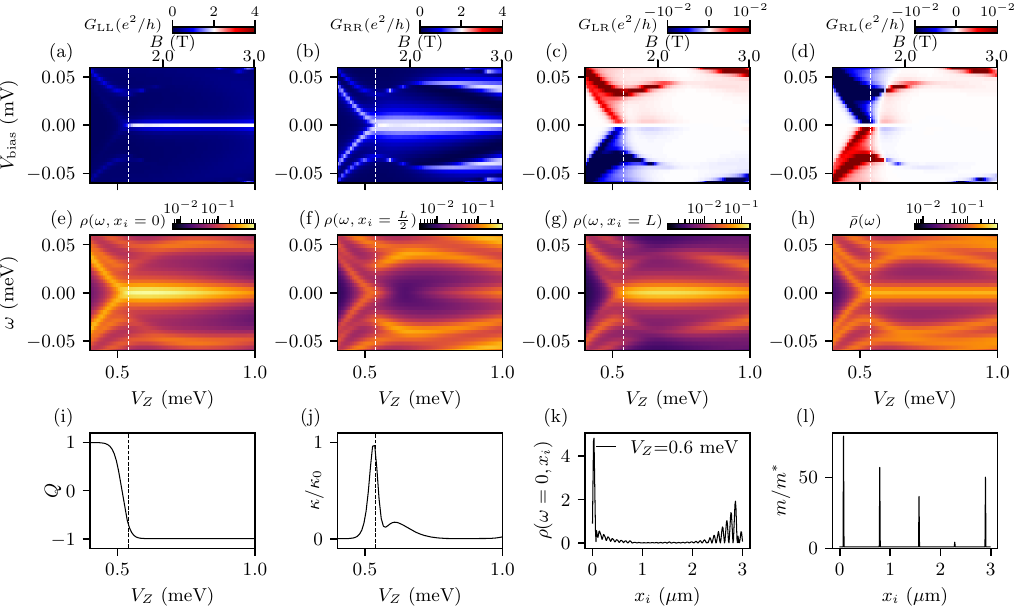}
    \caption{$n=5$, $k=80$, and $L=3~\mu$m. 
    Same captions as Fig.~\ref{fig:thy_positive}.
    }
    \label{fig:8038}
\end{figure*}

\begin{figure}[ht]
    \centering
    \includegraphics[width=6.8in]{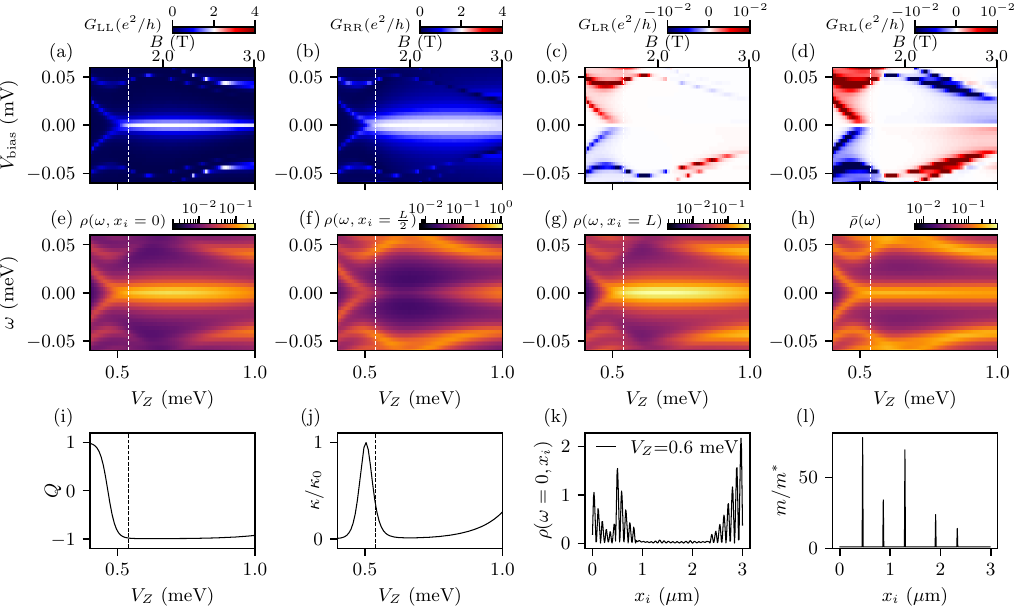}
    \caption{$n=5$, $k=80$, and $L=3~\mu$m. 
    Same captions as Fig.~\ref{fig:thy_positive}.
    }
    \label{fig:8067}
\end{figure}

\begin{figure*}[ht]
    \centering
    \includegraphics[width=6.8in]{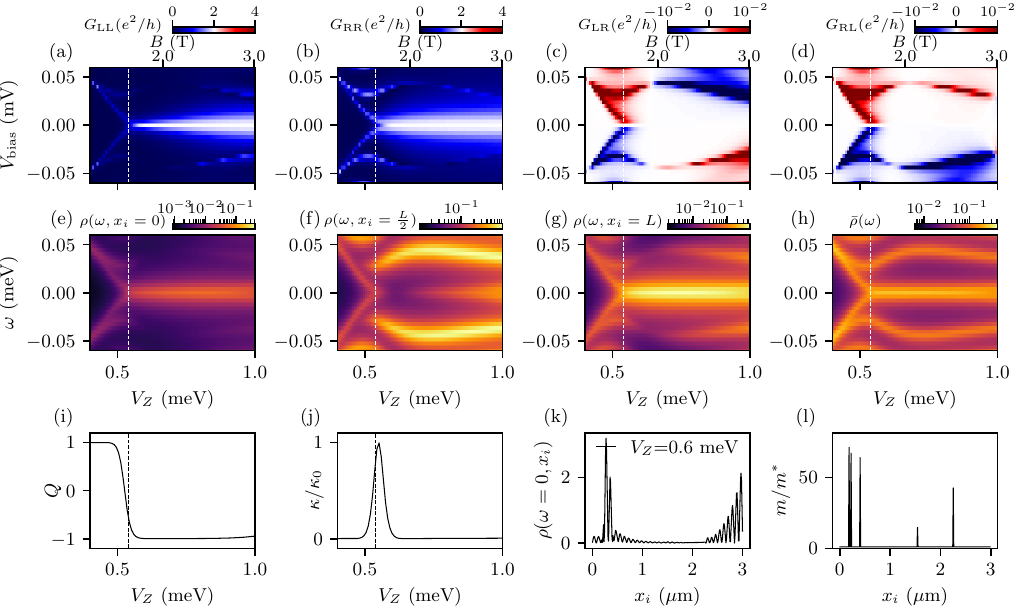}
    \caption{$n=5$, $k=80$, and $L=3~\mu$m. 
    Same captions as Fig.~\ref{fig:thy_positive}.
    }
    \label{fig:8073}
\end{figure*}

\begin{figure*}[ht]
    \centering
    \includegraphics[width=6.8in]{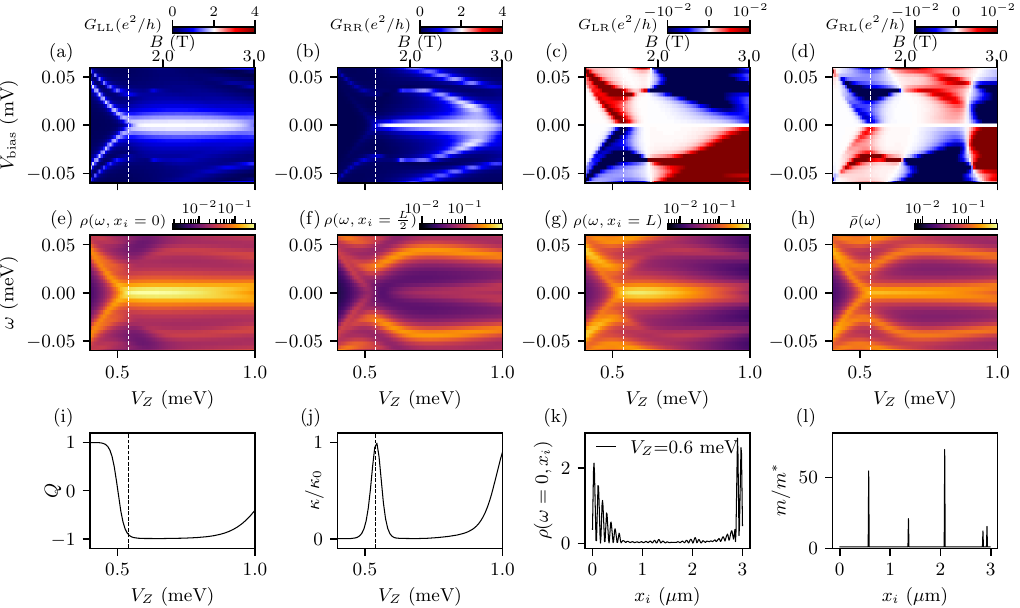}
    \caption{$n=5$, $k=80$, and $L=3~\mu$m. 
    Same captions as Fig.~\ref{fig:thy_positive}.
    }
    \label{fig:8079}
\end{figure*}

\begin{figure*}[ht]
    \centering
    \includegraphics[width=6.8in]{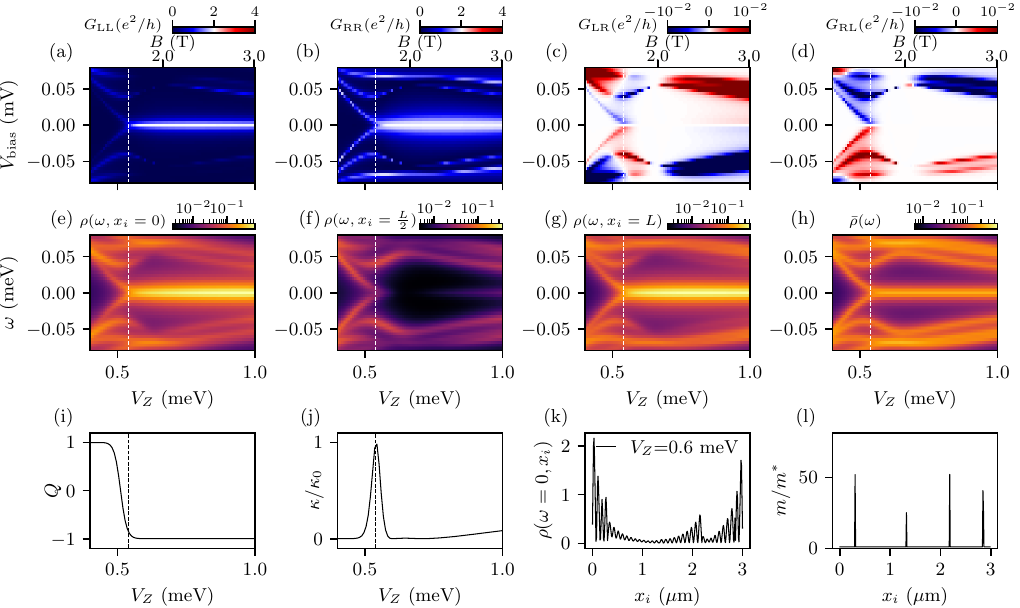}
    \caption{$n=5$, $k=80$, and $L=3~\mu$m. 
    Same captions as Fig.~\ref{fig:thy_positive}.
    }
    \label{fig:8096}
\end{figure*}

\begin{figure*}[ht]
    \centering
    \includegraphics[width=6.8in]{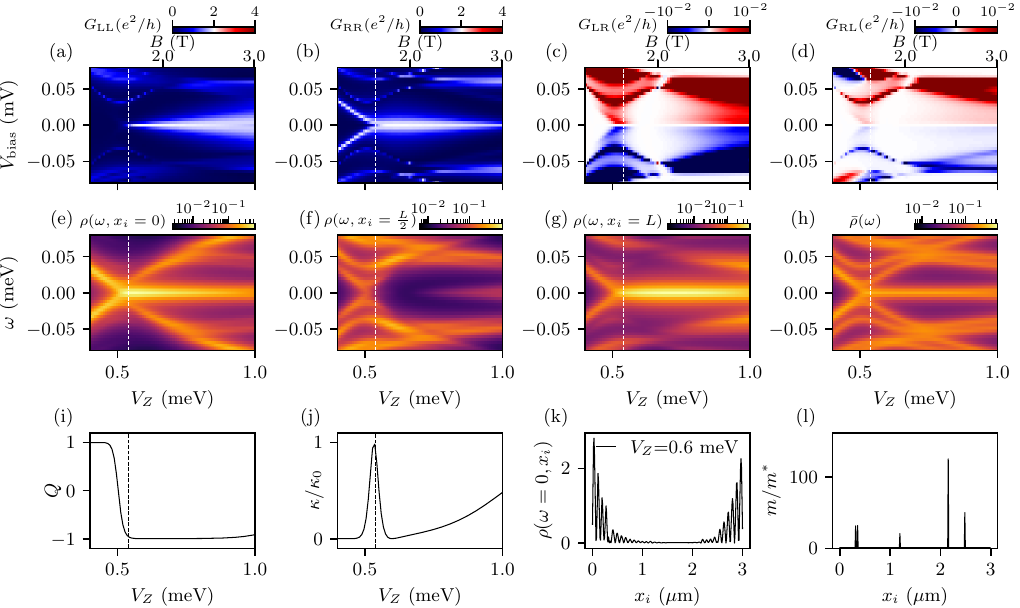}
    \caption{$n=5$, $k=160$, and $L=3~\mu$m. 
    Same captions as Fig.~\ref{fig:thy_positive}.
    }
    \label{fig:16017}
\end{figure*}

\begin{figure*}[ht]
    \centering
    \includegraphics[width=6.8in]{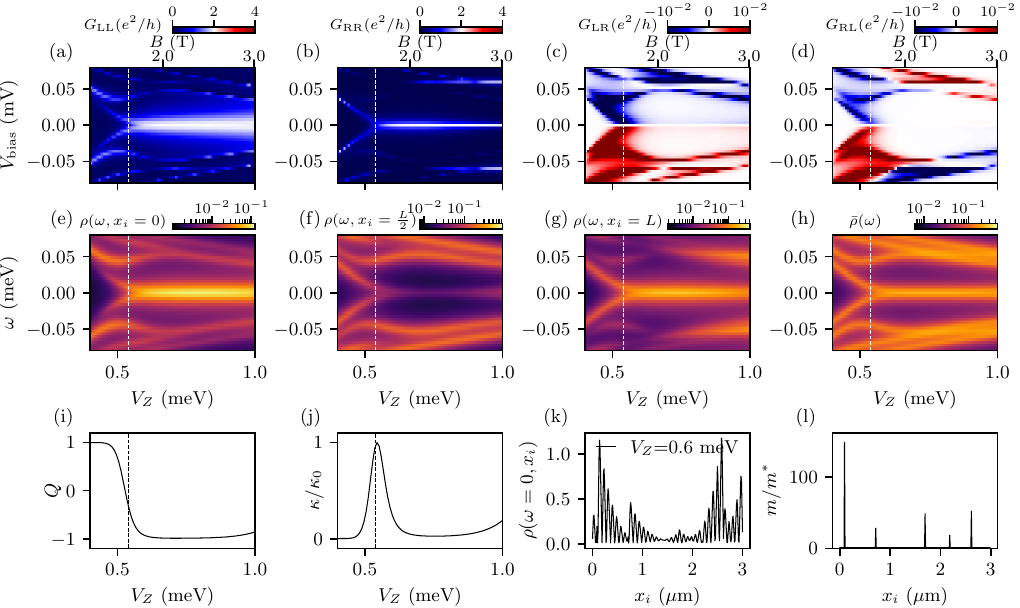}
    \caption{$n=5$, $k=160$, and $L=3~\mu$m. 
    Same captions as Fig.~\ref{fig:thy_positive}.
    }
    \label{fig:16019}
\end{figure*}

\begin{figure*}[ht]
    \centering
    \includegraphics[width=6.8in]{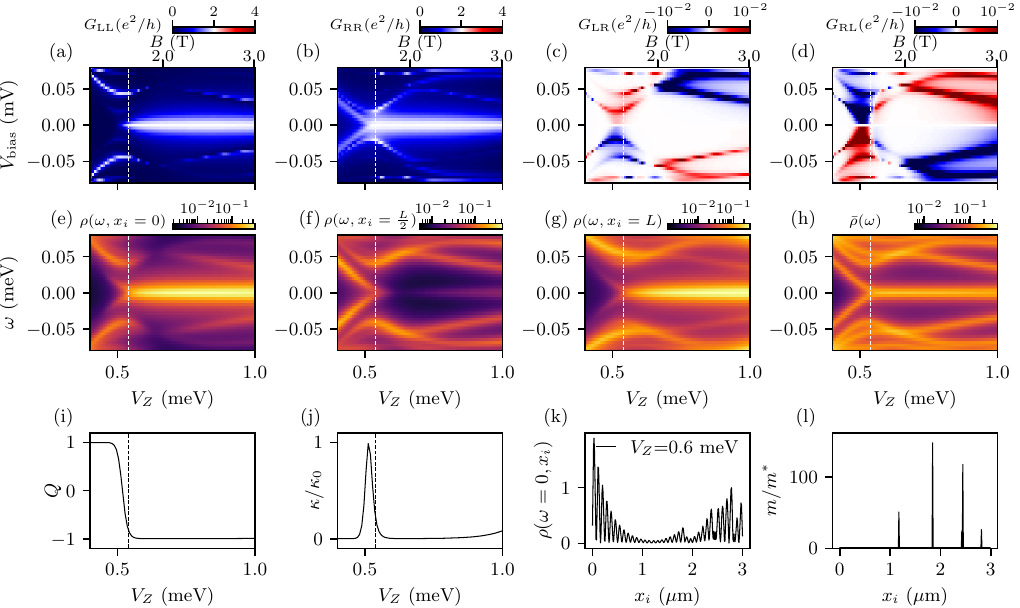}
    \caption{$n=5$, $k=160$, and $L=3~\mu$m. 
    Same captions as Fig.~\ref{fig:thy_positive}.
    }
    \label{fig:16085}
\end{figure*}

In this section, we present more results that resemble the positive results in Fig.~\ref{fig:exp_positive} with an increasing $k$. 
Figure~\ref{fig:2015} to Figure~\ref{fig:2063} show the results for $k=20$.
Figure~\ref{fig:4001} to Figure~\ref{fig:4068} show the results for $k=40$.
Figure~\ref{fig:8038} to Figure~\ref{fig:8096} show the results for $k=80$.
Figure~\ref{fig:16017} to Figure~\ref{fig:16085} show the results for $k=160$.




\clearpage
\section{More results showing the dependence of $L$}\label{app:dependence_of_L}
\begin{figure*}[ht]
    \centering
    \includegraphics[width=6.8in]{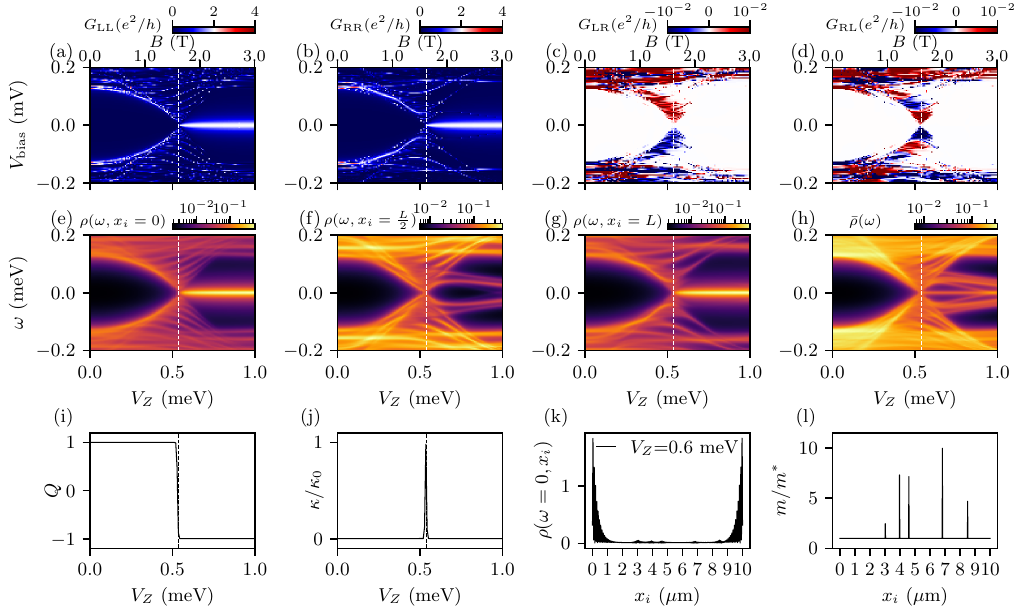}
    \caption{
    Long wire limit $L=10~\mu$m with $n$=5 and $k$=9.
    Same captions as Fig.~\ref{fig:n5L10}.
    }
    \label{fig:512}
\end{figure*}

\begin{figure*}[ht]
    \centering
    \includegraphics[width=6.8in]{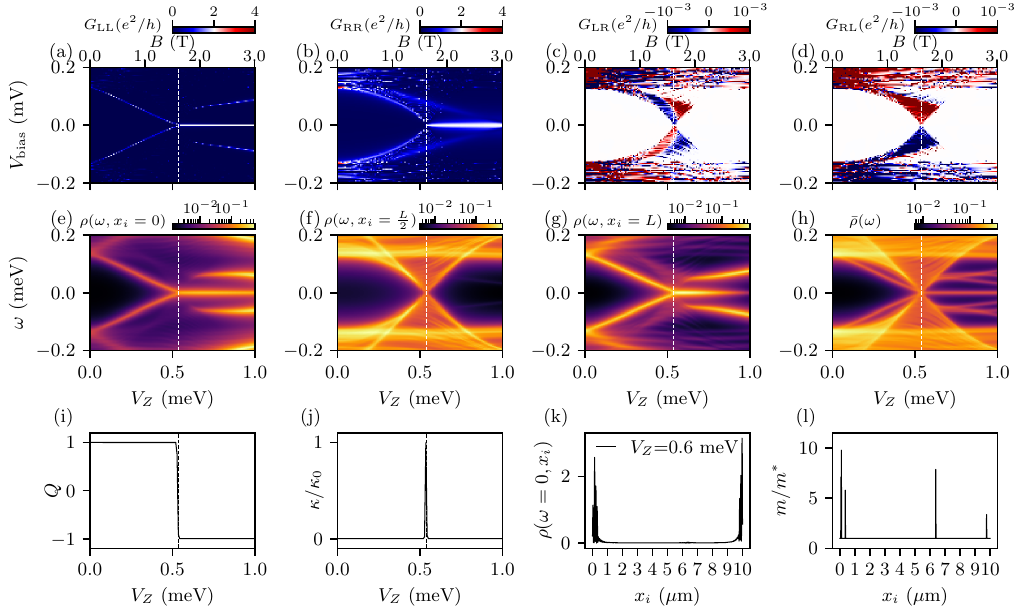}
    \caption{
    Long wire limit $L=10~\mu$m with $n$=5 and $k$=9.
    Same captions as Fig.~\ref{fig:n5L10}.
    }
    \label{fig:584}
\end{figure*}

\begin{figure*}[ht]
    \centering
    \includegraphics[width=6.8in]{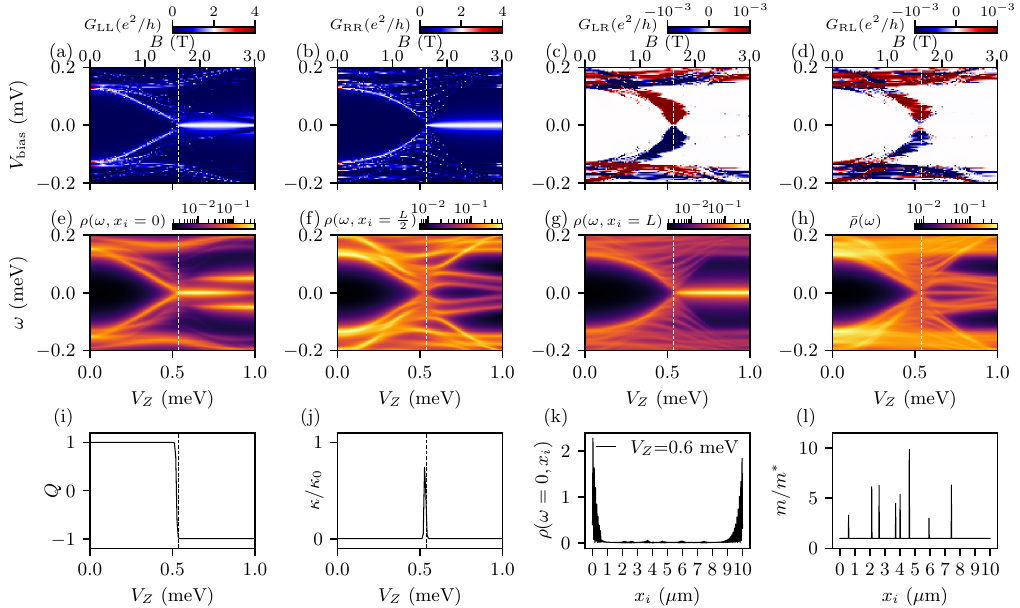}
    \caption{
    Long wire limit $L=10~\mu$m with $n$=9 and $k$=9.
    Same captions as Fig.~\ref{fig:n9L10}.
    }
    \label{fig:922}
\end{figure*}

\begin{figure*}[ht]
    \centering
    \includegraphics[width=6.8in]{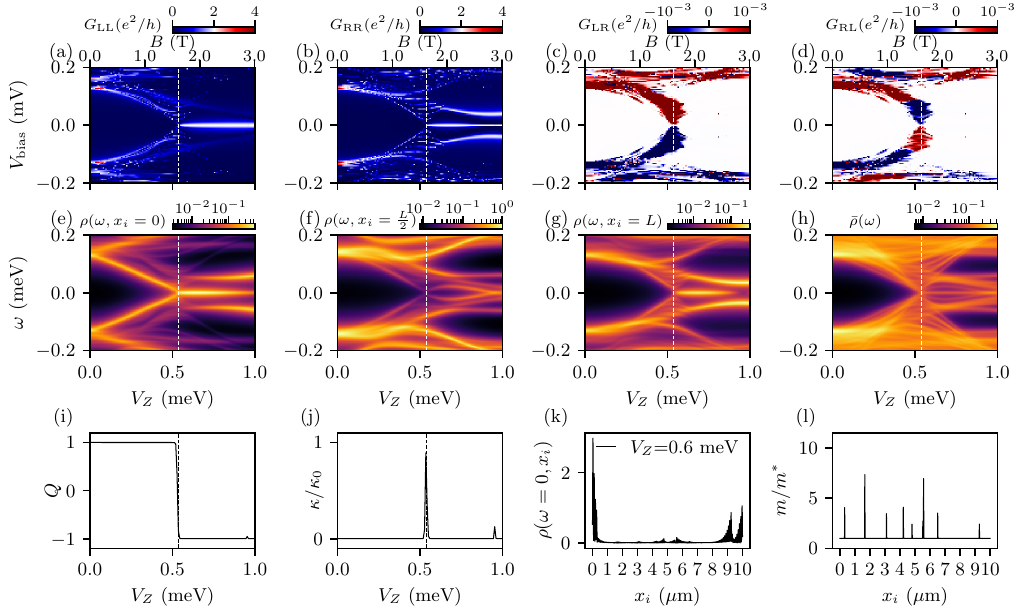}
    \caption{
    Long wire limit $L=10~\mu$m with $n$=9 and $k$=9.
    Same captions as Fig.~\ref{fig:n9L10}.
    }
    \label{fig:933}
\end{figure*}

\begin{figure*}[ht]
    \centering
    \includegraphics[width=6.8in]{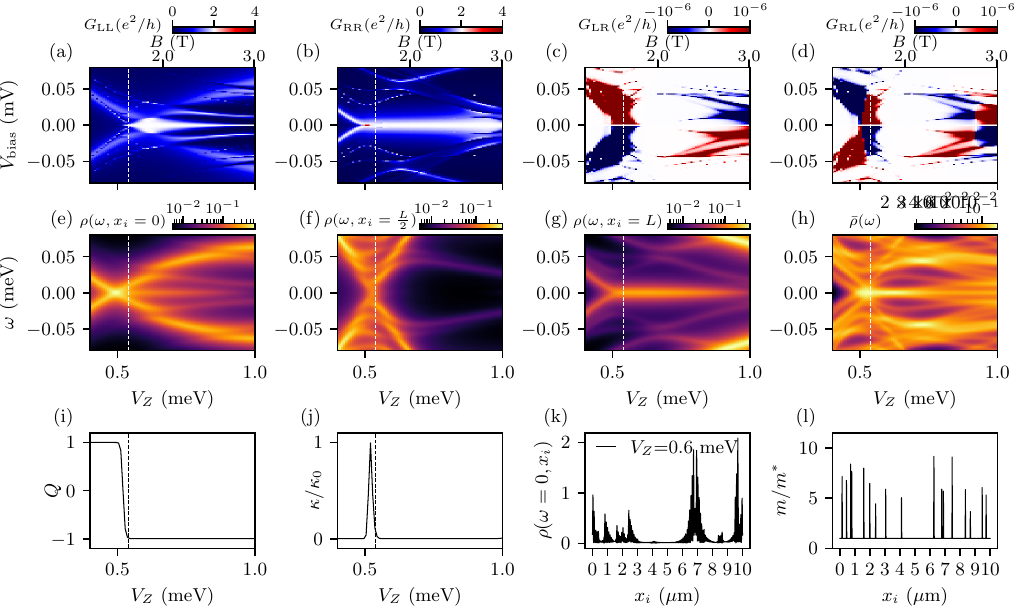}
    \caption{
        Long wire limit $L=10~\mu$m with $n$=19 and $k$=9.
    Same captions as Fig.~\ref{fig:n19L10}.
    }
    \label{fig:1935}
\end{figure*}

\begin{figure*}[ht]
    \centering
    \includegraphics[width=6.8in]{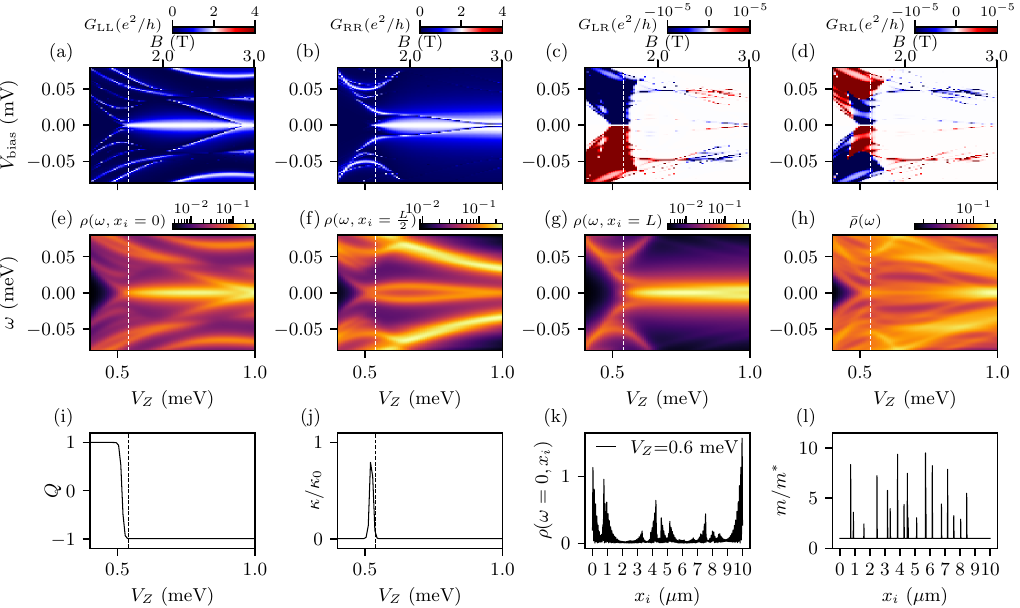}
    \caption{
    Long wire limit $L=10~\mu$m with $n$=19 and $k$=9.
    Same captions as Fig.~\ref{fig:n19L10}.
    }
    \label{fig:1974}
\end{figure*}

\begin{figure*}[ht]
    \centering
    \includegraphics[width=6.8in]{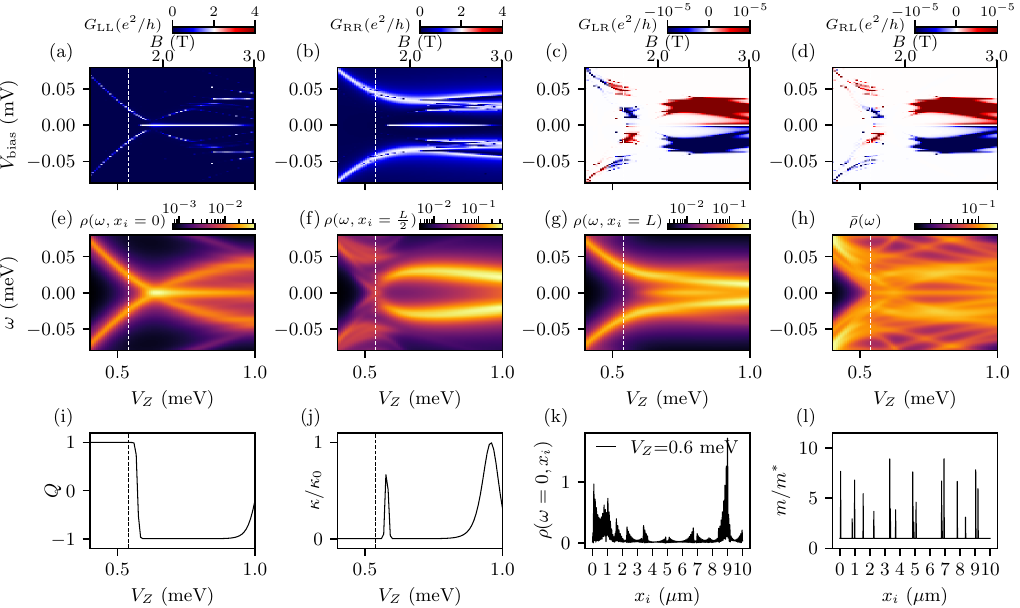}
    \caption{
        Long wire limit $L=10~\mu$m with $n$=19 and $k$=9.
    Same captions as Fig.~\ref{fig:n19L10}.
    }
    \label{fig:1977}
\end{figure*}

\begin{figure*}[ht]
    \centering
    \includegraphics[width=6.8in]{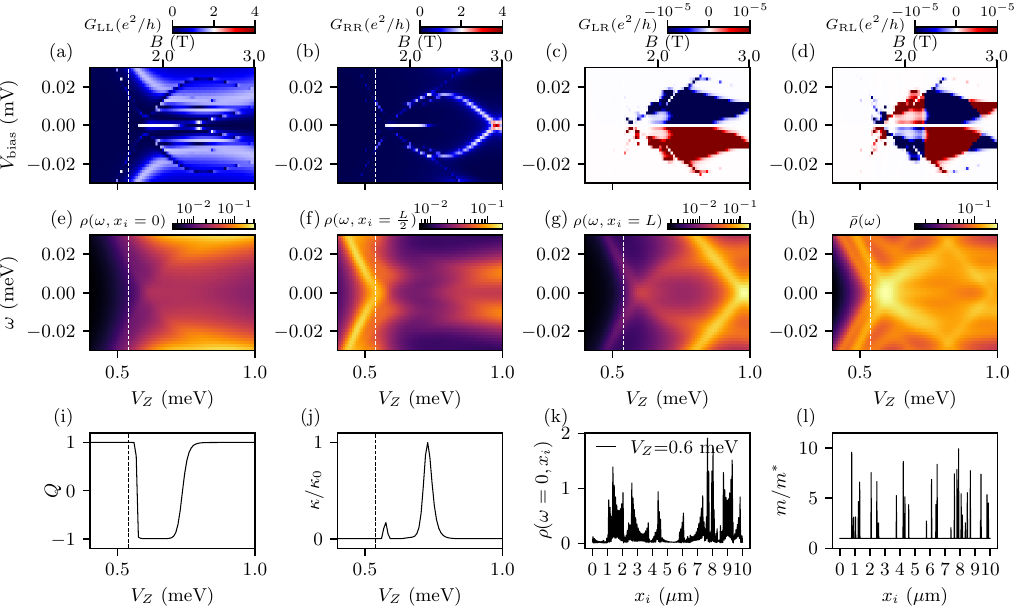}
    \caption{
        Long wire limit $L=10~\mu$m with $n$=32 and $k$=9.
    Same captions as Fig.~\ref{fig:n32L10}.
    }
    \label{fig:3229}
\end{figure*}

\begin{figure*}[ht]
    \centering
    \includegraphics[width=6.8in]{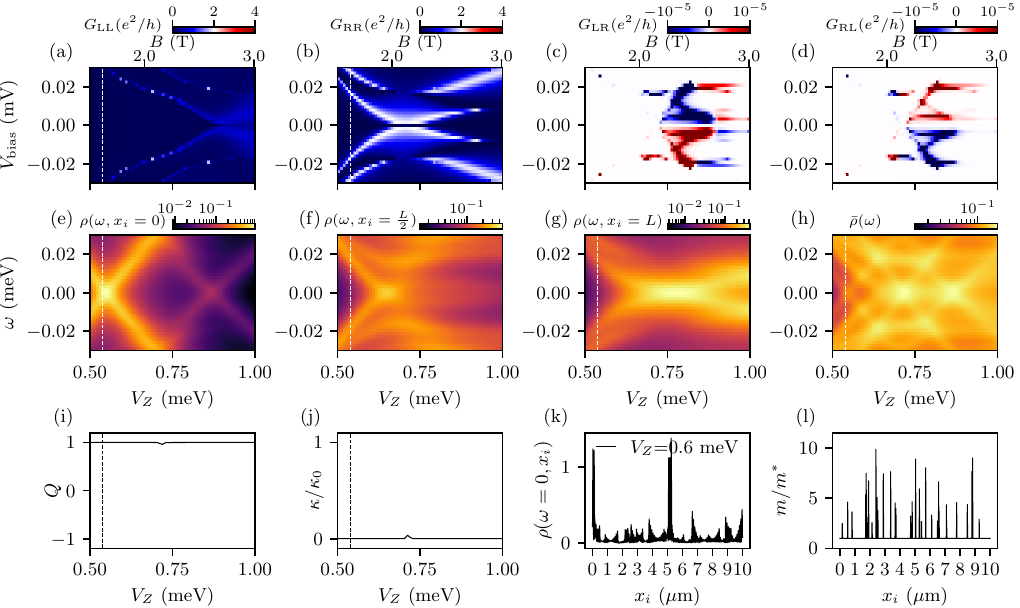}
    \caption{
        Long wire limit $L=10~\mu$m with $n$=32 and $k$=9.
    Same captions as Fig.~\ref{fig:n32L10}.
    }
    \label{fig:3268}
\end{figure*}

We present more results in a 10-micron wire 
in (1) the pristine limit with $n=5$ (Fig.~\ref{fig:512} - Fig.~\ref{fig:584});
(2) weak disorder regime with $n=9$ (Fig.~\ref{fig:922} - Fig.~\ref{fig:933}); 
(3) weak disorder regime with $n=19$ (Fig.~\ref{fig:1935} - Fig.~\ref{fig:1977});
(4) intermediate disorder regime with $n=32$ (Fig.~\ref{fig:3229} - Fig.~\ref{fig:3268});




\clearpage
\section{More benchmark results on the four indicators}\label{app:indicator}
\begin{figure*}[ht]
    \centering
    \includegraphics[width=6.8in]{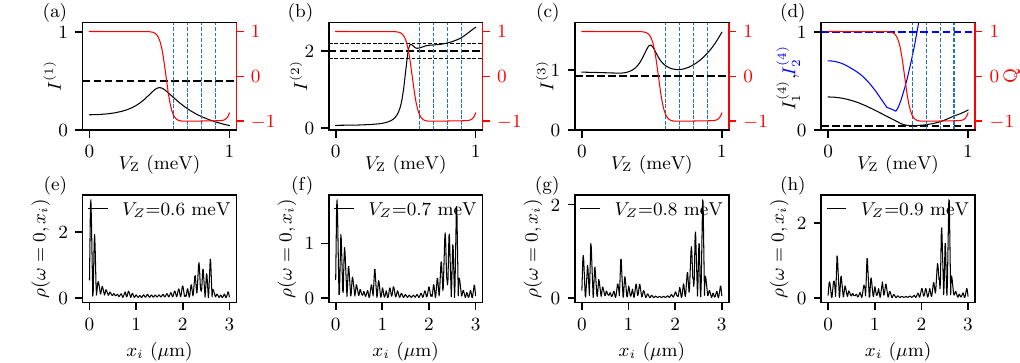}
    \caption{
    The benchmark of four indicators on the weak disorder 3-micron wire with $n=6$ and $k=9$ in Fig.~\ref{fig:649}. 
    Same captions as Fig.~\ref{fig:pristine_metrics}.
    }
    \label{fig:649_metrics}
\end{figure*}

\begin{figure*}[ht]
    \centering
    \includegraphics[width=6.8in]{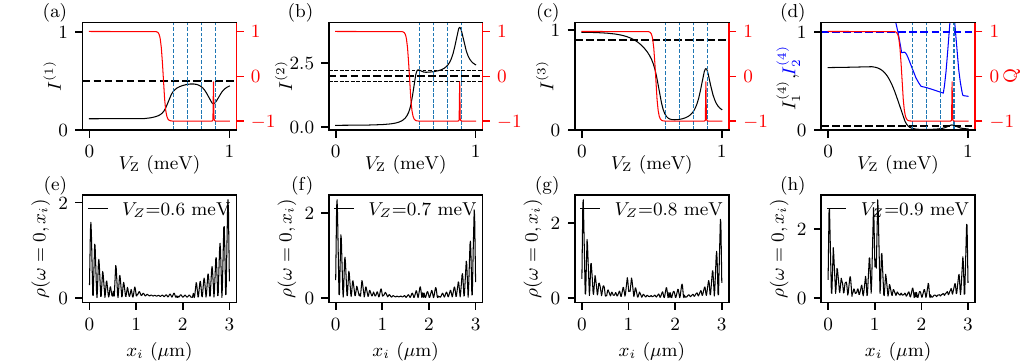}
    \caption{
    The benchmark of four indicators on the weak disorder 3-micron wire with $n=6$ and $k=9$ in Fig.~\ref{fig:658}. 
    Same captions as Fig.~\ref{fig:pristine_metrics}.
    }
    \label{fig:658_metrics}
\end{figure*}

\begin{figure*}[ht]
    \centering
    \includegraphics[width=6.8in]{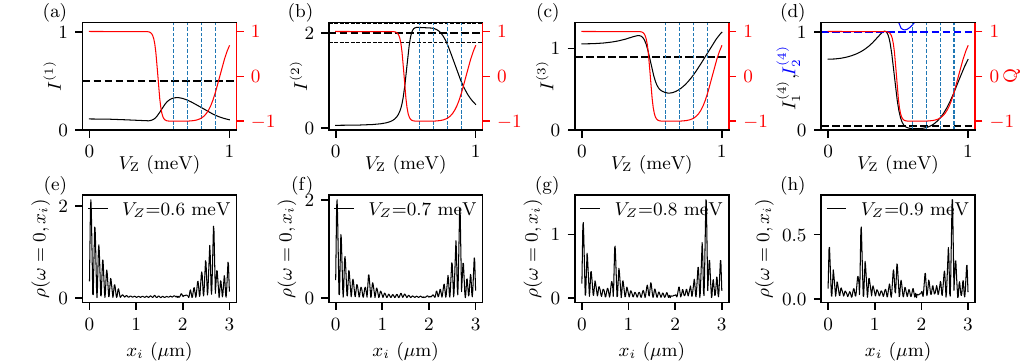}
    \caption{
    The benchmark of four indicators on the weak disorder 3-micron wire with $n=7$ and $k=9$ in Fig.~\ref{fig:703}. 
    Same captions as Fig.~\ref{fig:pristine_metrics}.
    }
    \label{fig:703_metrics}
\end{figure*}

\begin{figure*}[ht]
    \centering
    \includegraphics[width=6.8in]{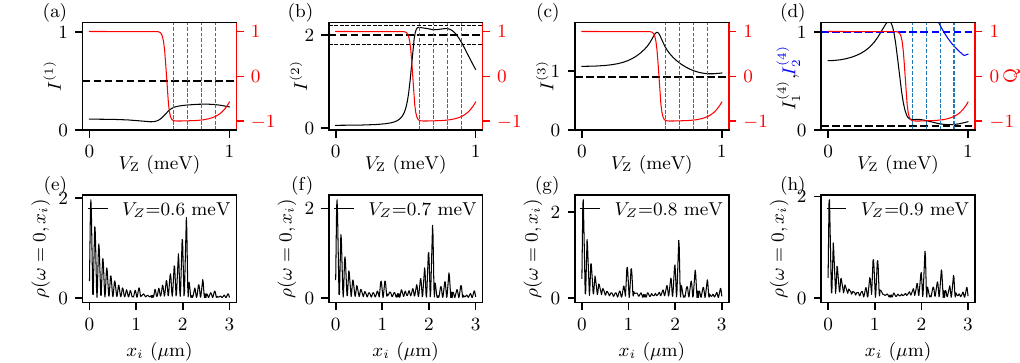}
    \caption{
    The benchmark of four indicators on the weak disorder 3-micron wire with $n=9$ and $k=9$ in Fig.~\ref{fig:971}. 
    Same captions as Fig.~\ref{fig:pristine_metrics}.
    }
    \label{fig:971_metrics}
\end{figure*}

\begin{figure*}[ht]
    \centering
    \includegraphics[width=6.8in]{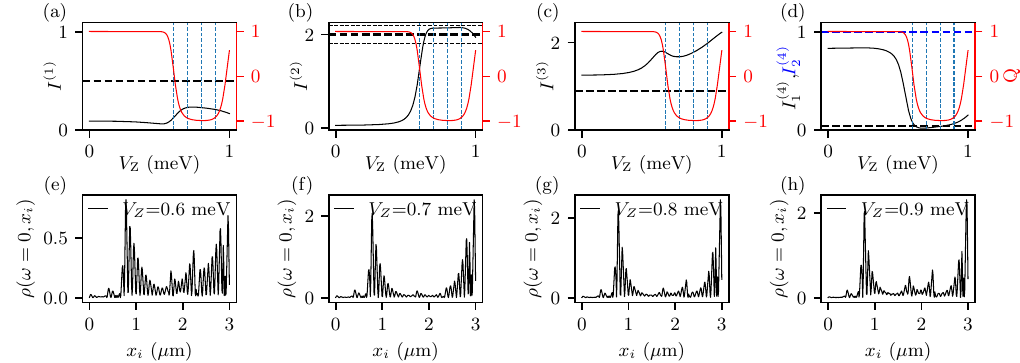}
    \caption{
    The benchmark of four indicators on the weak disorder 3-micron wire with $n=10$ and $k=9$ in Fig.~\ref{fig:1001}. 
    Same captions as Fig.~\ref{fig:pristine_metrics}.
    }
    \label{fig:1001_metrics}
\end{figure*}

In this section, we provide more benchmark results for the four indicators from the weak disorder regime with $n=6$ (Figs.~\ref{fig:649_metrics}-\ref{fig:703_metrics}) to the intermediate disorder regime (Fig.~\ref{fig:971_metrics}), and strong disorder regime with $n=10$ (Fig.~\ref{fig:1001_metrics}).

\clearpage
\section{More results showing localized MZMs without quantized ZBCP}\label{app:localized_MZM}
\begin{figure*}[ht]
    \centering
    \includegraphics[width=6.8in]{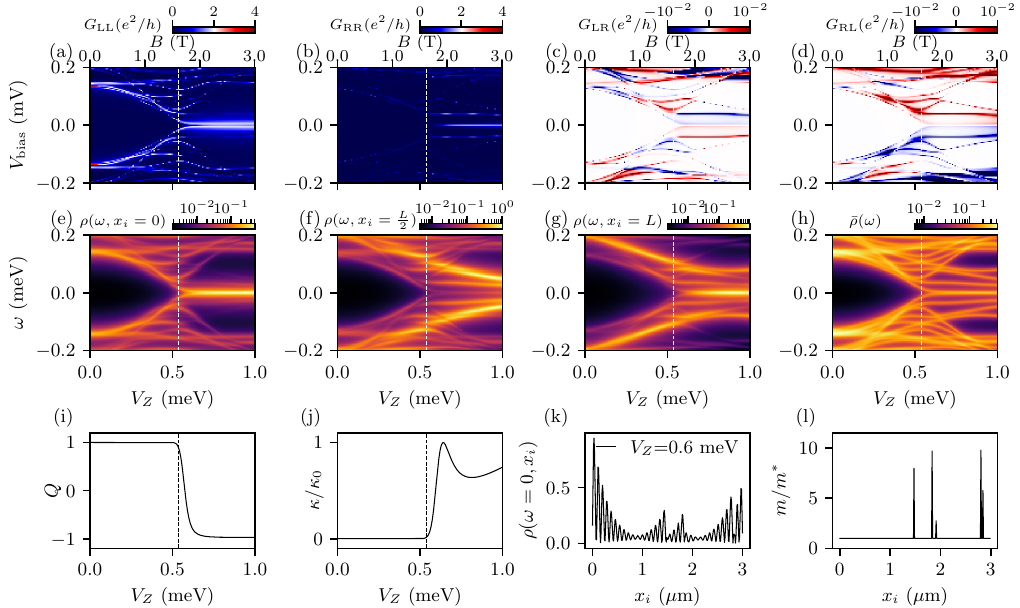}
    \caption{
    An example of localized MZMs without quantized ZBCP in a 3-micron wire with $n=5$ and $k=9$.
    Same captions as Fig.~\ref{fig:746}.
    }
    \label{fig:550}
\end{figure*}

\begin{figure*}[ht]
    \centering
    \includegraphics[width=6.8in]{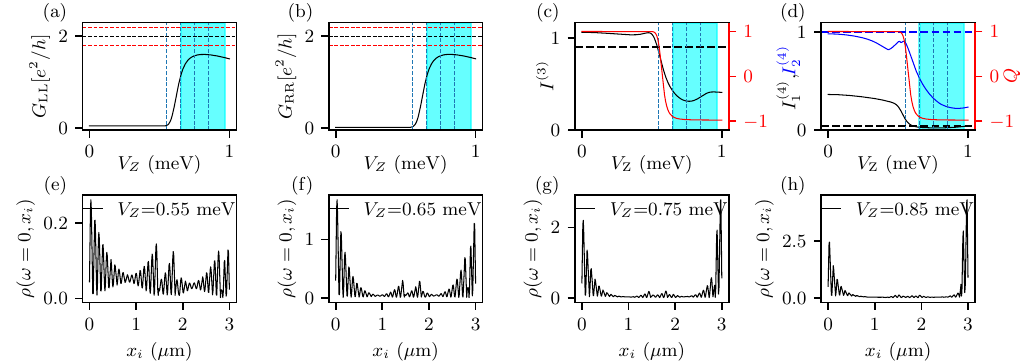}
    \caption{
    The benchmark of $I^{(3)}$ and $I_{1,2}^{(4)}$ on an example of localized MZMs without quantized ZBCP (cyan-shaded regions) in a 3-micron wire with $n=5$ and $k=9$ in Fig.~\ref{fig:550}.
    Same captions as Fig.~\ref{fig:746_cond_metrics}.
    }
    \label{fig:550_cond_metrics}
\end{figure*}

\begin{figure*}[ht]
    \centering
    \includegraphics[width=6.8in]{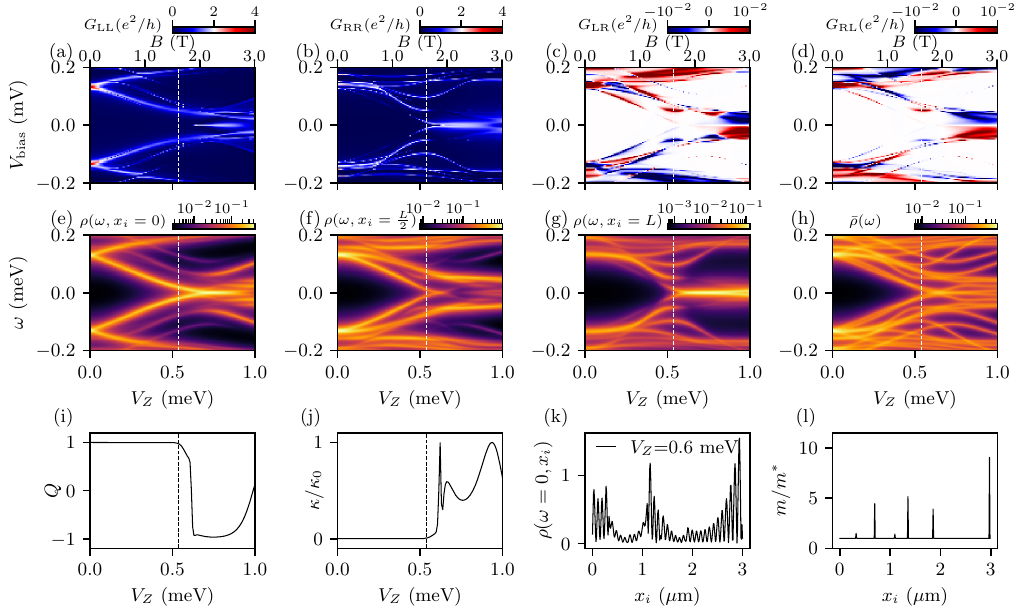}
    \caption{
    An example of localized MZMs without quantized ZBCP in a 3-micron wire with $n=6$ and $k=9$.
    Same captions as Fig.~\ref{fig:746}.
    }
    \label{fig:604}
\end{figure*}

\begin{figure*}[ht]
    \centering
    \includegraphics[width=6.8in]{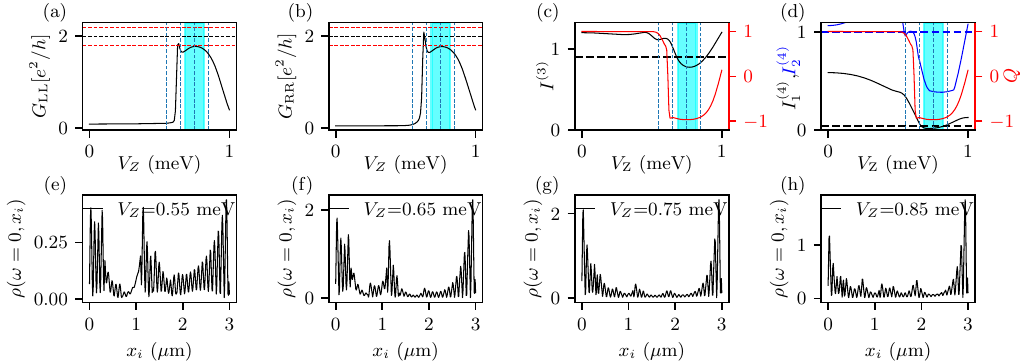}
    \caption{
    The benchmark of $I^{(3)}$ and $I_{1,2}^{(4)}$ on an example of localized MZMs without quantized ZBCP (cyan shaded regions) in a 3-micron wire with $n=6$ and $k=9$ in Fig.~\ref{fig:604}.
    Same captions as Fig.~\ref{fig:746_cond_metrics}.
    }
    \label{fig:604_cond_metrics}
\end{figure*}

In this section, we present more results that show isolated localized MZMs but without manifesting quantized ZBCP in the strong disorder limit from Fig.~\ref{fig:550} to Fig.~\ref{fig:604_cond_metrics}.

\clearpage
\section{More results showing quantized ZBCP without localized MZMs}\label{app:quantized_ZBCP}
\begin{figure*}[ht]
    \centering
    \includegraphics[width=6.8in]{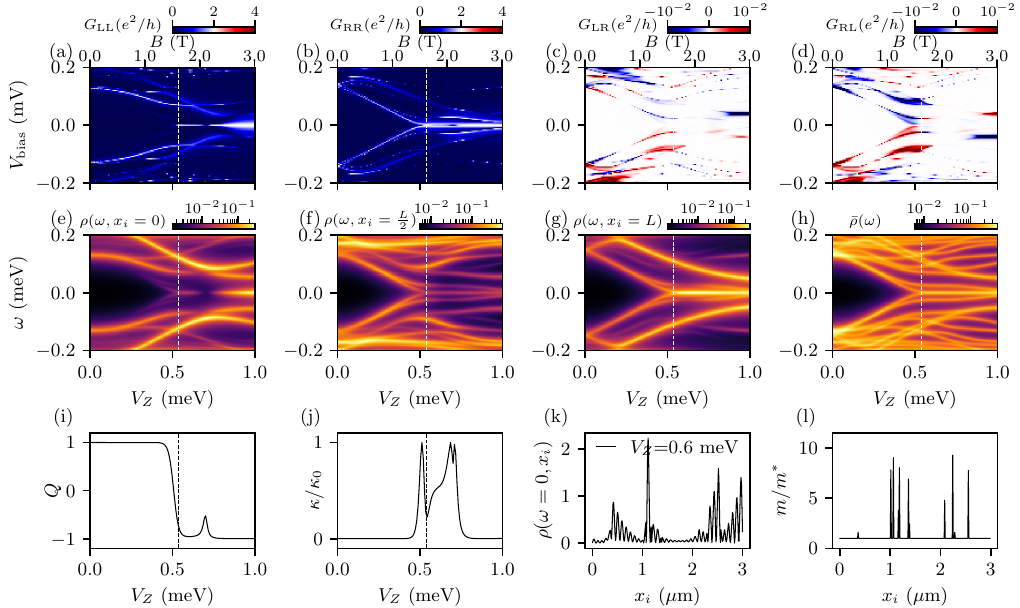}
    \caption{
    An example of quantized ZBCP without localized MZMs in a 3-micron wire with $n=11$ and $k=9$.
    Same captions as Fig.~\ref{fig:622}.
    }
    \label{fig:1163}
\end{figure*}

\begin{figure*}[ht]
    \centering
    \includegraphics[width=6.8in]{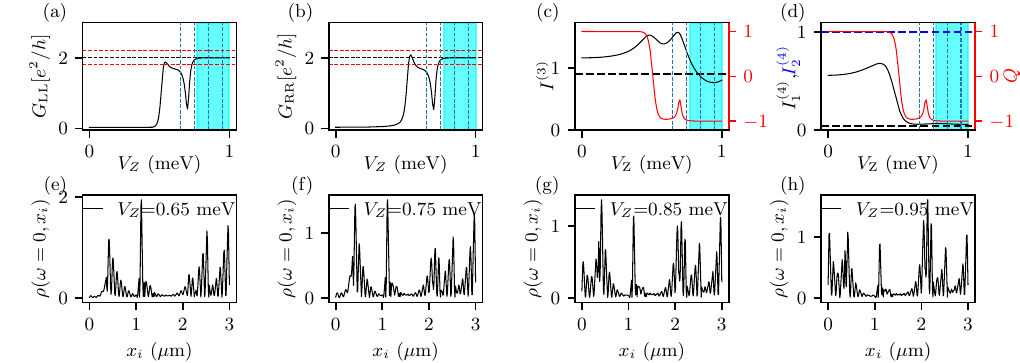}
    \caption{
    The benchmark of $I^{(3)}$ and $I_{1,2}^{(4)}$ on an example of quantized ZBCP without localized MZMs (cyan-shaded regions) in a 3-micron wire with $n=11$ and $k=9$ in Fig.~\ref{fig:1163}.
    Same captions as Fig.~\ref{fig:622_cond_metrics}.
    }
    \label{fig:1163_cond_metrics}
\end{figure*}

\begin{figure*}[ht]
    \centering
    \includegraphics[width=6.8in]{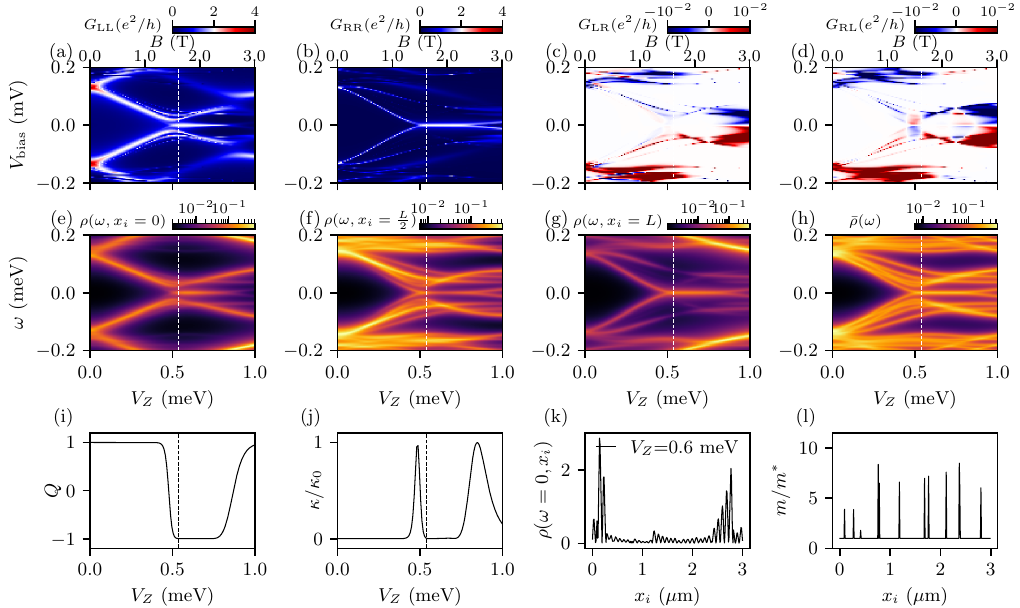}
    \caption{
    An example of quantized ZBCP without localized MZMs in a 3-micron wire with $n=12$ and $k=9$.
    Same captions as Fig.~\ref{fig:622}.
    }
    \label{fig:1207}
\end{figure*}

\begin{figure*}[ht]
    \centering
    \includegraphics[width=6.8in]{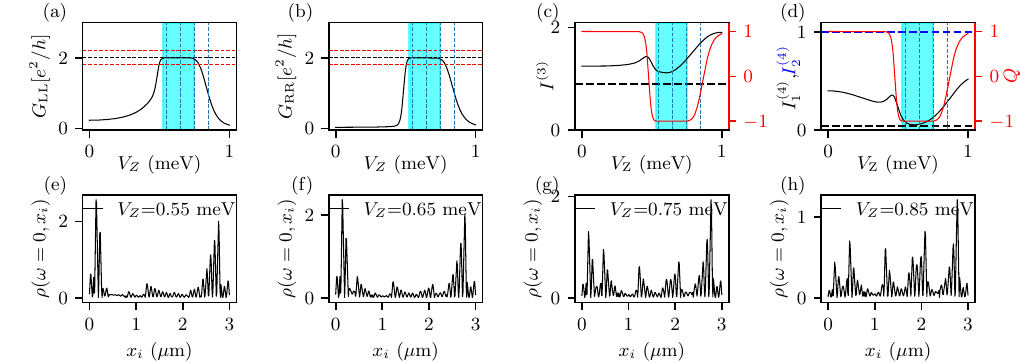}
    \caption{
    The benchmark of $I^{(3)}$ and $I_{1,2}^{(4)}$ on an example of quantized ZBCP without localized MZMs (cyan-shaded regions) in a 3-micron wire with $n=12$ and $k=9$ in Fig.~\ref{fig:1207}.
    Same captions as Fig.~\ref{fig:622_cond_metrics}.
    }
    \label{fig:1207_cond_metrics}
\end{figure*}

In this section, we present more results that manifest quantized ZBCP but without showing isolated localized MZMs in the strong disorder limit from Fig.~\ref{fig:1163} to Fig.~\ref{fig:1207_cond_metrics}.

\end{document}